\documentclass[cernpreprint, texlive=2023, UKenglish, texmf, orcidlogo]{atlasdoc}

\usepackage[subfigure=true, backend=biber]{atlaspackage}
\usepackage{atlasbiblatex}
 
\usepackage[process,jetetmiss]{atlasphysics}

\graphicspath{{logos/}{figures/}}
 
\usepackage{ANA-TOPQ-2019-01-PAPER-defs}

\AtlasTitle{Measurements of observables sensitive to colour reconnection in \ttbar events with the ATLAS detector at $\sqrt{s} = \SI{13}{\tev}$}
 
\AtlasVersion{2.3}
\AtlasAbstract{
A measurement of observables sensitive to effects
of colour reconnection in top-quark pair-production events is presented using \SI{139}{\ifb} of \SI{13}{\TeV} proton--proton collision data collected by the ATLAS detector at the LHC.
Events are selected by requiring exactly one isolated electron and one
isolated muon with opposite charge and two or three jets, where exactly two jets are required to be $b$-tagged.
For the selected events, measurements are presented for the charged-particle multiplicity, the scalar sum of the transverse momenta of the charged particles, and the same scalar sum
in bins of charged-particle multiplicity.  These observables are unfolded to the stable-particle level, thereby correcting for migration effects due to finite detector resolution, acceptance and efficiency effects. The particle-level measurements are compared with different colour reconnection models in Monte Carlo generators.
These measurements disfavour some of the colour reconnection models and provide inputs to future optimisation of the parameters in Monte Carlo generators.
}
 
\author{The ATLAS Collaboration}
 
\AtlasRefCode{TOPQ-2019-01}

\AtlasJournal{Eur. Phys. J. C}
\AtlasJournalRef{Eur. Phys. J. C 83, 518 (2023)}
\AtlasDOI{10.1140/epjc/s10052-023-11479-x}
\PreprintIdNumber{CERN-EP-2022-167}

\hypersetup{pdftitle={ATLAS document},pdfauthor={The ATLAS Collaboration}}
 
\begin{document}
 
\maketitle
 
\tableofcontents

\section{Introduction}
\label{sec:intro}

Monte Carlo event generators are of the utmost importance for experimental particle
physics as a tool providing a source of simulated events for analysis
design, signal and background modelling, unfolding detector effects, and training of multivariate discriminants.
A rich set of measurements provided by the Large Hadron Collider (LHC) experiments can be used to probe both soft and hard quantum chromodynamics (QCD), and thereby test many features of the Monte Carlo (MC) event generators. Even though the overall performance of the event generators is good, the understanding of some phenomena is insufficiently understood or can be significantly improved. One of these is the modelling of colour reconnection (CR). None of the models describing CR in hadron collisions is derived from first principles of QCD.
Measurements are required to exclude certain models or to constrain their parameter space.
 
The simulation of the hard-scattering process is accompanied by the so-called underlying event (UE) consisting of multiple parton interactions (MPIs), initial- and final-state radiation (ISR and FSR), and remnants of the interacting protons.
According to QCD the strong interactions between quarks and gluons
are related to their colour charges.
In order to trace the colour information in an event, MC event generators make use of the leading-colour (LC) approximation~\cite{Nagy:2008ns,BUCKLEY:2011145}. Here, each quark is colour-connected to only one other parton (quark or gluon), while gluons are each connected to two other partons, since they carry both a colour and an anticolour charge.
Additional colour lines are added to the final state by each MPI system, leading to
a non-negligible possibility of phase-space overlaps between final states
from different MPI systems and the hard-scattering process~\cite{Ortiz:2013yxa, Christiansen:2015yqa}. Colour reconnection~\cite{Sjostrand:2013cya} models introduce a mechanism for reassigning colour connections between partons to resolve ambiguities between overlapping colour lines in space and time, and thus allow colour topologies different from those in the simple LC approach.
The interactions and interference among the different processes leading to a given generated event are poorly understood and several models are implemented in the three parton-shower MC programs, \PYTHIA[8]~\cite{Sjostrand:2007gs}, \HERWIG[7]~\cite{Bellm:2015jjp}, and  \SHERPA[2]~\cite{Bothmann:2019yzt}.

In general, these models define some momentum-based measure of the colour configuration of partons in the event. In their workflow, the programs execute changes to the colour connections to reduce the defined measure and thereby correct for deficiencies of the LC approximation and allow colour connections between different MPIs.
The modified colour configuration effectively leads to a reduction of the energy available to produce new particle--antiparticle pairs in the space between partons. Thus, the particle multiplicity is reduced and the produced particles carry, on average, more momentum than they would have in events without CR.
Details can be found in Refs.~\cite{Christiansen:2015yqa, Bellm:2019wrh} and references therein.
 
An unresolved question is the involvement of the top quark and its decay products in CR.
The typical hadronisation scale is about \SI{1}{\femto\metre}, while
the average transverse decay length of top quarks is \SI{0.2}{\femto\metre}~\cite{Zyla:2020zbs}.
Thus, it is possible that top quarks interact with the colour fields stretching
between partons in the final state~\cite{Argyropoulos:2014zoa}.
However, in the current implementation of CR models, the reconnection algorithm is always performed before
the top quark decays, i.e.\ only the top quark itself takes part in the CR machinery, not its decay products.
 
In recent top-quark mass (\mtop) measurements, which reach a precision well below $1\,\GeV$~\cite{TOPQ-2017-03, CMS-TOP-17-008}, the uncertainty in the CR modelling
is estimated to be as large as \SI{400}{\MeV}~\cite{CMS-TOP-17-008}. Thus, this uncertainty is becoming increasingly important as direct measurements of \mtop become more precise.
 
Several measurements sensitive to the modelling of the UE and (partially) to CR in different scattering processes, e.g.\ minimum-bias events~\cite{STDM-2016-07}, \Zboson boson production~\cite{CMS-FSQ-16-008, STDM-2011-42, STDM-2017-28} and top-quark--antiquark (\ttbar) production~\cite{CMS-TOP-17-015} events,
have been performed at the LHC. However, dedicated measurements of CR in top-quark events have not been performed yet. This is particularly relevant since a well-defined prescription of the uncertainty due to CR is needed for top-quark mass measurements.
 
This article presents measurements of normalised differential cross-sections of \ttbar production as a function of observables sensitive to CR effects, namely the charged-particle multiplicity outside of jets (\nch), excluding leptons from the top-quark decay; the scalar sum of the transverse momenta of these charged particles (\sumptch); and the double-differential measurement of the two quantities (\sumptch in bins of \nch). The last of these measurements contains partial information about the `standard' observable which is often used in CR and UE measurements: the average transverse momentum of charged particles as a function of the charged-particle multiplicity.
The observables are corrected for tracks from additional inelastic $pp$ collisions in the same or neighbouring bunch crossings (\pileup), for tracks from secondary vertices, and for the effect of tracking inefficiencies.
The distributions of the corrected observables are unfolded to the particle level in a fiducial phase space chosen to maximise the sensitivity to CR effects. The fiducial cross-section is obtained by integrating the differential cross-section.
 
In proton--proton collisions, the dominant top-quark production process is \ttbar production via the strong interaction. As a result of top-quark and top-antiquark decays, two on-shell $W$ bosons emerge in \ttbar events.
In order to select a nearly background-free sample of \ttbar events, this analysis
targets events in which one $W$ boson decays in the $W\rightarrow e\nu$ channel and
the second $W$ boson decays via $W\rightarrow \mu\nu$.\footnote{
Events involving $W\rightarrow \tau\nu$ decays with a subsequent decay of the
$\tau$-lepton to either $e\nu_e\nu_\tau$ or $\mu\nu_\mu\nu_\tau$ are included
in the signal.}
Excluding dilepton final states with two charged leptons of the same flavour ($e^+e^-$ or $\mu^+\mu^-$)
essentially eliminates the background arising from $Z$+jets production. Thus, the only relevant backgrounds are single top-quark production in the $tW$ channel and events where jets are misidentified as leptons.
The experimental signature of the selected events is given by exactly one prompt isolated electron, exactly one prompt isolated muon, and two or three hadronic jets -- two of which each originate from a $b$-quark and are called \bjets.
The selected events were recorded with the ATLAS detector~\cite{PERF-2007-01} during \RunTwo of the LHC in proton--proton collisions at a centre-of-mass energy of $\SI{13}{\tev}$. The resulting data set corresponds to an integrated luminosity of \SI{139}{\ifb}.


\section{Colour reconnection in \ttbar events}
\label{sec:theory}


In the following, the CR models implemented in the two multipurpose MC generators,
\PYTHIA[8] and \HERWIG[7], are described. The \SHERPA[2] event generator has an MPI model based on the \PYTHIA one, but without any CR model. CR is currently being implemented in the upcoming \SHERPA[3].
 
\subsection{Colour reconnection models in \PYTHIA[8]}
\label{subsec:CR_Pythia8}
 
\PYTHIA[8]~\cite{Bierlich:2022pfr} uses the so-called Lund string model for the hadronisation, which is based on the observation that at large distances the potential energy of colour sources increases linearly with their separation. As soon as the potential energy reaches values similar to hadron masses, the string breaks and creates a new quark--antiquark pair. The two string segments then begin to stretch and break again until all the energy has been converted into quark--antiquark pairs connected by short string segments, identified as hadrons.
 
Three general CR models are currently implemented in \PYTHIA[8]:
\begin{enumerate}
\item \textbf{MPI-based model (CR0)}\\
Starting from the lowest-\pT interaction in a set of MPIs
the reconnection probability for an MPI-system with transverse
momentum \pT is given by:
\begin{eqnarray}
P_{\mathrm{rec}}(\pT) &=& \frac{(R_{\mathrm{rec}}p_{\mathrm{T}0})^2}{(R_{\mathrm{rec}}p_{\mathrm{T}0})^2 + \pT^2} \,, \nonumber
\label{eq:rec_prob}
\end{eqnarray}
where $R_{\mathrm{rec}}$ is a phenomenological parameter with a value in the `colour reconnection range', i.e.\ $0 \leq R_{\mathrm{rec}} \leq 10 $, and $p_{\text{T}0}$ is an energy-dependent dampening parameter used for MPIs. The latter parameter regularises the partonic cross-section to avoid divergence at low \pt and it depends on the centre-of-mass energy, $\sqrt{s} = E_{\text{CM}}$:
\begin{equation*}
p_{\text{T}0} = p_{\mathrm{T}0}^{\text{ref}} \left( \frac{E_{\text{CM}}}{E_{\text{CM}}^{\text{ref}}} \right)^{E_{\text{CM}}^{\text{pow}}} \,,
\label{eq:MPI_pt}
\end{equation*}
where $p_{\mathrm{T}0}^{\text{ref}}$ is the value of $p_{\text{T}0}$ at a reference energy $E_{\text{CM}}^{\text{ref}}$, and $E_{\text{CM}}^{\text{pow}}$ is a tunable parameter.
For each MPI system that undergoes a reconnection, partons from lower-\pT MPI systems are added to the colour dipoles defined by the higher-\pT MPI system, in a way that minimises the total colour-string length,
defined as the sum of the logarithm of normalised squared invariant masses of all two parton combinations in the event.
This model is used in all ATLAS simulated samples generated by \PYTHIA[8] with the A14 set of tuned parameters~\cite{ATL-PHYS-PUB-2014-021}.

\item \textbf{QCD-based model (CR1)}\\
This model is based on the CR0 model but uses a more complete treatment of the QCD multiplet structure which allows
reconnections of dipoles to produce structures with three (anti)colour indices, referred to as `junctions', thereby enhancing the production of (anti)baryons. Only reconnections which lower the string length are performed.
 
\item \textbf{Gluon-move model (CR2)}\\
In the gluon-move model, reconnections are performed in the same way as in the CR0 model, but only gluons are considered
for reconnection. For each gluon, all the reconnections to all MPI systems are considered
(not only the ones for softer MPIs), so in principle the colour connections from the hard interaction can be affected more significantly than in the default model. Again, only reconnections which lower the string length are performed.
\end{enumerate}
 
In all three models the top-quark lifetime is long enough to avoid colour reconnection
of the top-quark decay products with the rest of the event.
Only the top quark itself is involved in the CR process and the top-quark decays are performed after the colour reconnections have taken place. Thus, there is also no CR among the top-quark decay products.
The first tuning of these models to ATLAS data is documented in Ref.~\cite{ATL-PHYS-PUB-2017-008} and a recent tuning to CMS data is presented in Ref.~\cite{CMS-GEN-17-002}.
 
Additionally, a class of CR models that is designed to affect the top-quark decay products
separately is considered. These models are described in detail in Ref.~\cite{Argyropoulos:2014zoa}. Two different sets of gluons are considered:
gluons radiated from the top-quark decay products ($\{g_{t}\}$) and gluons from the rest of the event ($\{ g_{r} \}$). Performing colour exchanges in different ways leads to five models:
\begin{enumerate}
\item \textbf{Forced random (TCR1)}\\
A gluon from $\{g_{t}\}$ is forced to exchange colours with a random gluon from the other set, $\{ g_{r} \}$.
\item \textbf{Forced nearest (TCR2)}\\
A gluon from $\{g_{t}\}$ is forced to exchange colours with the gluon from $\{g_{r}\}$ that minimises $m^2(g_t,g_r)$.
\item \textbf{Forced farthest (TCR3)}\\
A gluon from $\{g_{t}\}$ is forced to exchange colours with the gluon from $\{g_{r}\}$ that maximises $m^2(g_t,g_r)$.
\item \textbf{Forced smallest $\Delta\lambda$ (TCR4)}\\
A gluon from $\{g_{t}\}$ is forced to exchange colours with the gluon from $\{g_{r}\}$ for which the change
in $\lambda$ (available rapidity range of particle production) is smallest.
\item \textbf{Smallest $\Delta\lambda$ (TCR5)}\\
This is the same as the previous model, except that gluons exchange colours only if $\Delta\lambda < 0$.
\end{enumerate}
For all of these models a phenomenological strength parameter $s$ is used, where each gluon from the set $\{g_t\}$ is tested for reconnection with a probability $s$.

\subsection{Colour reconnection models in \HERWIG[7]}
\label{subsec:CR_Herwig7}

\HERWIG[7] uses the cluster model~\cite{Webber:1983if} to model the hadronisation of quarks into hadrons. After the parton-shower calculation, gluons are
split into quark--antiquark pairs, and a cluster is formed from
each colour-connected pair of quarks. Before hadrons are produced
from clusters, CR can modify the configuration of the clusters.
 
Three general CR models are currently implemented in \HERWIG[7]:

\begin{itemize}
\item \textbf{Plain CR}\\
In the plain CR model~\cite{Gieseke:2012ft}, quarks from two clusters can be rearranged
into two alternative clusters. A cluster is randomly chosen from the list of clusters and compared with all other clusters on the list.
The rearrangement is performed for the combination of clusters with the
lowest sum of the masses. The reconnection is accepted with a probability $p_{\mathrm{R}}$, which is the only parameter of this model.
This model typically leads to clusters with smaller invariant
mass and thus the overall activity in the UE is reduced. This model is used in all ATLAS samples generated by \HERWIG[7], with different versions using different sets of tuned parameters (`tunes') for the UE and the CR. The H7-UE-MMHT tune is used in \HERWIG[7.0.4], the H7.1-Default tune is used in \HERWIG[7.1.3], and the H7.2-Default tune is used in \HERWIG[7.2.1].
Also, in the latest version the MPI model was improved~\cite{Bellm:2019icn}.
 
\item \textbf{Statistical CR}\\
The statistical CR model uses the Simulated Annealing algorithm~\cite{Kirkpatrick:1983zz} to find the configuration of clusters that results in the absolute lowest value of the colour length, defined as the sum of the squared invariant masses of all clusters in the event.
In this model the only possible reconnections which are not allowed are those connecting
quarks and antiquarks produced in the non-perturbative splitting of gluons, which
would lead to the production of a colour-singlet object~\cite{Reichelt:2017hts}.

\item \textbf{Baryonic CR}\\
In contrast to the other two models, the baryonic CR model~\cite{Gieseke:2017clv} uses
a simple geometric picture of nearest neighbours, and the algorithm tries to find quarks that populate approximately the same phase-space region, based on their rapidity $y$.
Suitable combinations of clusters are then categorised as mesonic or baryonic, and reconnected with probabilities $p_{\mathrm{R}}$ or $p_{\mathrm{B}}$ respectively. This model is only available since \HERWIG[7.1.5].

\end{itemize}


\section{ATLAS detector}
\label{sec:detector}

\newcommand{\AtlasCoordFootnote}{
ATLAS uses a right-handed coordinate system with its origin at the nominal interaction point (IP)
in the centre of the detector and the \(z\)-axis along the beam pipe.
The \(x\)-axis points from the IP to the centre of the LHC ring,
and the \(y\)-axis points upwards.
Cylindrical coordinates \((r,\phi)\) are used in the transverse plane,
\(\phi\) being the azimuthal angle around the \(z\)-axis.
The pseudorapidity is defined in terms of the polar angle \(\theta\) as \(\eta = -\ln \tan(\theta/2)\).
Angular distance is measured in units of \(\Delta R \equiv \sqrt{(\Delta\eta)^{2} + (\Delta\phi)^{2}}\).}

The ATLAS detector~\cite{PERF-2007-01} at the LHC covers nearly the entire solid angle around the collision point.\footnote{\AtlasCoordFootnote}
It consists of an inner tracking detector surrounded by a thin superconducting solenoid, electromagnetic and hadron calorimeters,
and a muon spectrometer incorporating three large superconducting air-core toroidal magnets.
 
The inner-detector system (ID) is immersed in a \SI{2}{\tesla} axial magnetic field
and provides charged-particle tracking in the range \(|\eta| < 2.5\).
The high-granularity silicon pixel detector covers the vertex region and typically provides four measurements per track,
the first hit normally being in the insertable B-layer (IBL) installed before Run~2~\cite{ATLAS-TDR-19,PIX-2018-001}.
It is followed by the silicon microstrip tracker (SCT), which usually provides eight measurements per track.
These silicon detectors are complemented by the transition radiation tracker (TRT),
which enables radially extended track reconstruction up to \(|\eta| = 2.0\).
The TRT also provides electron identification information
based on the fraction of hits (typically 30 in total) above a higher energy-deposit threshold corresponding to transition radiation.
 
The calorimeter system covers the pseudorapidity range \(|\eta| < 4.9\).
Within the region \(|\eta|< 3.2\), electromagnetic calorimetry is provided by barrel and
endcap high-granularity lead/liquid-argon (LAr) calorimeters,
with an additional thin LAr presampler covering \(|\eta| < 1.8\)
to correct for energy loss in material upstream of the calorimeters.
Hadron calorimetry is provided by the steel/scintillator-tile calorimeter,
segmented into three barrel structures within \(|\eta| < 1.7\), and two copper/LAr hadron endcap calorimeters.
The solid angle coverage is completed with forward copper/LAr and tungsten/LAr calorimeter modules
optimised for electromagnetic and hadronic energy measurements respectively.
 
The muon spectrometer (MS) comprises separate trigger and
high-precision tracking chambers measuring the deflection of muons in a magnetic field generated by the superconducting air-core toroidal magnets.
The field integral of the toroids ranges between \num{2.0} and \SI{6.0}{Tm}
across most of the detector.
A set of precision chambers covers the region \(|\eta| < 2.7\) with three layers of monitored drift tubes,
complemented by cathode-strip chambers in the forward region, where the background is highest.
The muon trigger system covers the range \(|\eta| < 2.4\) with resistive-plate chambers in the barrel, and thin-gap chambers in the endcap regions.
 
Interesting events are selected by the first-level trigger system implemented in custom hardware,
followed by selections made by algorithms implemented in software in the high-level trigger~\cite{TRIG-2016-01}.
The first-level trigger accepts events from the \SI{40}{\MHz} bunch crossings at a rate below \SI{100}{\kHz},
which the high-level trigger further reduces in order to record events to disk at about \SI{1}{\kHz}.
 
An extensive software suite~\cite{ATL-SOFT-PUB-2021-001} is used in data simulation, in the reconstruction
and analysis of real and simulated data, in detector operations, and in the trigger and data acquisition systems of the experiment.


\section{Samples of data and simulated events}
\label{sec:samples}

The analysis uses proton--proton ($pp$) collision data which were recorded with the ATLAS detector during Run~2 of the LHC in the years from 2015 to 2018 at a centre-of-mass energy of 13$\,\TeV$.
After applying data-quality requirements~\cite{DAPR-2018-01}, the combined data set corresponds to an integrated luminosity of $139\,\ifb$ with an uncertainty of $1.7\%$~\cite{ATLAS-CONF-2019-021}.
Split by years, the integrated luminosities are $3.22\,\ifb$ for 2015, $33.0\,\ifb$ for 2016, $44.3\,\ifb$ for 2017 and $58.5\,\ifb$ for 2018.
The integrated luminosity of the data set is derived following a methodology similar to that detailed in Ref.~\cite{DAPR-2013-01}. The LUCID-2 detector~\cite{LUCID-2} is used for the primary luminosity measurements.
At the high instantaneous luminosity reached at the LHC, there are additional \pileup collisions.
The average number of interactions per bunch crossing
increased from 13 to 36 from 2015 to 2018.
 
Events were selected online during data taking by single-electron or single-muon triggers~\cite{TRIG-2018-05,TRIG-2018-01}.
Multiple triggers were used to increase the selection efficiency. The lowest-threshold triggers utilised isolation requirements to reduce the trigger rate. The isolated-lepton triggers had \pT thresholds of $20\,\GeV$ for muons and $24\,\GeV$ for electrons in 2015 data, and $26\,\GeV$ for both lepton types in 2016, 2017 and 2018 data. They were complemented by other triggers with higher \pT thresholds, but no isolation requirements, to increase the trigger efficiency.
 
Large sets of events simulating \ttbar and background processes were produced with event generator programs based on the Monte Carlo (MC) method to model the recorded and selected data. After event generation, the response of the ATLAS detector was simulated using either the \GEANT toolkit~\cite{Agostinelli:2002hh} with a full detector model or a fast simulation~\cite{SOFT-2010-01} which employs a parameterisation of the calorimeter response~\cite{ATL-PHYS-PUB-2010-013,}. Samples using the fast simulation were employed to estimate systematic uncertainties associated with the event generators.
 
To account for pile-up effects, simulated minimum-bias interactions were overlayed on the hard-scattering events and the resulting events were weighted to reproduce the observed pile-up distribution. The minimum-bias events are simulated using \PYTHIA[8.186]~\cite{Sjostrand:2007gs} with a set of tuned parameters called the A3 tune~\cite{ATL-PHYS-PUB-2016-017} and the \NNPDF[2.3lo] PDF set~\cite{Ball:2012cx}. To account for differences in the pile-up profile between individual years in a suitable way, simulated events representing the detector conditions in the years 2015 and 2016 were weighted to reproduce the distribution of the number of interactions averaged over each bunch crossing
observed in that data-taking period, while simulated events modelling the detector conditions in the years 2017 and 2018 were weighted to reproduce the actual distribution of the number of interactions per bunch crossing observed in that time.
Finally, the simulated events were reconstructed using the same software as applied to the collision data, the same event selection requirements were applied and the selected events were passed through the same analysis chain. More details about the samples of simulated events are provided in the following subsections.
 
\subsection{Simulation of \ttbar, $tW$ and $t\tbar{+}V$ production}
\label{sec:samples_top_quark}
Samples of events simulating \ttbar and $tW$ production were generated with the \POWHEGBOX[v2]~\cite{Nason:2004rx,Frixione:2007nw,Frixione:2007vw,Alioli:2009je,Alioli:2010xd,Re:2010bp,Frederix:2012dh} next-to-leading-order (NLO) matrix-element generator,
setting  $\mtop = 172.5\,\GeV$. The \NNPDF[3.0nlo] PDF set~\cite{Ball:2014uwa} implementing the five-flavour scheme was used. Parton showers, hadronisation and the underlying event were modelled using \PYTHIA[8.230] with the A14 tune~\cite{ATL-PHYS-PUB-2014-021} and the \NNPDF[2.3lo] PDF set.
The {\POWHEGBOX}\,+\,\PYTHIA generator set-up implements a matching scheme for their modelling of hard emissions that avoids double counting them. The matrix-element--to--parton-shower matching is steered by the \hdamp parameter, which controls the $\pt$ of the first additional gluon emission beyond the \LO Feynman diagram and therefore regulates the high-\pt emission against which the \ttbar system recoils. Harder emissions in the parton shower are then vetoed. Event generation was run with $\hdamp =1.5\times \mtop$~\cite{ATL-PHYS-PUB-2016-020}.
The renormalisation and factorisation scales were set dynamically on an event-by-event basis, namely to $\muR = \muF = \sqrt{\mtop^2 + \pt^2(t)}$ for \ttbar production and to $\muR = \muF = \mtop$ for $tW$ production.
Decays of bottom and charm hadrons were simulated using the \EVTGEN[1.6.0] program~\cite{Lange:2001uf}.
When generating $tW$ events, the diagram-removal scheme~\cite{Frixione:2008yi} was employed to handle the interference with \ttbar\ production~\cite{ATL-PHYS-PUB-2016-020}.

The cross-section of \ttbar production was scaled to the next-to-next-to-leading-order (NNLO) predictions by the \TOPpp[2.0] program (see Ref.~\cite{Czakon:2011xx} and references therein), which include the resummation of next-to-next-to-leading logarithmic (NNLL) soft-gluon terms. A value of $\sigma(\ttbar)=832^{+47}_{-51}\,$pb was used, assuming $\mtop=172.5\,\GeV$. The total cross-section for $tW$ production was computed at NLO in QCD with the addition of third-order corrections of soft-gluon emissions by resuming NNLL terms~\cite{Kidonakis:2021vob} and the simulated event sample was scaled to $\sigma(tW+\tbar W)=71.7\pm3.8\,$pb.

The production of a $\ttbar$ pair in association with a vector boson $V$ was modelled at NLO using the \MGNLO[2.3.3] generator~\cite{Alwall:2014hca} with the \NNPDF[3.0nlo] set of PDFs. The events were interfaced to \PYTHIA[8.210] for modelling of the parton showers, using the A14 tune and the \NNPDF[2.3lo] PDF set. The decays of bottom and charm hadrons were simulated using \EVTGEN[1.2.0].
Cross-sections were calculated at NLO QCD and NLO EW accuracy using
\MGNLO as reported in Ref.~\cite{deFlorian:2016spz}.
The predicted values are $0.88^{+0.09}_{-0.11}\,$pb and
$0.60^{+0.08}_{-0.07}\,$pb for $\ttbar{+}Z$ and $\ttbar{+}W$, respectively, where the
uncertainties were estimated from variations of \alphas and the renormalisation and
factorisation scales.

\subsection{Simulation of $Z+$jets and diboson production}
\label{subsec:Zjets_diboson}
Backgrounds from $Z$ boson production in association with jets, especially heavy-flavour jets, and on-shell diboson production ($WW$, $WZ$ and $ZZ$) were simulated with the \SHERPA[2.2.1] generator~\cite{Bothmann:2019yzt}. To simulate $Z+$jets (diboson) production, NLO-accurate matrix elements for up to two additional partons (one additional parton) and LO-accurate matrix elements for up to four (three) additional parton emissions were calculated with the \textsc{Comix}~\cite{Gleisberg:2008fv} and \OPENLOOPS[1]~\cite{Buccioni:2019sur,Cascioli:2011va,Denner:2016kdg} libraries. The matrix elements used for diboson production contain all diagrams with four electroweak vertices.
The default \SHERPA parton shower~\cite{Schumann:2007mg}, based on Catani--Seymour dipole factorisation, and the cluster hadronisation model~\cite{Winter:2003tt} were used. The samples employ the dedicated set of tuned parameters developed by the \SHERPA authors and the \NNPDF[3.0nlo] PDF set.
The NLO matrix elements of a given jet multiplicity were matched to the parton shower using a colour-exact variant of the MC@NLO algorithm~\cite{Hoeche:2011fd}. Different jet multiplicities were then merged into an inclusive sample using an improved CKKW matching procedure~\cite{Catani:2001cc,Hoeche:2009rj} which was extended to NLO accuracy using the MEPS@NLO prescription~\cite{Hoeche:2012yf}. The merging threshold was set to $20$~\GeV.
 
The $Z$+jets sample was normalised to an NNLO prediction~\cite{Anastasiou:2003ds} of the total cross-section, obtained with the \textsc{FEWZ} package~\cite{Melnikov:2006kv}. Considering the dilepton event selection used in the analysis, the simulation efficiency of the $Z+$jets samples was increased by forcing the $Z$ bosons to decay into a pair of charged leptons ($e^+e^-$, $\mu^+\mu^-$ or $\tau^+\tau^-$). In diboson events, at least one of the two bosons was required to decay leptonically. The simulated diboson event samples were normalised to the total cross-sections provided by \SHERPA at NLO in QCD.


\section{Object reconstruction and definitions of observables}
\label{sec:objects_observables}

Event candidates are identified by means of isolated electrons and muons, and jets, some of which are \btagged. In the following, the definitions of these reconstructed objects at detector level, together with the corresponding
objects reconstructed at particle level, as well as the event selection are described. In addition the analysis observables
are introduced.
 
\subsection{Detector-level object definitions}
\label{subsec:detectorselection}
Vertices are reconstructed from at least two ID tracks with $\pT > \SI{500}{\MeV}$.
The primary vertex of an event is defined as the vertex with the highest sum of $\pT^2$ of all associated ID
tracks~\cite{ATL-PHYS-PUB-2015-026}.
 
Tracks are reconstructed from the measured positions of `hits' in the ID caused by the passage of charged particles~\cite{PERF-2015-08}.
The track reconstruction consists of an iterative track-finding algorithm seeded by combinations of at least three silicon-detector
hits followed by a combinatorial Kalman filter~\cite{Fruhwirth:1987fm} to build track candidates based on hits compatible with the
extrapolated trajectory. Ambiguities between the track candidates are then resolved
by applying quality criteria to suppress combinations of hits unlikely to originate from a single charged particle. Finally, a high-resolution track fit is performed, resulting in more precise track reconstruction parameters.
Track candidates are required to pass the \textit{Tight Primary} selection~\cite{ATL-PHYS-PUB-2015-051}. In particular, they must have $\pt>500$~\MeV\ and be within $|\eta|<2.5$.
At least one hit in the two innermost layers is required if the extrapolated track crosses the sensitive region of an active sensor module. The number of hits in the pixel and SCT
detectors must be larger than 9 for $|\eta|\leq1.65$ or larger than 11 for $|\eta|>1.65$, with no more than 2 missing SCT hits
on a track if the respective SCT modules are operational. Additionally, track-to-vertex-matching criteria are applied to
mitigate the effect of \pileup tracks and improve the rejection of tracks from secondary interactions: the transverse
impact parameter calculated relative to the beam line must satisfy $|d_0|<\SI{0.5}{\milli\metre}$, and the longitudinal impact
parameter calculated relative to the event primary vertex must satisfy $|z_{0}\sin(\theta)|<\SI{0.5}{\milli\metre}$, where
$\theta$ is the polar angle of the track.
 
Electron candidates are reconstructed from clusters of energy deposited in the electromagnetic calorimeter with a matched track reconstructed in the ID~\cite{EGAM-2018-01}. The pseudorapidity of clusters, $\eta_{\mathrm{cluster}}$, is required to be in the range $|\eta_{\mathrm{cluster}}|<2.47$. However, clusters are excluded if they are in the transition region $1.37 < |\eta_{\mathrm{cluster}}| < 1.52$
between the barrel and endcap electromagnetic calorimeters. A likelihood-based method is used to simultaneously evaluate several properties of electron candidates, including shower shapes in the electromagnetic calorimeter, track quality, and transition radiation produced in the TRT. Electron candidates must
have $\pt>\SI{25}{\GeV}$ and satisfy the \textit{Tight} identification criteria as defined in Ref.~\cite{EGAM-2018-01}. Electrons are required to be isolated from other activity by applying
the \textit{Tight} requirements on the sum of nearby energy in the calorimeter and the sum of the momenta of
nearby tracks~\cite{EGAM-2018-01}.
 
Muon candidates are reconstructed by combining tracks in the MS with tracks in the ID~\cite{PERF-2015-10,MUON-2018-03}
and must have $|\eta|<2.5$, $\pt>\SI{25}{\GeV}$, satisfy the \textit{Medium} identification criteria, and pass the \textit{Tight} isolation requirements as defined in Ref.~\cite{MUON-2018-03}.
 
The tracks associated with electron and muon candidates must satisfy slightly different requirements on $d_0$, i.e.\ $d_0$ significance requirements of $|d_{0}/\sigma_{d_{0}}|<5.0$ for electrons and $|d_{0}/\sigma_{d_{0}}|<3.0$ for muons are imposed.
The lepton energy and momentum scales and resolutions are calibrated in data
by using $Z\rightarrow ee$~\cite{EGAM-2018-01} or $Z\rightarrow \mu\mu$ events and inspecting the
dilepton mass spectrum near the $Z$-boson peak.
 
Jets are reconstructed from particle-flow objects~\cite{PERF-2015-09} with the \antikt\ clustering algorithm~\cite{Cacciari:2008gp,Fastjet}
using a radius parameter of 0.4. Their energy is calibrated~\cite{JETM-2018-05}, and they must fulfil  $\pt > \SI{25}{\GeV}$ and
$|\eta|<2.5$. Jets with $\pt < \SI{60}{\GeV}$ and $|\eta|<2.5$ are required to pass a requirement on the jet-vertex-tagger (JVT) discriminant~\cite{PERF-2014-03} to suppress jets originating from pile-up collisions. The JVT-discriminant is required to be above $0.5$, which corresponds to an efficiency of $92\%$ for non-pile-up jets.
Jets containing $b$-hadrons are identified (\btagged) with the DL1r algorithm~\cite{FTAG-2018-01},
which uses a deep feed-forward neural network with several $b$-tagging algorithms as input~\cite{PERF-2012-04}. The algorithms exploit the impact parameters of charged-particle tracks, parameters of reconstructed secondary vertices and the topology of $b$- and $c$-hadron decays inside the jets. In order to strongly reduce the misidentification rate of $c$-jets and light-flavour ($u$, $d$ or $s$) jets, a specific working point of the DL1r algorithm was defined and
calibrated, using Run~2 data and the standard calibration technique, i.e.\ a likelihood-based
method in a sample highly enriched in \ttbar events ~\cite{FTAG-2018-01}. The $b$-tagging efficiency for jets that originate from the hadronisation of $b$-quarks is 70\% in simulated $\ttbar$ events.
 
To avoid double-counting objects satisfying more than one set of selection criteria, a procedure called \textit{overlap removal} is applied. Reconstructed objects are removed in the following order:
electrons sharing an ID track with a muon; jets within $\Delta R = 0.2$ of an electron, thereby avoiding double-counting electron
energy deposits as jets; electrons within $\Delta R = 0.4$ of a remaining jet, to reduce the impact of non-prompt electrons;
jets within $\Delta R = 0.2$ of a muon if they have less than three associated tracks; muons within $\Delta R = 0.4$ of a remaining jet, reducing the rate of non-prompt muons.
 
In order to build observables in this analysis, tracks within $\Delta R = 0.4$ of a jet or within $\Delta R = 0.01$ of the selected electron or muon are discarded.
 
\subsection{Particle-level object definitions}
\label{subsec:particle_level_objcts}
Particle-level objects are defined in simulated events by using only stable particles with a mean lifetime $\tau > \SI{30}{\pico\second}$,
using the following criteria.
 
Charged particles are required to have $\pt>\SI{500}{\MeV}$ and $|\eta|<2.5$. Charged particles that are generated through interactions with matter, referred to as secondary particles, are excluded.
 
Electrons and muons are required not to originate from a generated hadron in the MC event, either directly or through a $\tau$-lepton decay. This ensures that they originate from a $W$-boson decay without requiring a direct match.
The four-momenta of the bare leptons are modified (dressed) by adding the four-momenta of prompt photons within $\Delta R = 0.1$. The dressed leptons must then have $\pt>\SI{25}{\GeV}$ and $|\eta|<2.5$.
 
Jets are clustered using all stable particles, excluding those used in the definition of dressed leptons and prompt
neutrinos, using the \antikt\ algorithm with a radius parameter of 0.4. Jets are required to have $\pt>\SI{25}{\GeV}$
and $|\eta|<2.5$ and are identified as $b$-jets via ghost-matching~\cite{Cacciari:2008gn} to a $b$-hadron with $\pt >\SI{5}{\GeV}$.
 
To avoid double-counting, particle-level objects are subject to overlap removal criteria. Dressed muons and electrons with a separation $\Delta R < 0.4$ from a jet are excluded.
 
In order to obtain a track collection outside of jets, charged particles within $\Delta R = 0.4$ of a jet or within $\Delta R = 0.01$ of the electron or muon from the top-quark decay are excluded.
 
\subsection{Event selection}
\label{sec:selection}
Events are required to have at least one primary vertex, exactly one electron and one muon with $\pt > \SI{25}{\GeV}$ for the 2015 data-taking period
and $\pt > \SI{27}{\GeV}$ for data taken from 2016 to 2018 to match the trigger thresholds. Furthermore, the events are required to
contain two or three jets with $\pt>\SI{25}{\GeV}$. Exactly two of the jets must be \btagged jets. The third jet is mainly accepted in order to increase the statistics of the selected sample while keeping the modelling uncertainties manageable. Hadronic resonances are suppressed by requiring the dilepton invariant mass to satisfy $m_{\ell\ell}>\SI{15}{\GeV}$. Finally, events with track multiplicities $\ntrkout$ (defined below) above 100 are excluded because the dominant contribution is due to \pileup tracks.\footnote{
The fraction of selected data events with track multiplicities above 100 is below 0.2\%.}
 
Events with an opposite-sign (OS) $e\mu$ pair define the nominal sample, whilst events with a same-sign (SS) $e\mu$ pair are used to estimate the background from fake leptons.
The particle-level events are required to pass the same selection requirements  as the data selection but applied to the particle-level objects. In the case of the lepton \pT the lower requirement of $\SI{25}{\GeV}$ is applied.
 
\subsection{Definition of the observables at particle level and detector level}
\label{subsec:observables}
 
\paragraph{Particle-level observables}$~$ \\
Particle-level observables are built from charged particles, as defined in~\Sect{\ref{subsec:particle_level_objcts}}. The three observables are defined as:
\begin{itemize}
\item \textbf{\nch}: Charged-particle multiplicity with the binning defined as $[0,10,20,30,40,50,60,80,100]$.
\item \textbf{\sumptch}: Scalar sum of the transverse momenta of the selected charged particles with the binning defined as $[0,10,20,30,40,50,60,70,80,90,100,\infty)$ in units of $\si{\GeV}$. This observable is highly correlated with the charged-particle multiplicity, but contains additional information about the kinematic configuration of the event.
\item \textbf{\sumptch in bins of \nch}: This two-dimensional observable is built from the above two variables, as the \sumptch in bins of \nch with the binning defined in~\cref{tab:binning_2d}.
\end{itemize}
 
\begin{table}[htbp]
\centering
\caption{Binning used for the measured \sumptch in bins of \nch observable.}
\label{tab:binning_2d}
\begin{tabular}{ll}
\nch bin		& \sumptch binning [\GeV] \\
\toprule
$[0,20)$		& $[0,10,\infty)$			  \\
$[20,40)$		& $[0,35,\infty)$			  \\
$[40,60)$		& $[0,60,\infty)$			  \\
$[60,80)$		& $[0,80,\infty)$			  \\
$[80,100]$	& $[0,\infty)$				  \\
\bottomrule
\end{tabular}
\end{table}
The binning of the observables was chosen according to a predefined procedure described in \Sect{\ref{sec:unfolding}}.
 
\paragraph{Detector-level observables}$~$ \\
At detector level, the observables are calculated using reconstructed tracks with the same binning as at particle level.
The corresponding variables are then defined as:
\begin{itemize}
\item \textbf{\ntrkout}: Number of selected tracks outside of jets.
\item \textbf{\sumpttrkout}: Scalar sum of the transverse momenta of these tracks.
\item \textbf{\sumpttrkout in bins of \ntrkout}: The \sumpttrkout in bins of \ntrkout.
\end{itemize}
Before these variables are used in the unfolding, track-based backgrounds are subtracted and
further corrections are applied, as explained in \Sect{\ref{sec:track_bkgs}}.


\section{Background estimation and corrections to observables}
\label{sec:backgrounds}
\renewcommand\arraystretch{1.4}

Two different sources can lead to background contamination in the measured observables: event-based contributions from background
processes with a final state similar to \ttbar dileptonic events, discussed in~\Sect{\ref{sec:event_bkgs}}, and track-based contributions originating from sources other than the hard-scattering process, discussed in~\Sect{\ref{sec:track_bkgs}}.
Since the track-based background has to be subtracted from the measured variables \ntrkout and \sumpttrkout the following procedure is applied:
\begin{itemize}
\item The \pileup and secondary-particle contribution per event is estimated in a stochastic way. This procedure is detailed in~\Sect{\ref{sec:track_bkgs}}.
\item A tracking-efficiency correction (TEC) is applied to the background-subtracted observables to ease the unfolding. This procedure is detailed in~\Sect{\ref{sec:observable_correction}}.
\end{itemize}

\subsection{Event-based backgrounds}
\label{sec:event_bkgs}
After event selection, several background processes with a final state similar to \ttbar dileptonic events pass the selection.
Sources that contain top quarks contribute to the background, with $tW$ events containing a single top quark being
the dominant contribution. The production of $t\tbar{+}V$ events, with $V$ being either a $W$ or a $Z$ boson, contributes an irreducible
background. Events that contain either two electroweak
bosons or a $Z$ boson produced in association with jets, $\Zboson (\ra \tau\tau)$+jets, can be misidentified as a \ttbar dilepton event.
Finally, there is a contribution from multijet processes when at least one of the reconstructed leptons is wrongly reconstructed as an isolated lepton,
arising from either a heavy-flavour hadron decay, an electron from a photon conversion, a jet misidentified
as an electron, or a muon produced from decay-in-flight of a pion or kaon. This last category is collectively
called the fake-lepton background.
 
The fake-lepton background contribution is estimated using a partially data-driven approach. Events with an enriched fake-lepton
contribution are obtained from the observed same-sign $e\mu$ sample.
The total number of fake-lepton background events, $N^{\mathrm{fake}}$, is given by:
\begin{eqnarray}
N^{\mathrm{fake}} &=& R \cdot \left( N^{\mathrm{data}, \mathrm{SS}} - N^{\mathrm{prompt}, \mathrm{SS}} \right), \nonumber \\*[1mm]
R &=& \frac{N^{ \mathrm{fake}, \mathrm{OS}}}{N^{ \mathrm{fake}, \mathrm{SS}}} ~  ~, \nonumber
\end{eqnarray}
where $N^{\mathrm{data}, \mathrm{SS}}$ is the number of same-sign events in data, $N^{\mathrm{prompt}, \mathrm{SS}}$ is the estimated number of prompt same-sign events from simulation, and $R$ is the ratio of opposite-sign ($N^{ \mathrm{fake}, \mathrm{OS}}$) to same-sign ($N^{ \mathrm{fake}, \mathrm{SS}}$) events with fake leptons in simulation, with $R$ being $2.4 \pm 0.1$.
The kinematic distributions of the fake-lepton background are determined from the simulated \ttbar events with single-lepton and dilepton final states.
 
The composition of the same-sign samples is illustrated in~\Fig{\ref{fig:sslept}}.
Electron and muon \pt distributions obtained from same-sign selected data events are shown together with the corresponding predictions from simulated events. The contributions are broken down into prompt leptons and various fake-lepton categories. The prompt contributions are about \SI{35}{\%} of the same-sign sample.
They include wrong-sign (WS) contributions, dominated by \ttbar dilepton events where the electron charge sign has
been wrongly reconstructed, and right-sign (RS) contributions, with same-sign prompt leptons, from $\ttbar{+}V$ events,
$Z+$jets and diboson production. The fake-lepton contributions are dominated by electrons from photon conversions,
which represent \SI{62}{\%} of the same-sign sample. Sub-leading contributions come from leptons produced in semileptonic decays of heavy-flavour hadrons, and other sources, such as misidentified hadrons or decays-in-flight of pions and kaons.
A conservative \SI{50}{\%} uncertainty is assigned to the fake-lepton background~\cite{TOPQ-2018-17}.
 
All other background processes are modelled by simulated events and the expected number of events is calculated using the theoretical cross-sections and the acceptance from the simulated samples scaled to the integrated luminosity.
The expected and the observed event yields are shown in \Tab{\ref{tab:yields}}. The estimated purity of \ttbar dilepton events is approximately \SI{96}{\%}, with the backgrounds from single top quarks being the largest impurities.
Observed distributions of the leading jet \pt and the jet multiplicity compared with the model obtained from
simulated events are shown in \Fig{\ref{fig:dataMC}}. In the leading jet \pt distribution, a slope can be seen in the
ratio of data to prediction. This is in agreement with previous differential cross-section measurements by ATLAS
~\cite{TOPQ-2018-15,TOPQ-2018-11,TOPQ-2018-17,TOPQ-2015-06}, and CMS~\cite{CMS-TOP-20-010,CMS-TOP-18-013,CMS-TOP-17-014,CMS-TOP-12-028}, which have
consistently observed a softer top-quark \pt spectrum (and related distributions) than what is predicted by NLO+PS
MC generators. This discrepancy is at least partially due to missing NNLO corrections~\cite{Catani:2019hip,Czakon:2017wor,Czakon:2016dgf,Czakon:2015owf,Czakon:2013goa}.
The difference between data and prediction is covered by the systematic uncertainties.
 
\begin{figure}[htbp]
\begin{center}
\subfigure[]{\includegraphics[width=0.46\textwidth]{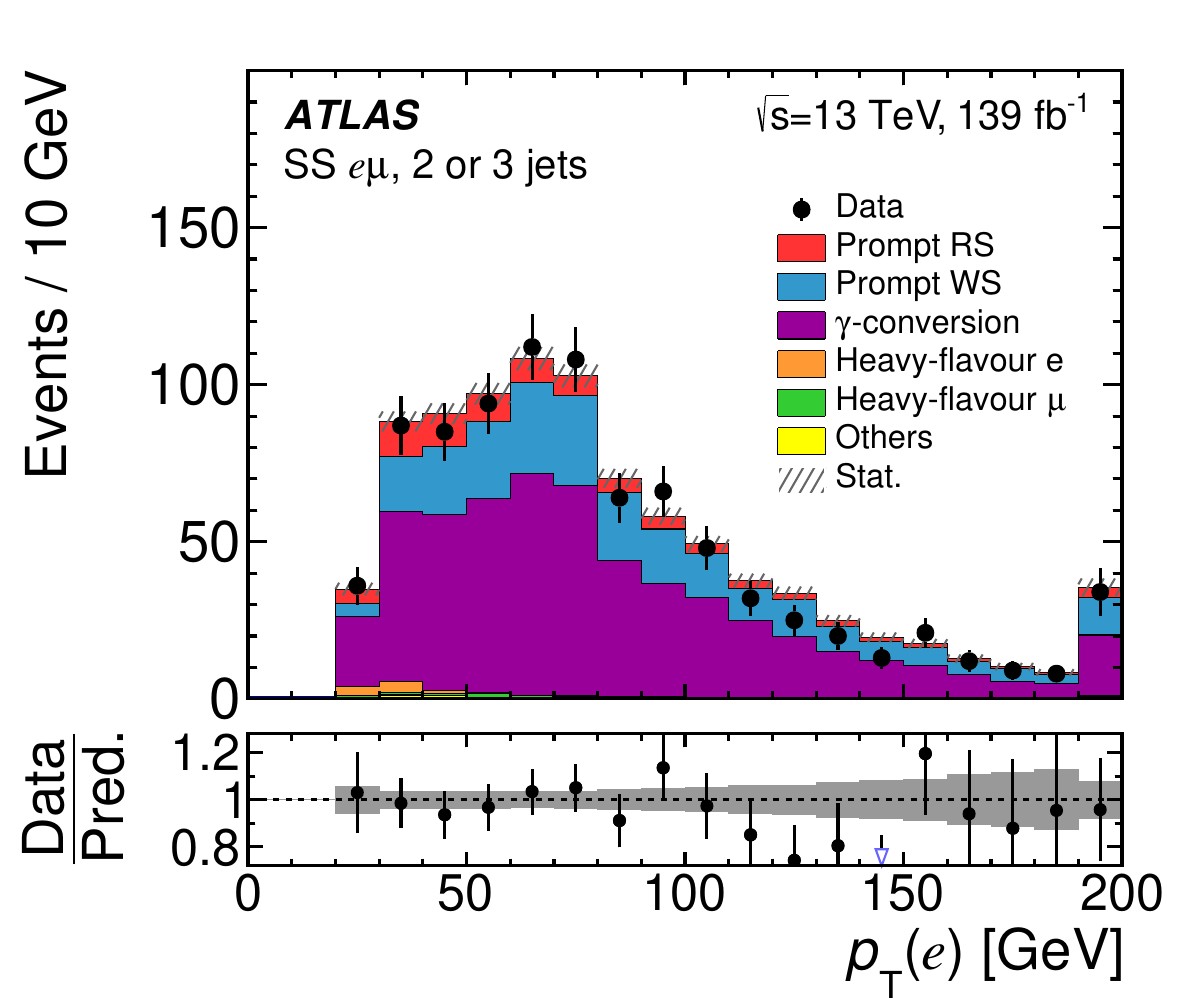}}
\subfigure[]{\includegraphics[width=0.46\textwidth]{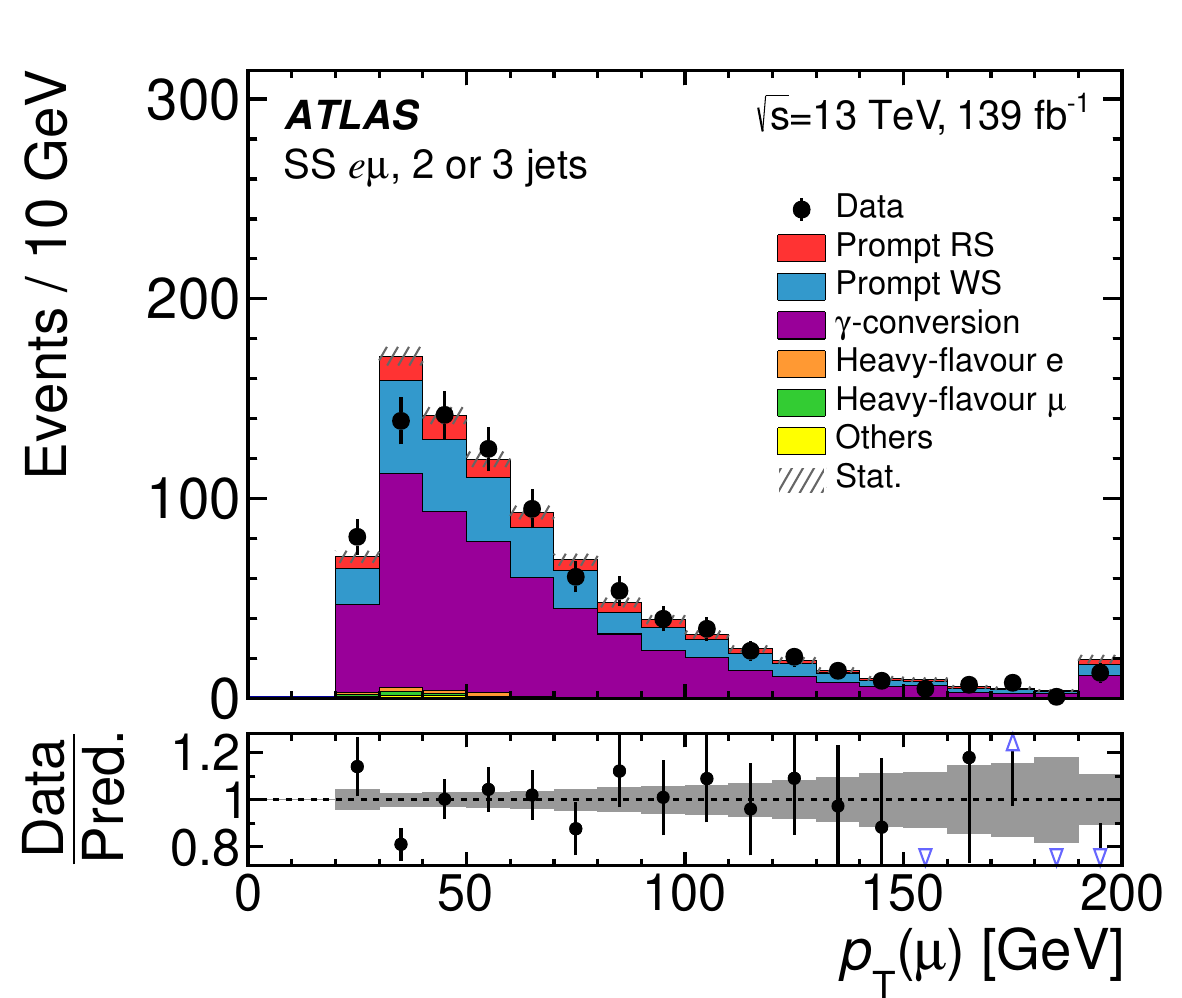}}
 
\caption{\label{fig:sslept} Distributions of (a) the electron \pt and (b) the muon \pt,
in events with a same-sign $e\mu$ pair and two or three jets, of which two are \btagged jets.
The simulation's prediction is normalised to the number of expected events and broken down into contributions
where both leptons are prompt RS or WS, or one is a fake lepton from a photon
conversion originating from a top-quark decay or from background, from
heavy-flavour decay or from other sources. The lower panel shows the
ratio of data to the prediction in each bin. The hatched (grey) uncertainty
band represents the statistical uncertainty due to limited size of the simulated samples.
Events beyond the $x$-axis range are included in the last bin.
The blue triangular markers in the lower panels point to data-to-prediction ratio values which lie
beyond the shown $y$-axis range.}
\end{center}
\end{figure}
 
\begin{table}[ht]
\hspace{-5mm}
\centering
\caption{\label{tab:yields} Event yields obtained after the event selection. The expected event yields from \ttbar production and the various
background processes are compared with the observed event yield. The fractional contributions from \ttbar production
and the background processes to the expected event yield is given in \%.
The processes labelled by `Others' include production of $Z$+jets and diboson background events.
The uncertainties include the MC statistical uncertainty and the normalisation uncertainty.
}
\begin{tabular}{l
r@{$\,\pm\,$}l
l}
\toprule
Process                         & \multicolumn{2}{c}{Events}    	     & Fraction [\%]  \\
\hline
\ttbar                          &  190\,600 	  & 9\,500       	 & \quad96       			\\
Single top                      &  6\,310       & 640          & \quad~~3.2       			\\
Fake leptons                    &  1\,320       & 660          & \quad~~0.66      			\\
$t\tbar{+}V$                      &  213        & 26         	 & \quad~~0.11           	    \\
Others                          &  91         & 21        	 & \quad~~0.046          	    \\
\hline
Total                           &  199\,000    & 11\,000        &  				    \\
\hline
Data                            &   \multicolumn{2}{l}{195\,507}  &        \\
\bottomrule
\end{tabular}
\end{table}
 
\begin{figure}[htbp]
\begin{center}
 
\subfigure[]{\includegraphics[width=0.46\textwidth]{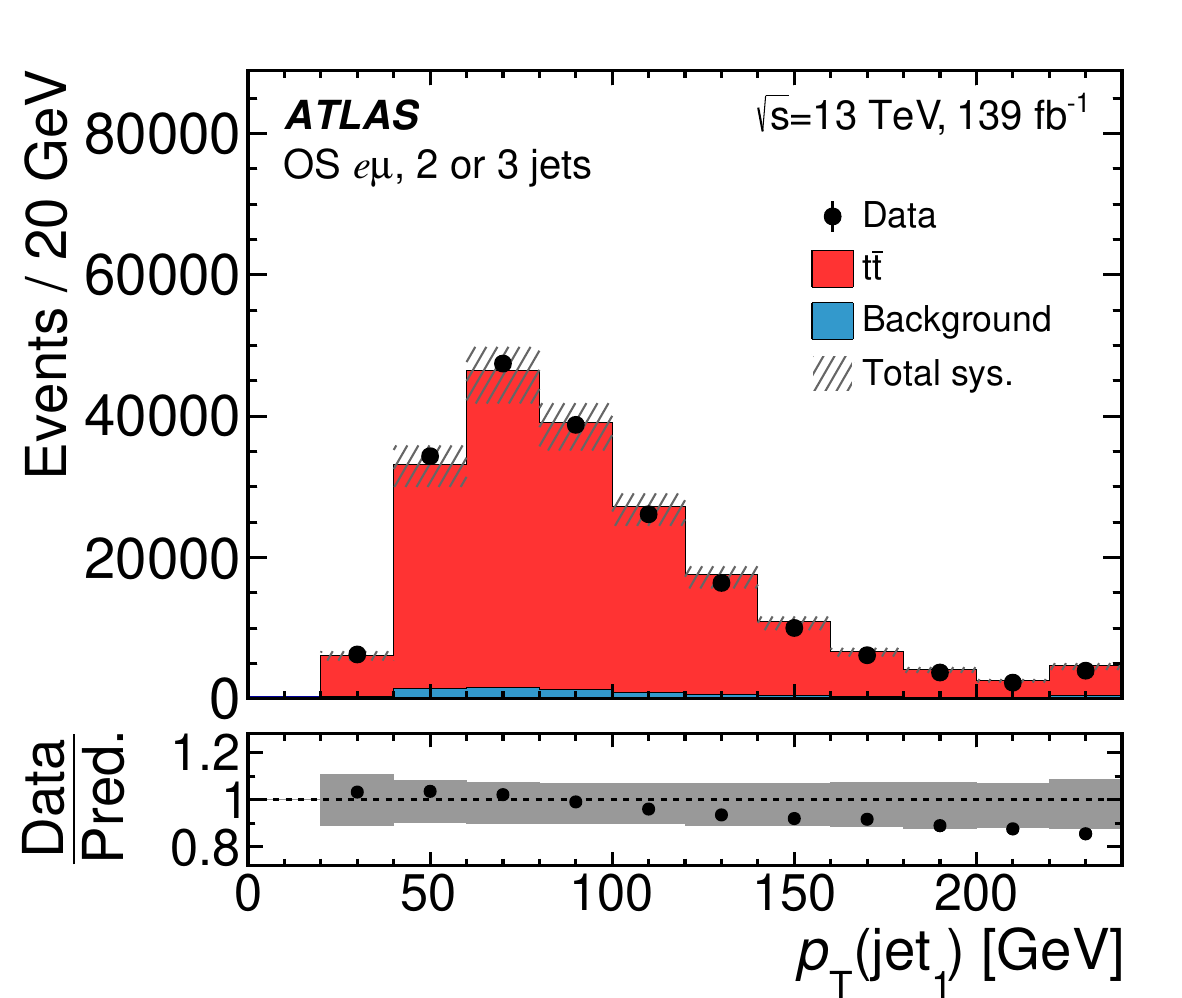}}
\subfigure[]{\includegraphics[width=0.46\textwidth]{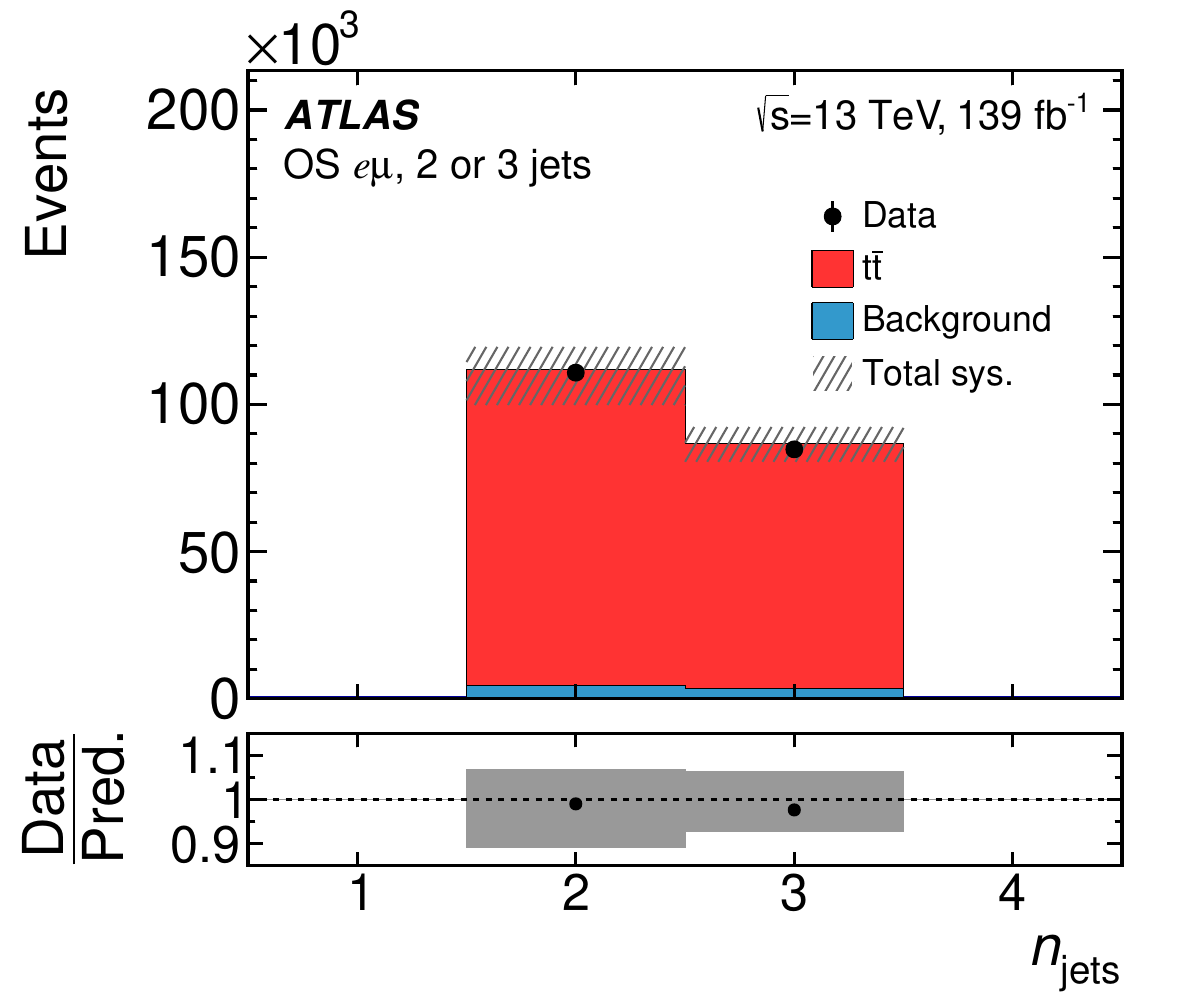}}
 
\caption{\label{fig:dataMC} Distributions of (a) the leading jet \pt and (b) the jet multiplicity in
events with an opposite-sign $e\mu$ pair and two or three jets, of which two are \btagged jets.
The observed data is compared with the expectation from simulated \ttbar and background events, where the background includes contributions from $tW$, $t\tbar{+}V$, $Z$+jets
and diboson processes and the estimated fake-lepton background.
The lower panel shows the ratio of data to the prediction in each bin.
The hatched (grey) uncertainty band includes the MC statistical uncertainty, background normalisation uncertainties, detector systematic uncertainties and
\ttbar modelling systematic uncertainties.
Events beyond the $x$-axis range are included in the last bin.}
\end{center}
\end{figure}
 
\FloatBarrier
 
\subsection{Track-based backgrounds}
\label{sec:track_bkgs}
Selected tracks other than the hard-scatter primary tracks can arise from additional sources.
Tracks originating from additional $pp$ collisions, referred to as \pileup, are the dominant background source. In addition, secondary particles produced through interactions with matter rather than as a result of direct fragmentation processes form a small contribution, while secondary particles produced in heavy-flavour decays are negligible, since tracks within jets are excluded.
Finally, tracks that are formed by a random combination of hits or from a combination of hits from several particles, referred to as fake tracks, are considered.
Corrected quantities \nprim and \sumptPrim are obtained from the measured observables \ntrkout and \sumpttrkout at detector level by subtracting these contributions on an event-by-event basis in the following way:
\begin{eqnarray}
\nprim &=& \ntrkout - c_{\mathrm{PU}}(\mu,\ntrkout)\cdot\npu - c_{\mathrm{sec}}\cdot \nsec  \label{eqn:nprim}\,, \\
\sumptPrim &=& \sumpttrkout - \sum_{i=1}^{c_{\mathrm{PU}}\cdot\npu} \pT^{\mathrm{corr}} - \sum_{i=1}^{c_{\mathrm{sec}}\cdot\nsec} \pT, \label{eqn:sumptprim}
\end{eqnarray}
where $\mu$ is the number of interactions per bunch crossing, \npu is the number of \pileup tracks, \nsec is the number of secondary-particle tracks including the contribution from fake tracks, and $c_{\mathrm{PU}}$, binned in $\mu$ and \ntrkout, and $c_{\mathrm{sec}}$ are scale factors to correct for differences between simulation and observed data. The corrected \sumptPrim in bins of the \nprim observable is built from Eqs.~(\ref{eqn:nprim}) and~(\ref{eqn:sumptprim}), and thus no additional background estimation is needed.

The \pileup and secondary-particle contamination (\npu and \nsec) per event is subtracted in the measured observables in a stochastic way based on a method using MC templates. The determination of the two contributions is based on the assumption that they are uncorrelated, which was also checked to be true in simulation. The procedure is performed as follows:
\begin{itemize}
\item \nprim: Templates are created from simulated \ttbar events for both the \pileup and secondary-particle track multiplicities, in bins of  \ntrkout. The background contamination in \ntrkout is expected to be higher for higher track multiplicities, since more tracks are likely to be selected from additional \pileup vertices overlapping with the primary vertex. For each event a random number is drawn from each template, corresponding to the \ntrkout value of the event. In this way the contamination is subtracted in the measured observables in a stochastic way.
 
\item \sumptPrim: A procedure similar to that for the \nprim observable is used, where templates are created from simulated \ttbar events for \pileup and secondary-particle track \pT. In order to compensate for deficiencies in the modelling of \pileup in simulation a reweighting method is used. To obtain a corrected \pT distribution (\ptcorr), tracks matched to \pileup vertices, in contrast to nominal tracks matched to the primary vertex, are compared between simulation and observed data and their ratio is used as a reweighting function. This function is used to reweight the \pT distribution obtained from \pileup tracks matched to the primary vertex. Since the contribution of tracks from secondary particles is small, no correction is applied in this case. These templates, together with the previously estimated \npu (\nsec), are then used to estimate the \pileup (secondary-particle) track contamination in the measured \sumpttrkout observable, by drawing random numbers \npu (\nsec) times from the corresponding track \pT template for each event.
\end{itemize}
In order to validate the procedure a closure test is performed, comparing the obtained distributions with distributions built from true \pileup tracks in MC events.
Differences between the expected and observed rates of \pileup tracks and secondary-particle tracks are accounted for by the scale factors $c_{\mathrm{PU}}$ and $c_{\mathrm{sec}}$.
Templates of the \zzero and \dzerosig distributions are created from simulated \ttbar events for hard-scatter primary tracks, \pileup tracks and secondary-particle tracks. The distributions for \pileup tracks are corrected for differences between collision data and simulation using high-purity \pileup tracks originating from \pileup vertices.
Two separate binned maximum-likelihood fits are performed to estimate $c_{\mathrm{PU}}$ and $c_{\mathrm{sec}}$. Studies have shown that these two contribution can be determined independently
in the following way: to obtain $c_{\mathrm{PU}}$,
hard-scatter primary-track and secondary-particle-track templates of the \zzero distribution are used,
while to obtain $c_{\mathrm{sec}}$, hard-scatter primary-track and \pileup-track templates
of the \dzerosig distribution are used. The scale factors of each contribution in the two fits are kept free-floating.
Studies have shown that $c_{\mathrm{PU}}$ depends on $\mu$ and the track multiplicity. Therefore, $c_{\mathrm{PU}}$ is determined for different intervals of $\mu$ and \ntrkout. Examples of  fitted \zzero and \dzerosig distributions
are shown in \Fig{\ref{fig:tracks_sf_fit}} and a summary of the estimated $c_{\mathrm{PU}}(\mu,\ntrkout)$ is shown in~\Tab{\ref{tab:pileup_sf}}. The obtained scale factors range from 0.9 to 1.4, where higher values are obtained for higher $\mu$ and $\ntrkout$ values. A value of $2.34\pm 0.02$ is estimated for  $c_{\mathrm{sec}}$.
The estimation of systematic uncertainties of $c_{\mathrm{PU}}$ is described in \Sect{\ref{sec:pileupunc}}.
\begin{figure}[htbp]
\begin{center}
\subfigure[]{\includegraphics[width=0.46\textwidth]{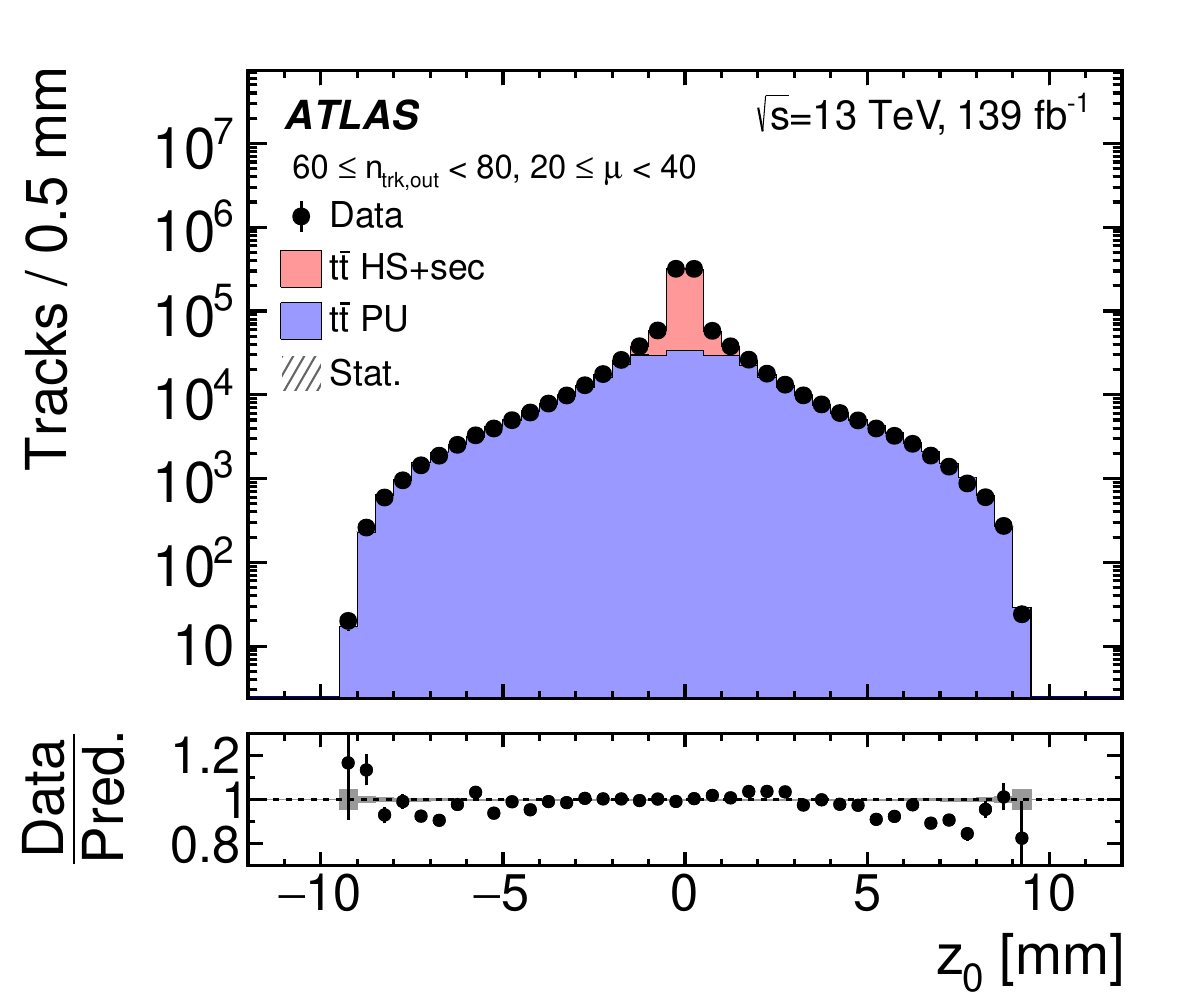}}
\subfigure[]{\includegraphics[width=0.46\textwidth]{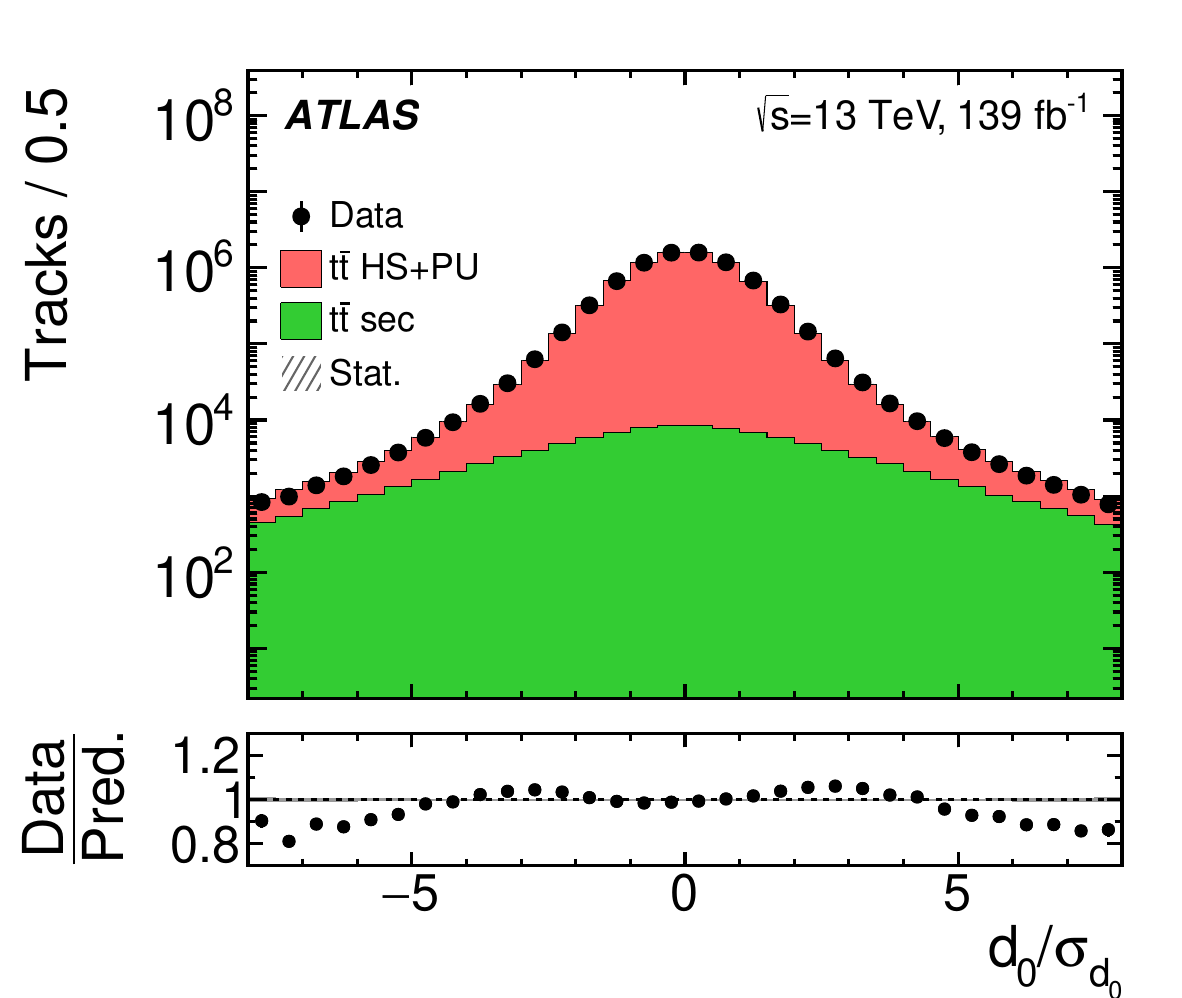}}
\caption{\label{fig:tracks_sf_fit} Illustration of the estimation of the \pileup (secondary-particle) track background by
fitting the \zzero (\dzerosig) distribution. As representative examples, the \zzero distribution is shown in the region with
$20 \leq \mu < 40$ and $60 \leq \ntrkout < 80$ in (a) and the fitted \dzerosig distribution is shown in (b). The stacked
histograms are normalised to the fit result. The lower panel shows the ratio of data to the prediction in each bin.
The hatched (grey) uncertainty band represents the statistical uncertainty due to the limited size of the simulated samples.}
\end{center}
\end{figure}
 
\begin{table}[ht]
\sisetup{round-mode=figures}
\centering
\caption{\label{tab:pileup_sf} Summary of the estimated \pileup scale factors $c_{\mathrm{PU}}$, parameterised in $\mu$ and \ntrkout.
All values have a statistical precision of 0.01.}
\begin{tabular}{l|
ccccc
}
\toprule
Region                          & \multicolumn{1}{c}{\ntrkout$<20$}   & \multicolumn{1}{c}{$20 \leq \ntrkout < 40$}    & \multicolumn{1}{c}{$40 \leq \ntrkout < 60$}     & \multicolumn{1}{c}{$60 \leq \ntrkout < 80$}         & \multicolumn{1}{c}{$80 \leq \ntrkout \leq 100$}   \\
\hline
$\mu < 20$ 						&0.91     &1.04     & 0.97   &1.05    &1.08    \\
$20 \leq \mu < 40$ 				&0.91     &1.08     & 1.08   &1.07    &1.11  \\
$\mu  \geq 40$ 					&0.95     &1.15     & 1.23   &1.27    &1.36  \\
\bottomrule
\end{tabular}
\end{table}
 
\subsection{Corrections to observables}
\label{sec:observable_correction}
To place the leading elements of the migration matrix on the principal diagonal, the background-subtracted quantities,
$\nprim$ and $\sumptPrim$, are scaled with correction factors $\cntrk$ and $\csumtrk$, respectively, accounting for
the limited track-reconstruction efficiency:
\begin{equation}
\ntrk = \frac{\nprim}{\cntrk(\nprim)}
\hspace*{35mm}
\sumpttrk = \frac{\sumptPrim}{\csumtrk(\sumptPrim)}
\label{eq:track_correction_factors}
\end{equation}
The TEC factors are calculated in bins of the background-subtracted quantity
by averaging the ratio of the quantity at the background-subtracted level to its value at
particle level across all selected events of the nominal simulated \ttbar sample.
The function $\cntrk(\nprim)$ has values between $0.68$ and $0.89$ in its range from $0$ to $100$,
while the function $\csumtrk(\sumptPrim)$ rises stepwise from $0.66$ to $0.89$ across the bins of its full range.
The corrections are applied to the observables in measured data events,
in simulated events from all scattering processes (i.e.\ the \ttbar process and all background processes), and in simulated events of the samples with systematically varied parameters.
The corrected quantities, \ntrk and \sumpttrk, are the final corrected-level observables
forming the distributions to be unfolded, and are shown in~\Fig{\ref{fig:dataMC_meas}}. The corrected \sumpttrk in bins of the \ntrk observable, shown in~\Fig{\ref{fig:dataMC_meas_2d}}, is built from \Eqn({\ref{eq:track_correction_factors}}) without any additional correction.
The uncertainty band includes detector systematic uncertainties and the \pileup background-track-rate uncertainty.
Within these experimental uncertainties there are considerable differences between the observed data and the nominal
prediction, reflecting non-optimal tuning of the UE and MPI.
\begin{figure}[htbp]
\begin{center}
\subfigure[]{\includegraphics[width=0.46\textwidth]{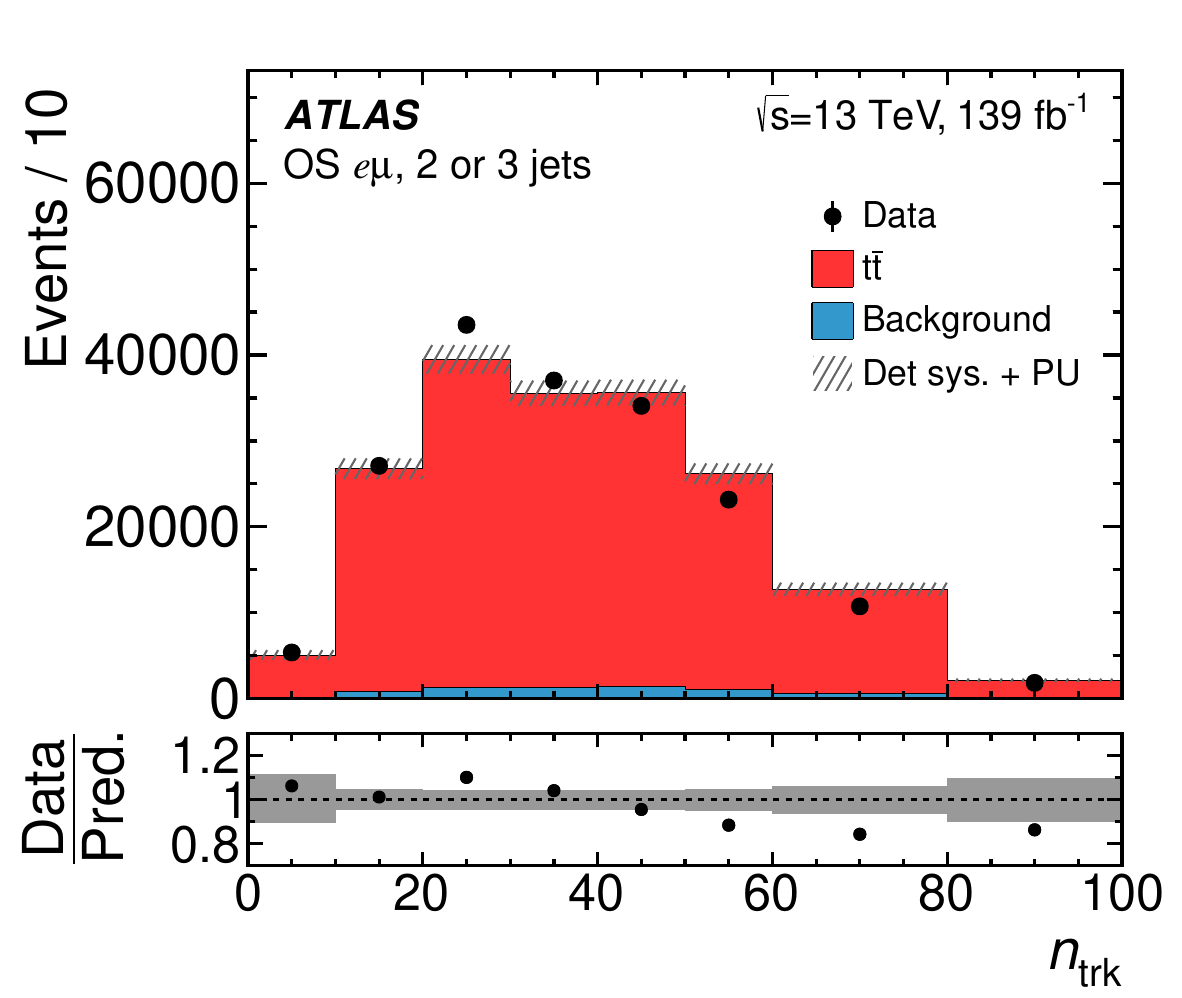}}
\subfigure[]{\includegraphics[width=0.46\textwidth]{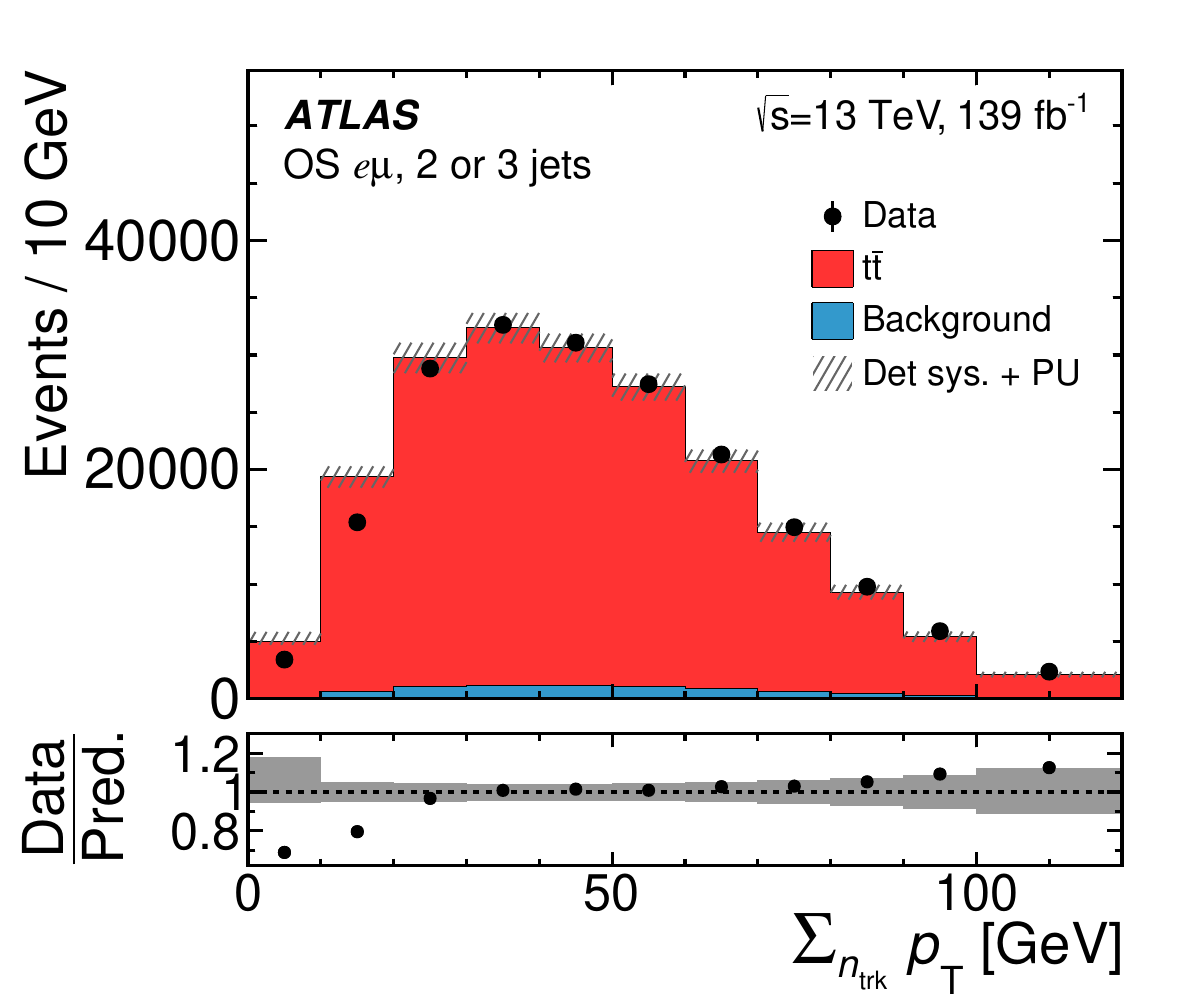}}
\subfigure[]{\hspace{-1cm}\includegraphics[width=\textwidth]{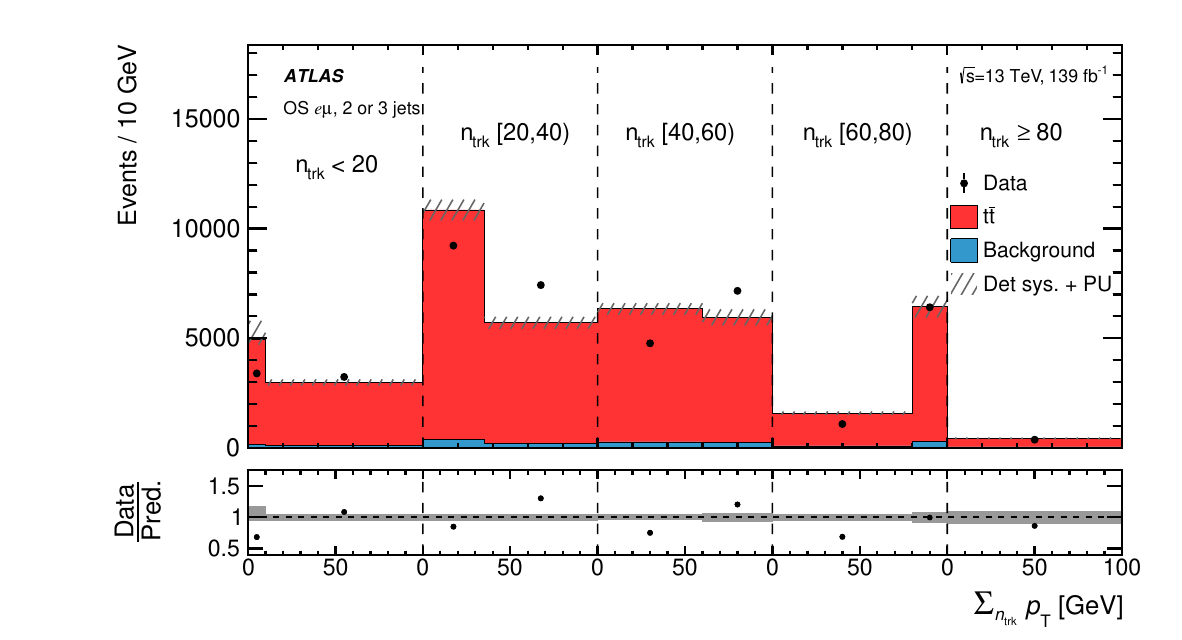}
\label{fig:dataMC_meas_2d}}
\caption{\label{fig:dataMC_meas} Distributions of (a) \ntrk, (b) \sumpttrk and (c) \sumpttrk in bins of \ntrk
for events with an opposite-sign $e\mu$ pair and two or three jets, of which two are \btagged jets. The corrected-level observed distributions are compared with the expectation from simulated events, broken down into contributions from \ttbar and the background processes $tW$, $t\tbar{+}V$,
$Z$+jets and diboson, together with the estimated fake-lepton background.  The estimated contributions from \pileup tracks and secondary particles is subtracted from measured data events and simulated \ttbar events.
The TEC is applied to all processes.
The lower panel shows the ratio of observed data to the prediction in each bin.
The hatched (grey) uncertainty band includes detector systematic uncertainties and the \pileup background-track-rate uncertainty.
Events beyond the $x$-axis range are included in the last bin.}
\end{center}
\end{figure}
Table~\ref{tab:vars_at_levels} summarises the naming convention for the observables at different levels of the analysis.
\begin{table}[htbp]
\centering
\caption{\label{tab:vars_at_levels}Naming convention for the observables at different levels of the analysis. At the background-subtracted level the contributions of tracks from pile-up collisions and tracks from secondary vertices are subtracted.
At the corrected level the TEC is applied. The observables at particle level are the analysis results.}
\begin{tabular}{lcc}
\toprule
Analysis level &
\begin{minipage}[c]{50mm}\vspace*{1mm}Number of charged particles \\outside of jets \vspace*{1mm} \end{minipage}  &
Scalar sum of track $\pt$ \\ \midrule
Detector level & \ntrkout & \sumpttrkout \\
Background-subtracted level & \nprim  & \sumptPrim \\
Corrected level & \ntrk & \sumpttrk \\
Particle level & \nch & \sumptch \\ \bottomrule
\end{tabular}
\end{table}


\FloatBarrier
\renewcommand\arraystretch{1.0}
\section{Unfolding}
\label{sec:unfolding}


The differential cross-section distributions in the fiducial phase space defined in Section~\ref{sec:selection}
are obtained from the corrected-level events
using an unfolding technique that corrects for acceptance and certain detector effects. The iterative Bayesian method~\cite{DAgostini:1995} as implemented in RooUnfold~\cite{Adye:2011gm} is used for this unfolding.
 
The unfolding procedure starts from the observables described in the previous section, i.e.\ after subtracting the track-based backgrounds and applying the TEC, and consists of two stages.
First, the event-based background contributions $N_\mathrm{bkg}$, summarised in \Tab{\ref{tab:yields}}, are subtracted bin-by-bin from the track-based background-subtracted and tracking-efficiency-corrected distribution $N_\mathrm{corr}$. The contributing
processes are normalised either to the corresponding theory predictions or, in the case of the fake-lepton background, to the partially data-driven estimate.
Second, the distributions are unfolded using a detector response model, the migration matrix $\mathcal{M}$, obtained from simulated \ttbar events.
As part of this second step, two correction factors are applied to correct for non-overlap of the fiducial
phase space at corrected level and particle level. The corrections account for events that fall within the fiducial
phase space of one level but not the other.
The extraction of the absolute differential cross-section for an observable $X$, $X$ being \ntrk, \sumpttrk or \sumpttrk in bins of \ntrk, is then summarised by the equation:
\begin{equation*}
\frac{\mathrm{d}\sigma^\mathrm{fid}}{\mathrm{d}X^i} \equiv \frac{1}{\mathcal{L} \cdot \Delta X^i} \cdot  \frac{1}{\epsilon^i} \cdot \sum_j \mathcal{M}_{ij}^{-1} \cdot f_\mathrm{acc}^j \cdot  \left(N_\mathrm{corr}^j - N_\mathrm{bkg}^j\right)
\end{equation*}
where the index $i$ ($j$) iterates over the bins of observable $X$ at particle (corrected) level, $\Delta X^i$ is the bin width, $\mathcal{L}$ is the integrated luminosity, the inversion of $\mathcal{M}$
symbolises unfolding using the iterative Bayesian method, and $f_\mathrm{acc}$ and $\epsilon$ are the phase-space correction
factors. These last two parameters are defined as:
 
\begin{equation*}
f_{\mathrm{acc}} = \left( \frac{N_{\mathrm{corr} \wedge \mathrm{part}}}{N_{\mathrm{corr}}} \right)
\hspace*{35mm}
\epsilon = \left( \frac{N_{\mathrm{corr} \wedge \mathrm{part}}}{N_{\mathrm{part}}} \right)
\end{equation*}
 
The number $N_{\mathrm{corr} \wedge \mathrm{part}}$ indicates the number of corrected-level events that satisfy
the particle-level selection and $N_{\mathrm{part}}$ is the number of events passing the particle-level selection.
The response model and phase-space correction factors are obtained from simulated \ttbar events.
 
The inclusive fiducial cross-section is obtained by integrating the unfolded differential cross-section over the kinematic
bins, and its value is used to compute the normalised differential cross-section $1/\sigma \cdot \mathrm{d}\sigma^\mathrm{fid}/\mathrm{d}X^i$.
 
The binning is chosen according to the resolution of each observable such that the fraction of events in the diagonal bins
is greater than \SI{50}{\%} and the data statistical uncertainty in each bin is below \SI{10}{\%}.
The unfolding is performed using three iterations.
This value is chosen because using more iterations produces only small event-count changes in each of the bins and gives lower correlation factors between them.
 
The unfolding procedure was validated by performing four tests.
To ensure that the results are not biased by the MC generator used in the unfolding procedure, a study was
performed in which the particle-level spectra from the default $\POWHEGBOX\,{+}\,\PYTHIA$ simulation are
altered by changing the shape of distributions using either a linear or sinusoidal function. Here, maximal variations of \SI{40}{\%} are considered. Alternatively, the shapes are also changed by a smooth function that describes the ratio of the background-subtracted data distribution to the distribution obtained from simulated \ttbar events. Here, variations of up to \SI{30}{\%} are applied.
The reweighted corrected-level \ttbar distributions are
then unfolded using the nominal migration matrices and unfolding corrections to see if the reweighted particle-level distributions can be recovered. These tests are referred to as `stress tests'.
The altered shapes are recovered within statistical uncertainties by the unfolding procedure in the case of the linear reweighting, while deviations of  \SI{5}{\%} in the tails are observed for the sinusoidal reweighting and the MC-to-data reweighting.
In order to explore the impact of discrepancies between data and MC simulations in variables other than the unfolded observables, a cross-reweighting scenario is also considered. This uses the same procedure as in the stress tests above, but the reweighting function is obtained from one of the one-dimensional observables, either \nch or \sumptch, and applied to the other two observables.
The reweighting function is defined as a smooth function that describes the ratio of the measured distribution (unfolded data) to the distribution obtained from simulated \ttbar events at the particle level.
Similarly to the nominal stress tests, the altered shapes are recovered with a maximum deviation of \SI{3.6}{\%}. Non-closure deviations after applying the background-subtracted data to \ttbar MC ratio reweighting and cross-reweighting scenario are added as systematic uncertainties (see Section~\ref{sec:syst:unfolding}).


\section{Systematic uncertainties}
\label{sec:sys}

Systematic uncertainties from several sources affect the measured observables. Experimental uncertainties are associated with the reconstruction of the four-momenta of final-state objects: electrons, muons, jets, and tracks.
Systematic uncertainties in the estimation of the background event rates and uncertainties related to the unfolding procedure are taken into account. Modelling uncertainties are related to parameters and methods used in MC event generation, and the statistical uncertainties are also considered.
The total uncertainty is then calculated, assuming all the systematic uncertainties are uncorrelated. Finally, the sensitivity of the measured observables to variations of inital- and final-state radiation and variation of \hdamp is tested. In the following, the different uncertainty components are explained in more detail.
 
\subsection{Detector systematic uncertainties}
Detector systematic uncertainties, due to residual differences between data and MC simulations after calibration, and uncertainties in
corrective scale factors are propagated through the analysis and evaluated for \ttbar production and the $tW$ background.
Non-significant uncertainties with negligible differences between the systematically varied distribution and the nominal distribution at the corrected level are not considered further.
The uncertainties are evaluated by building distributions from events contributed by the
systematically varied \ttbar production process and the systematically varied background processes such as $tW$.
The resulting distribution is then unfolded after subtracting the nominal background, using the nominal migration matrix and unfolding corrections.
The relative deviation from the nominal unfolded distribution is then assigned as a systematic uncertainty. The following sources are considered:
 
\begin{itemize}
 
\item \textbf{Jet uncertainties} \\
The jet energy scale (JES) uncertainties are derived from test-beam data,
in situ calibration measurements and simulation, and are parameterised in bins
of jet \pt and $\eta$~\cite{JETM-2018-05}. Its uncertainty is decomposed into a set of 30 uncorrelated components.
Sources of uncertainty contributing to the JES uncertainties include the $\eta$ intercalibration, jet flavour
composition and response, differences between jets induced by $b$-quarks and those from gluons or light-quarks, detector modelling, non-closure, punch-through losses and several \pileup properties. The effect of single high-\pt particles is also included.
Furthermore, uncertainties related to the jet energy resolution (JER) are considered. The uncertainty of the JVT efficiency~\cite{PERF-2014-03} is found to be negligible.
 
Since this analysis makes use of $b$-tagging, the associated uncertainties in
tagging efficiency for $b$-jets, $c$-jets, and light-flavour jets are considered by using an uncertainty model containing
nine/four/four independent variations for the $b$-/$c$-/light-flavour jet calibrations and two components for the MC-based extrapolation
to high-\pt jets~\cite{FTAG-2018-01}.
 
The combination of these uncertainties is labelled `Jet' in \Fig{\ref{fig:totalSyst_detector}}, with the main contribution coming from the jet flavour response.
 
\item \textbf{Track uncertainties} \\
Systematic uncertainties associated with \dzero and \zzero impact parameter resolutions are applied, based on measurements of minimum-bias data and $Z\rightarrow\mu\mu$ events and parameterised in \pT and $\eta$.
In addition, the impact parameter resolution is particularly sensitive to the
accuracy of the simulation's modelling of the number of inactive modules in the pixel detector~\cite{ATL-PHYS-PUB-2015-051}. An uncertainty is estimated by randomly disabling \SI{5}{\%} of the pixel modules.
 
Weak modes in the alignment uncertainties are accounted for by applying bias corrections to tracks as a function of their angular position to simulate the impact of such modes. Variations biasing \dzero and \zzero are taken into account, as well as sagitta biases~\cite{IDTR-2019-05}.
 
Finally, uncertainties in the track reconstruction efficiency as well as
track fake-rates are evaluated.
Two main sources of uncertainty are considered for the track reconstruction efficiency:  amount-of-material uncertainties and the physics model used in simulation. The effects of these
uncertainties are assessed by comparing the efficiency in samples with the different configurations: the passive material of the ID is scaled by \SI{5}{\%}, the passive material
of the IBL is scaled by \SI{10}{\%}, the passive material in the Patch Panel 0 (PP0) region is scaled by \SI{25}{\%}, and finally the \GEANT physics model is varied~\cite{PERF-2015-07}.
The estimation of the track fake-rate is based on observations of differences in the non-linear component of the evolution of track multiplicity as a function of $\langle\mu\rangle$ as
described in Ref.~\cite{ATL-PHYS-PUB-2015-051} and the uncertainty is assumed to be \SI{100}{\%}.
 
The combination of these uncertainties is labelled `Track' in \Fig{\ref{fig:totalSyst_detector}}, with the main contribution coming from the global tracking efficiency.
 
\item \textbf{Lepton uncertainties} \\
Uncertainties due to the lepton trigger, reconstruction and selection
efficiencies in simulation are estimated from a tag-and-probe method applied to electrons and muons from leptonic decays of $Z$ bosons and \Jpsi mesons using methods similar to those in Refs.~\cite{PERF-2017-01,PERF-2015-10}.
The same processes, using methods described in Refs.~\cite{EGAM-2018-01,PERF-2015-10}, are used to estimate uncertainties in the lepton momentum scale and resolution.
The combination of these uncertainties is labelled `Lepton' in \Fig{\ref{fig:totalSyst_detector}}.

\item \textbf{Average number of additional \pileup events} \\
Scale factors are applied to reweight simulated events in order to obtain the distribution of the average number of additional \pileup events corresponding to the distribution obtained in data. An uncertainty in these reweighting scale factors is considered, based on the difference
between the instantaneous luminosities in data and simulation, and is labelled as `$\mu$-reweight'.
 
\end{itemize}
\Fig{\ref{fig:totalSyst_detector}} shows the fractional uncertainty from each of the above-mentioned categories for the three observables. The dominant uncertainties are track and jet uncertainties, and they are observed to be largest at low or high values of the measured distributions.
\begin{figure}[htpb]
\subfigure[]{\includegraphics[width=0.48\textwidth]{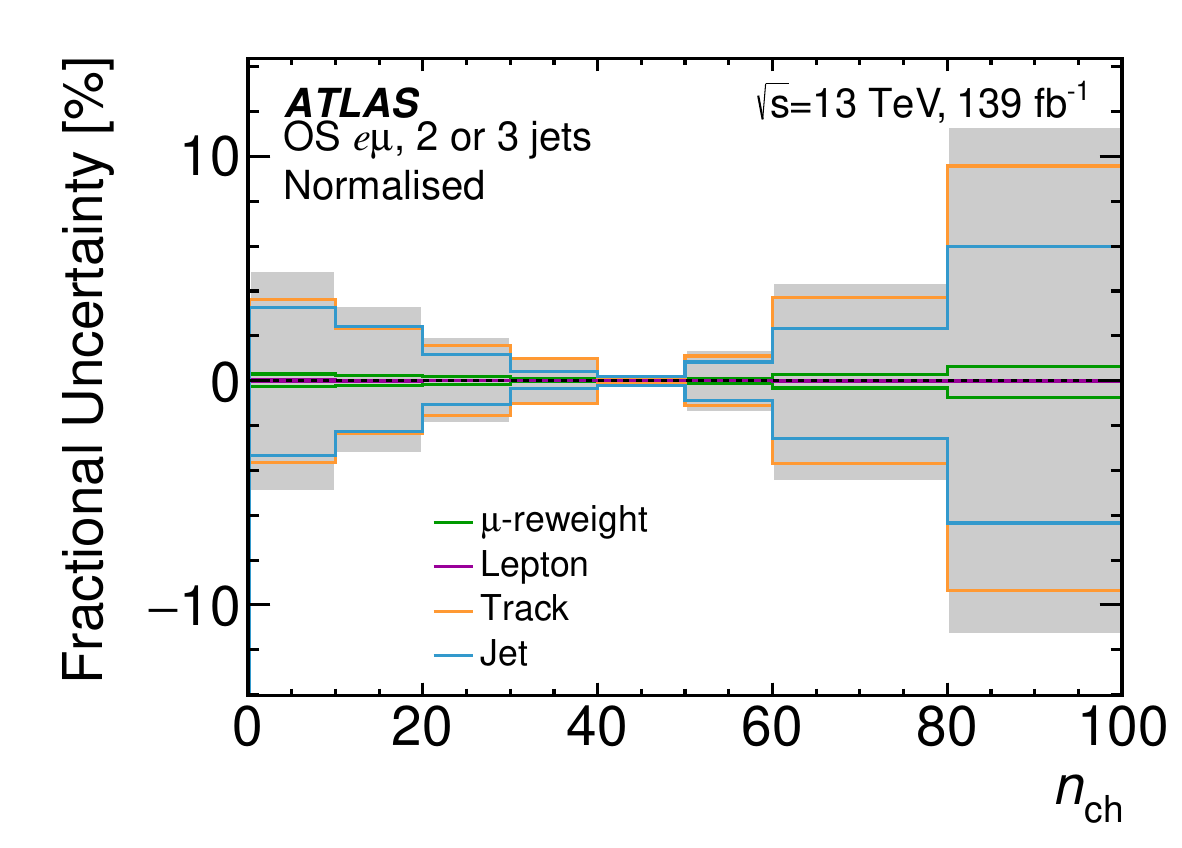}}
\subfigure[]{\includegraphics[width=0.48\textwidth]{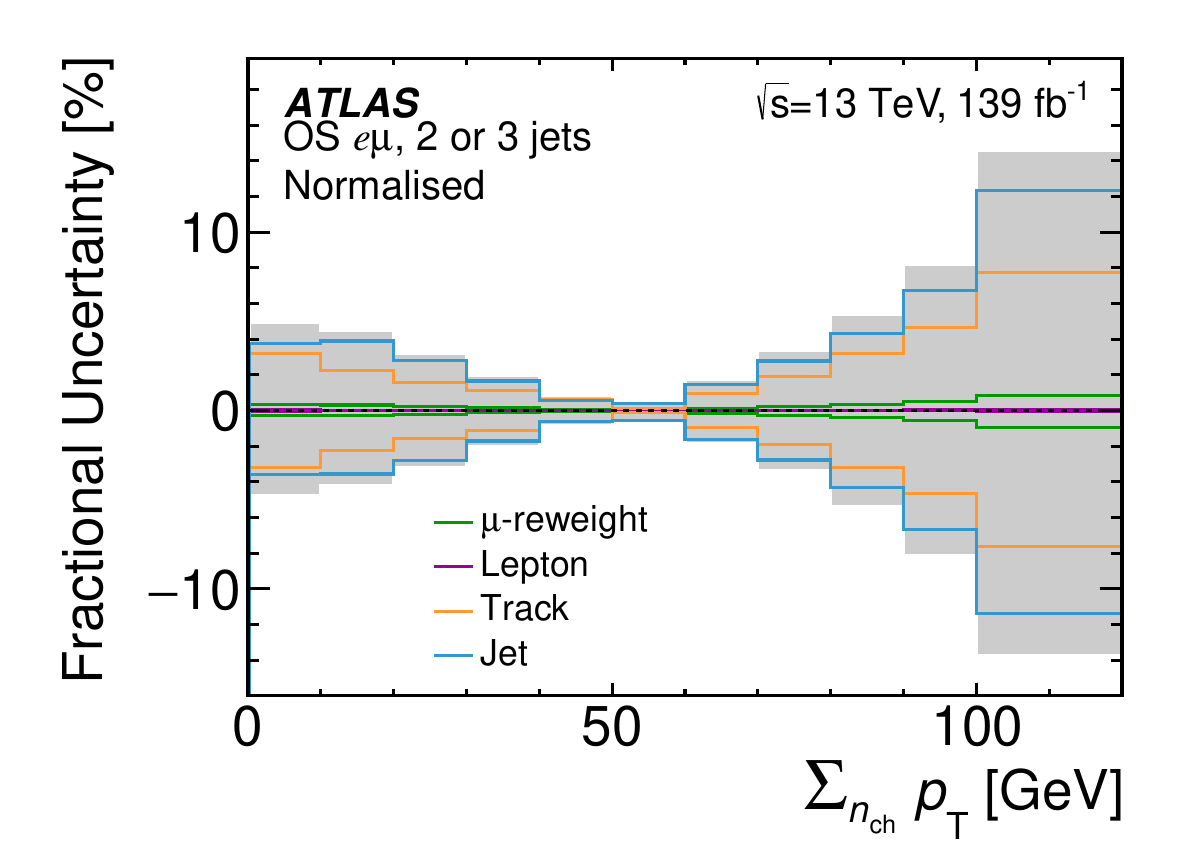}}
 
\subfigure[]{\hspace{-1cm}\includegraphics[width=\textwidth]{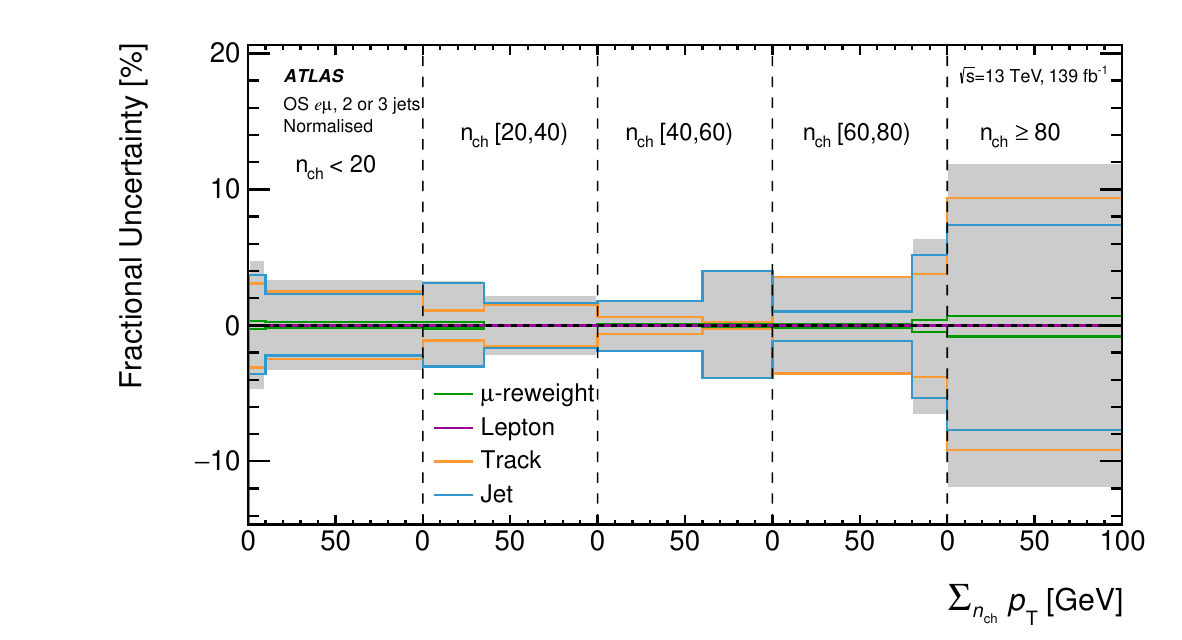}}
\caption{\label{fig:totalSyst_detector} Fractional uncertainties in the measured normalised differential cross-section as a function of (a) \nch , (b) \sumptch , and (c) \sumptch in bins of \nch for all experimental systematic uncertainties.
The $x$-axis in (c) is split into five bins of \nch by the dashed vertical lines and \sumptch is presented in each bin.
The grey bands represent the presented components added in quadrature. Events beyond the $x$-axis range are included in the last bin in (a) and (b), and in the corresponding last bin of each slice in (c).}
\end{figure}
 
\FloatBarrier
 
\subsection{Track-based-background uncertainties}
\label{sec:pileupunc}
The uncertainty in the rate of background \pileup tracks is evaluated by subtracting
a varied number of \pileup tracks from the number of reconstructed tracks in \ttbar simulated events.
The variation is given by a re-evaluated \pileup scale factor $c_{\mathrm{PU}}$, described in~\Sect{\ref{sec:track_bkgs}},
from either systematically varied samples, alternative samples, or sideband fits.
The resulting distribution is then unfolded using the nominal migration matrix and efficiency. The relative deviation from the nominal
unfolded distribution is assigned as a systematic uncertainty.
The following uncertainties in the scale factor are considered:
\begin{itemize}
\item The uncertainty associated with the $z_{0}$ impact parameter resolution is evaluated using systematically varied \ttbar samples and it is found to be smaller than \SI{2}{\%}.
 
\item  The uncertainty in the reweighting procedure used to correct the modelling of \pileup tracks in simulation is evaluated by varying the selection of \pileup vertices and leads to an uncertainty of \SI{1.4}{\%} in the \pileup scale factor.

\item Uncertainties due to the influence of hard-scatter primary tracks are estimated by fitting a sideband region with $|z_{0}|>\SI{1}{\milli\metre}$. The resulting uncertainty ranges from less than \SI{1}{\%} for most regions to as much as \SI{14}{\%} in the high \ntrkout and low $\mu$ region.
 
\item The uncertainty due to different beam conditions and alignment settings
is evaluated by re-fitting the rate of background \pileup tracks for different collision-data periods and leads to an uncertainty of \SI{0.4}{\%} in the \pileup scale factor.

\item Uncertainties in the modelling of parton showers, hadronisation and the underlying event, are evaluated by comparing the nominal samples to alternative samples for which
\POWHEGBOX is interfaced to \HERWIG[7.1.3] instead of \PYTHIA[8.230]. The main difference is caused by the different underlying-event models and thus by a modified $z_0$
distribution of the hard-scatter primary tracks used in the fit.
The largest and dominant uncertainty is found for low \ntrkout values, being at a level of \SI{17}{\%}, while otherwise the uncertainty is a few percent.
 
\end{itemize}
All \pileup scale-factor uncertainties are added in quadrature and the resulting relative uncertainties are presented in~\Tab{\ref{tab:pileup_sf_unc}}.
The largest uncertainty is observed for events with $\ntrkout<20$, mainly due to differences in the modelling of the UE between \PYTHIA and \HERWIG.
Finally, the uncertainty due to the non-closure of the subtraction method in a few bins is added in quadrature to the combined \pileup scale-factor uncertainty
and labelled as `PU subtraction' in \Fig{\ref{fig:totalSyst_and_Statistical}}.
\begin{table}[htpb]
\centering
\caption{\label{tab:pileup_sf_unc} The total \pileup scale-factor relative uncertainty parameterised in $\mu$ and \ntrk and expressed in percent.}
\begin{tabular}{l|ccccc
}
\toprule
Region                          & $\ntrkout<20$   & $20 \leq \ntrkout < 40$   & $40 \leq \ntrkout < 60$     & $60 \leq \ntrkout < 80$         & $80 \leq \ntrkout \leq 100$   \\
\hline
$\mu < 20$ 						&16     &3  	&3    &5		&15~~ \\
$20 \leq \mu < 40$ 				&16	    &1      &1    &3		&9 \\
$\mu  \geq 40$ 					&18 	&2	    &2    &4		&5\\
\bottomrule
\end{tabular}
\end{table}
 
\subsection{Event-based-background uncertainties}
Uncertainties due to the normalisation of the background processes are assessed by varying the normalisation
of each background process according to its uncertainty. The following uncertainties, based on inclusive cross-section uncertainties, are
considered: single-top-quark and $\ttbar{+}V$ processes \SI{10}{\%}, fake leptons \SI{50}{\%}, and other processes \SI{20}{\%}.
The combination of these uncertainties is labelled `Background rates'.
 
Uncertainties related to $tW$ background modelling are assessed using samples where the renormalisation and factorisation scales in the matrix-element calculations and the intensity of initial-state and final-state radiation are varied. These uncertainties are labelled `$tW$ Scale $\muR$', `$tW$ Scale $\muF$, `$tW$ ISR \alphas' and `$tW$ FSR', respectively.
The parameter variations in the generation of these samples are identical to those used for the nominal \ttbar process, described in \Sect{\ref{subsec:ttbarSyst_modelling}}. The uncertainty from the scheme for removing the overlap of the $tW$ process with \ttbar production is evaluated by comparing the nominal $tW$ sample, using the diagram-removal scheme, with a sample produced with an alternative scheme (diagram subtraction)~\cite{Frixione:2008yi} and is labelled as `$tW$ DS vs.\ DR'.
 
The above uncertainties are evaluated by building distributions where the \ttbar and systematically varied background contributions are added together. Each resulting distribution is then unfolded after subtracting the nominal background, using the nominal migration
matrix and efficiency. Again the relative deviation from the measured unfolded distribution is assigned as a systematic uncertainty.
 
\Fig{\ref{fig:totalSyst_background}} shows the fractional uncertainty from each of the above-mentioned categories for the three observables and the combination of these uncertainties is labelled `Event background' in \Fig{\ref{fig:totalSyst_and_Statistical}}.
 
\begin{figure}[htpb]
 
\subfigure[]{\includegraphics[width=0.48\textwidth]{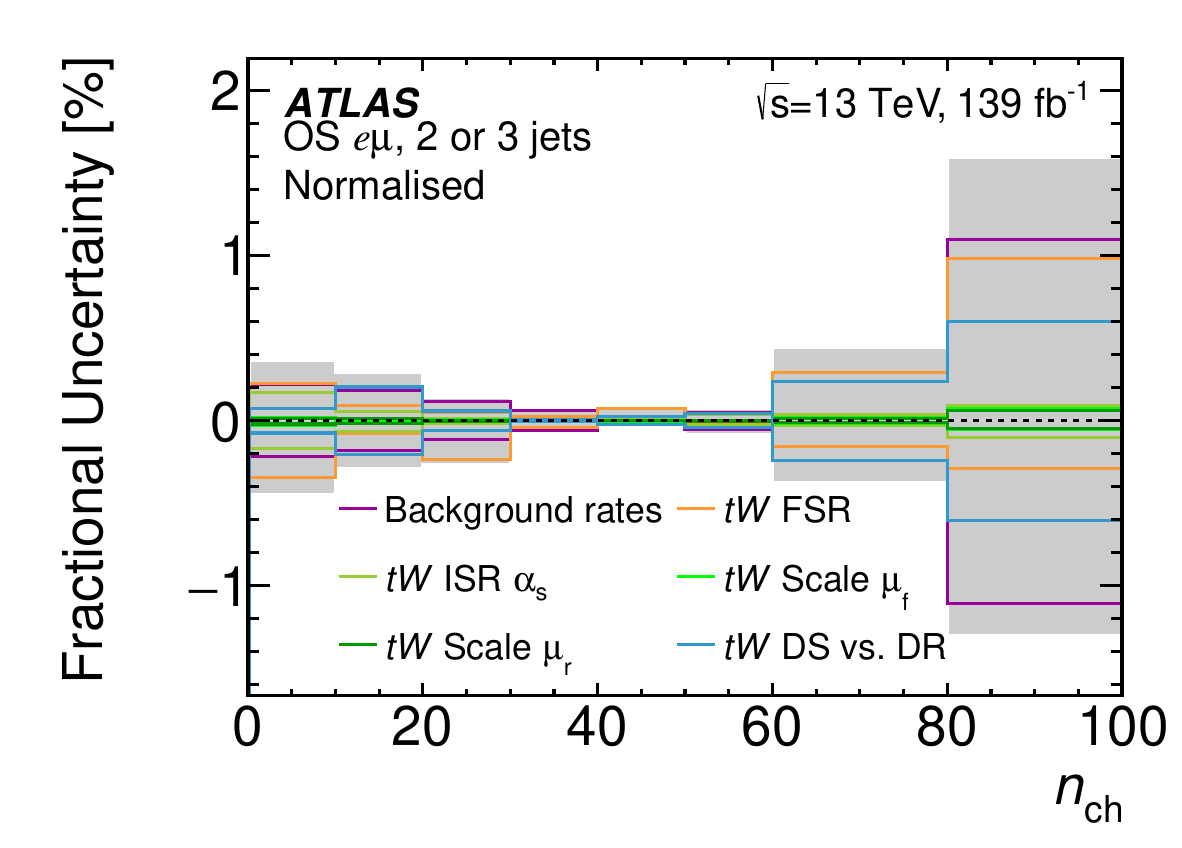}}
\subfigure[]{\includegraphics[width=0.48\textwidth]{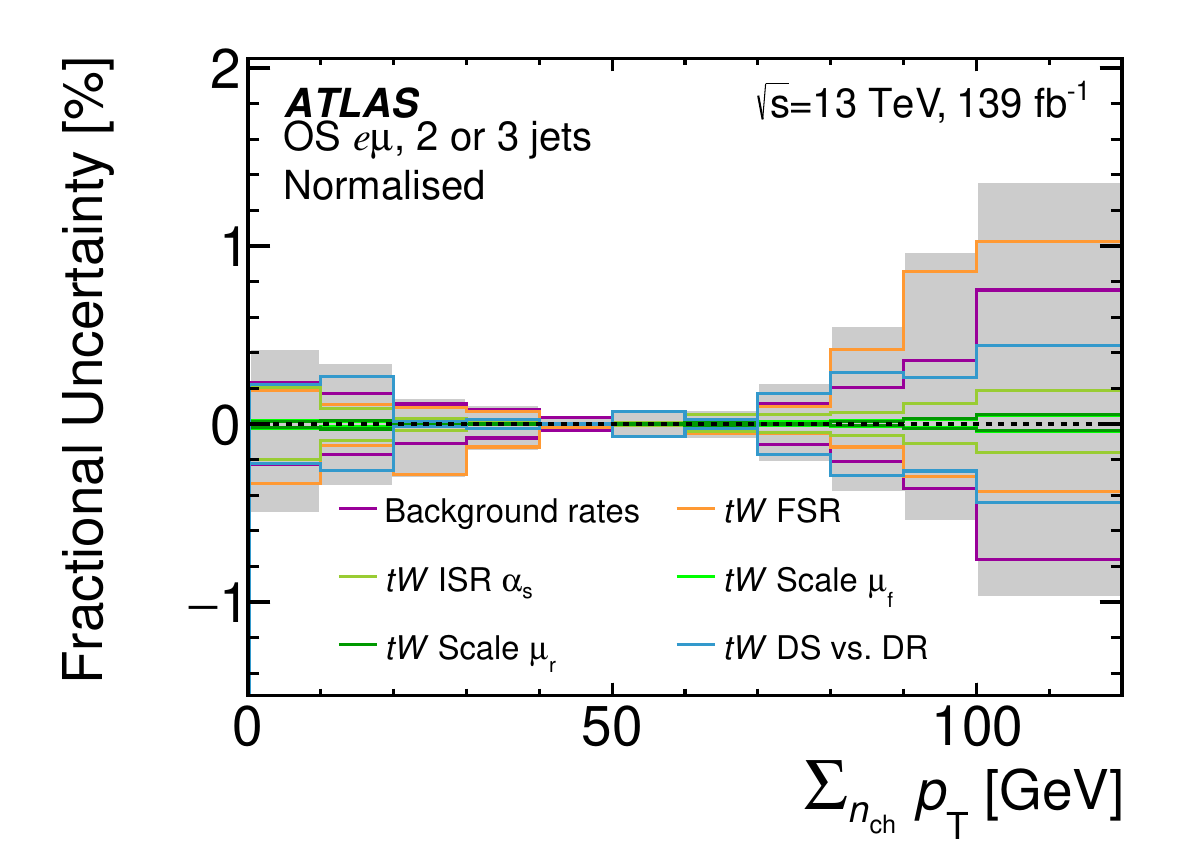}}
\subfigure[]{\hspace{-1cm}\includegraphics[width=\textwidth]{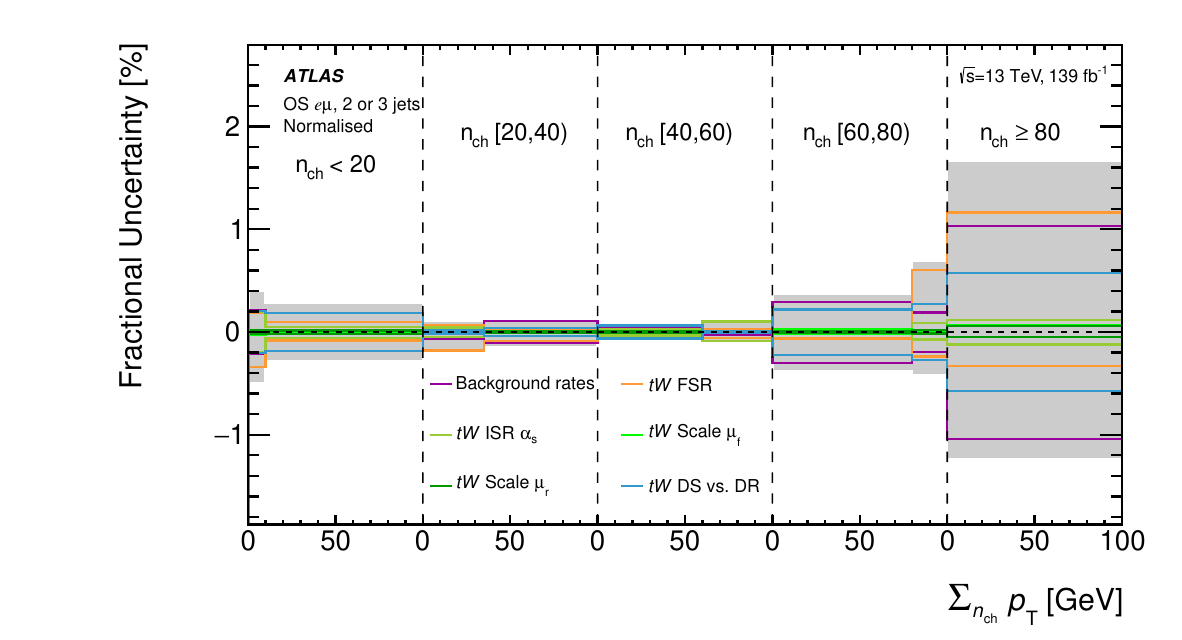}}
 
\caption{\label{fig:totalSyst_background} Fractional uncertainties in the measured normalised differential cross-section as a function of (a) \nch , (b) \sumptch, (c)
\sumptch in bins of \nch for event-based background rate uncertainties and modelling uncertainties for the $tW$ process.
The $x$-axis in (c) is split into five bins of \nch by the dashed vertical lines and \sumptch is presented in each bin.
The grey bands represent the sum in quadrature of the presented components. Events beyond the $x$-axis range are included in the last bin in (a) and (b), and in the corresponding last bin of each slice in (c).}
\end{figure}
 
\subsection{Unfolding-technique uncertainties}
\label{sec:syst:unfolding}
Based on the non-closure of the stress tests using the MC distribution reweighted to background-subtracted data and the cross-reweighting
scenario, an unfolding-technique uncertainty is defined.
It was checked that the cross-reweighting scenario together with generator uncertainties
cover potential differences in the responses of the measured observables due to
track-\pT differences between simulation and data.
The reweighted corrected-level \ttbar distribution is unfolded using the nominal migration matrix and unfolding corrections.
The envelope of relative differences between these unfolded distributions is assigned as a systematic uncertainty, referred to as
the `Unfolding technique' in \Fig{\ref{fig:totalSyst_and_Statistical}}.
For the distributions of \nch and \sumptch, the unfolding-technique uncertainty ranges between $0.1$\% and $2.2$\%, and is less than the statistical uncertainty of the data in all bins except the first one in each of these distributions.
In the two-dimensional measurement of \sumptch in bins of \nch the unfolding-technique uncertainty ranges between $0.3$\% and $3.6$\%, and exceeds the statistical uncertainty of the data in five of the nine bins.

\subsection{\ttbar modelling uncertainties}
\label{subsec:ttbarSyst_modelling}
The choice of MC generator and its settings determines the kinematic properties of simulated \ttbar events. In order to assess uncertainties in different detector responses or different modellings of the event kinematics, an alternative MC sample is unfolded using the nominal \PYTHIA[8] migration matrix
and corrections and then compared with the generated `truth' spectrum of that sample. The systematic uncertainty is then given by the size of the non-closure, which is propagated as a relative uncertainty. This procedure is applied to all following uncertainties if not stated otherwise.
 
Uncertainties in modelling parton showers, hadronisation, the underlying event, and colour reconnections are assigned by comparing the nominal samples with alternative samples for which \POWHEGBOX was interfaced to \HERWIG[7.1.3] instead of \PYTHIA[8.230]. When generating parton showers with \HERWIG, the \MMHT[lo]~\cite{Harland-Lang:2014zoa} PDF set was used with the H7-UE-MMHT~\cite{Bellm:2015jjp} set of tuned parameters. This uncertainty is labelled as `H7' in the following.
 
Uncertainties related to the choice of renormalisation and factorisation scales in the matrix-element calculations, labelled as `Scale $\muR$' and `Scale $\muF$', are evaluated by varying the scales in a correlated way by factors of $2$ and $0.5$.
The scale variations are implemented as generator weights in the nominal sample.
 
The uncertainty due to the scale choice for matching the matrix-element calculation of the \ttbar process to the parton shower is estimated using an additional \ttbar sample produced with the \hdamp parameter set to $3\times \mtop$ instead of $1.5\times \mtop$, while keeping all other generator settings the same as for the nominal sample of \ttbar events. This uncertainty is labelled as `\hdamp' in the following.
 
Uncertainties in the intensity of initial-state and final-state radiation, labelled as `ISR \alphas' and `FSR', are assessed by varying the parameter Var3c of the A14 parton-shower tune within the uncertainties of the tune and by varying the renormalisation scale $\muR$ of final-state radiation, at which the strong coupling constant \alphas is evaluated, by factors of $0.5$ and $2.0$. The two variations, of Var3c and $\muR$, are handled independently.
 
Uncertainties in the modelling of colour reconnection effects are studied using samples
with modified CR settings or using different CR models. A detailed description of the tuned parameter
settings is given in Ref.~\cite{ATL-PHYS-PUB-2017-008}, and the CR models are described in~\Sect{\ref{subsec:CR_Pythia8}}.
The final CR uncertainty is then given by the envelope of the non-closure differences of CR0--CR2 comparisons and labelled as `CR'.
 
The uncertainty due to the choice of PDF is evaluated using the PDF4LHC15 combined PDF set~\cite{Butterworth:2015oua} with 30 symmetric eigenvectors. The \ttbar production sample is reweighted with the central value and eigenvectors of the combined PDF set to construct a set of systematically varied templates.  This uncertainty is labelled as `PDF4LHC'.
 
In all uncertainty evaluations mentioned above the alternative samples or reweighted samples are normalised to the total cross-section of the nominal samples.
 
\Fig{\ref{fig:totalSyst_modelling}} shows the fractional uncertainty from each of the above-mentioned categories for the three observables.
The dominant uncertainties are uncertainties in the modelling of the parton shower and hadronisation and in the choice of colour reconnection model, especially in the tails of the distributions.
The combination of all \ttbar-modelling-related uncertainties is labelled `Signal modelling' in \Fig{\ref{fig:totalSyst_and_Statistical}}.
 
\begin{figure}[htpb]
 
\subfigure[]{\includegraphics[width=0.48\textwidth]{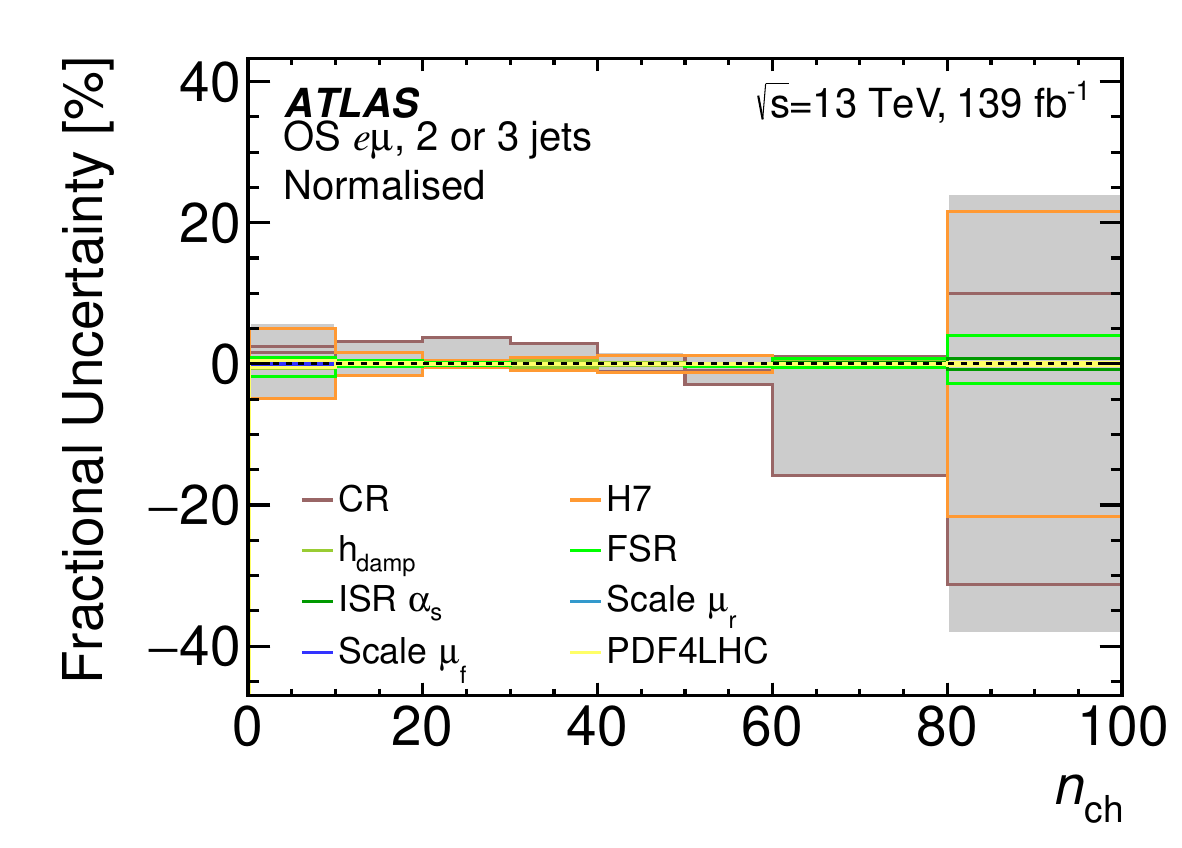}}
\subfigure[]{\includegraphics[width=0.48\textwidth]{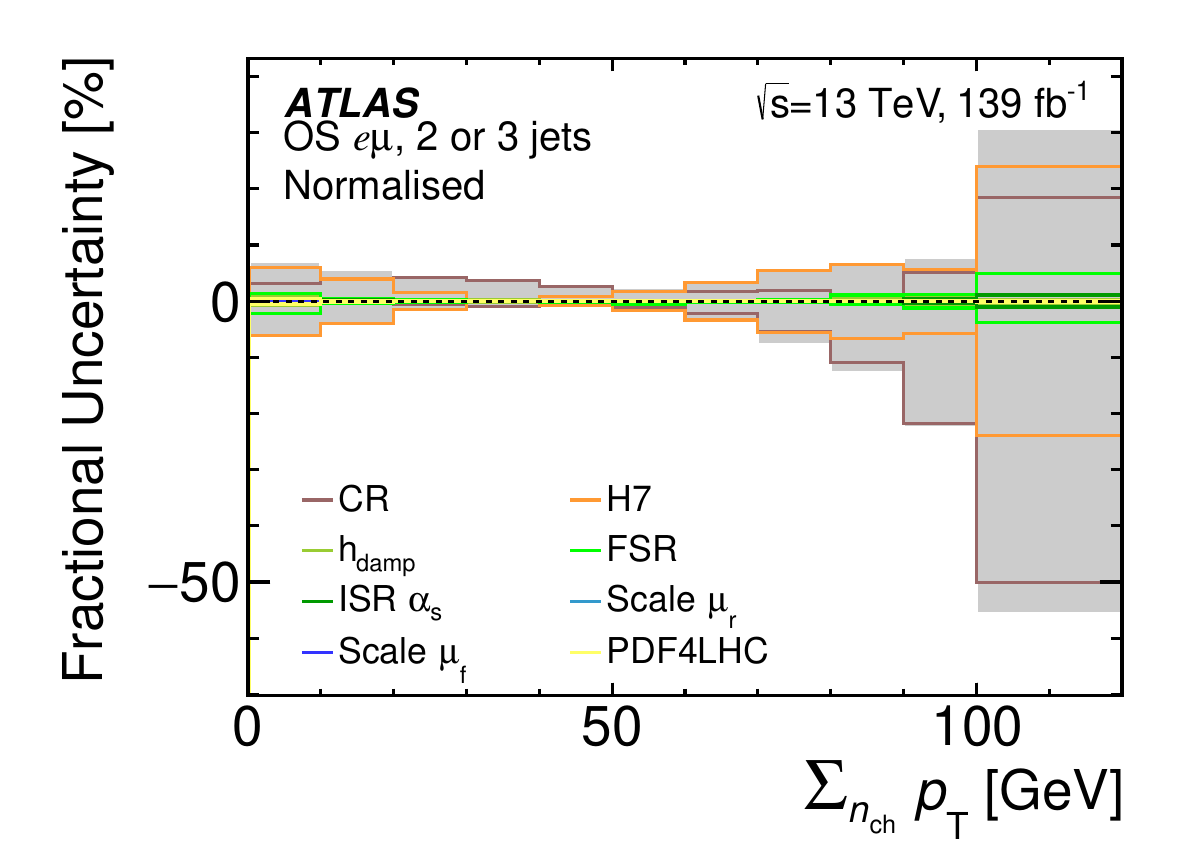}}
\subfigure[]{\hspace{-1cm}\includegraphics[width=\textwidth]{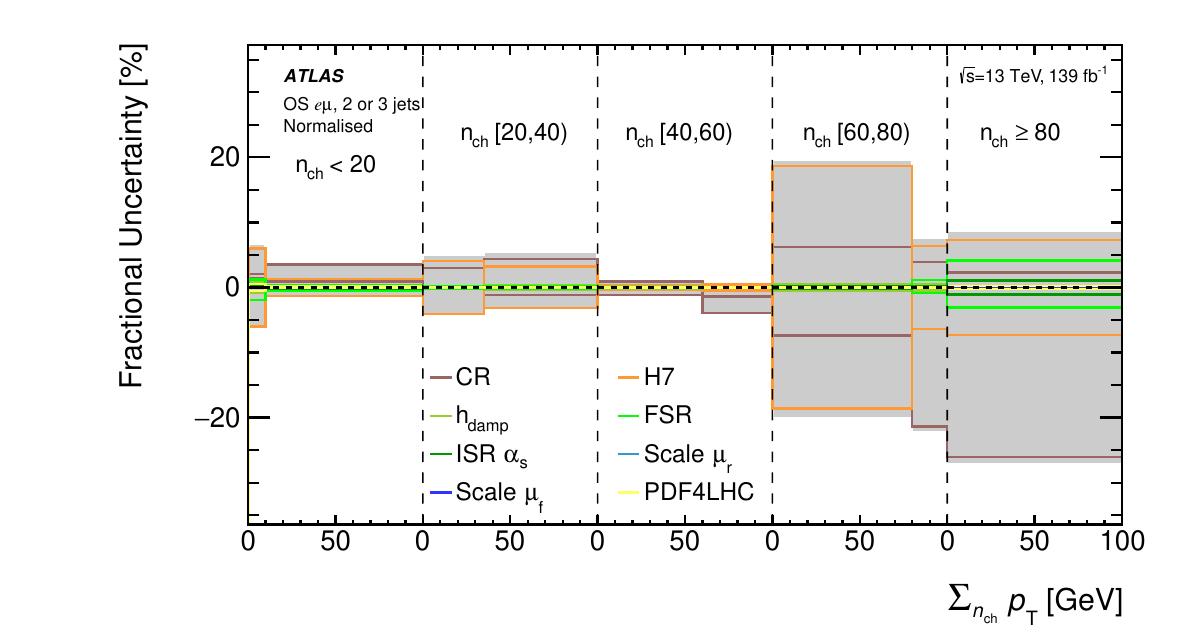}}
 
\caption{\label{fig:totalSyst_modelling} Fractional uncertainties in the measured normalised differential cross-section as a function of (a) \nch , (b) \sumptch, (c) \sumptch in bins of \nch for all \ttbar modelling uncertainties.
The $x$-axis in (c) is split into five bins of \nch by the dashed vertical lines and \sumptch is presented in each bin.
The grey bands represent the sum in quadrature of the presented components. Events beyond the $x$-axis range are included in the last bin in (a) and (b), and in the corresponding last bin of each slice in (c).}
\end{figure}

\subsection{Statistical uncertainties}
Statistical uncertainties are evaluated using pseudo-experiments, where the input distributions are varied and then unfolded.
The RMS of the resulting distribution for each bin is used as an estimator of the uncertainty. Two sets of pseudo-experiments
are used, one to account for the statistical uncertainty of the data and one to account for the limited size of the MC samples.
The combination of these uncertainties is labelled `Statistical' in \Fig{\ref{fig:totalSyst_and_Statistical}}.
 
\subsection{Luminosity}
The uncertainty in the combined 2015--2018 integrated luminosity is \SI{1.7}{\%}~\cite{ATLAS-CONF-2019-021},
obtained using the LUCID-2 detector~\cite{LUCID-2} for the primary luminosity measurements.
The combination of the uncertainty in the luminosity and all detector uncertainties is labelled `Detector' in \Fig{\ref{fig:totalSyst_and_Statistical}}.
 
A summary of the fractional uncertainty from each of the previously mentioned categories is shown for the three observables in \Fig{\ref{fig:totalSyst_and_Statistical}}.
The \pileup tracks background and the modelling uncertainties generally have the largest impact on the measurement.
 
\begin{figure}[htpb]
 
\subfigure[]{\includegraphics[width=0.48\textwidth]{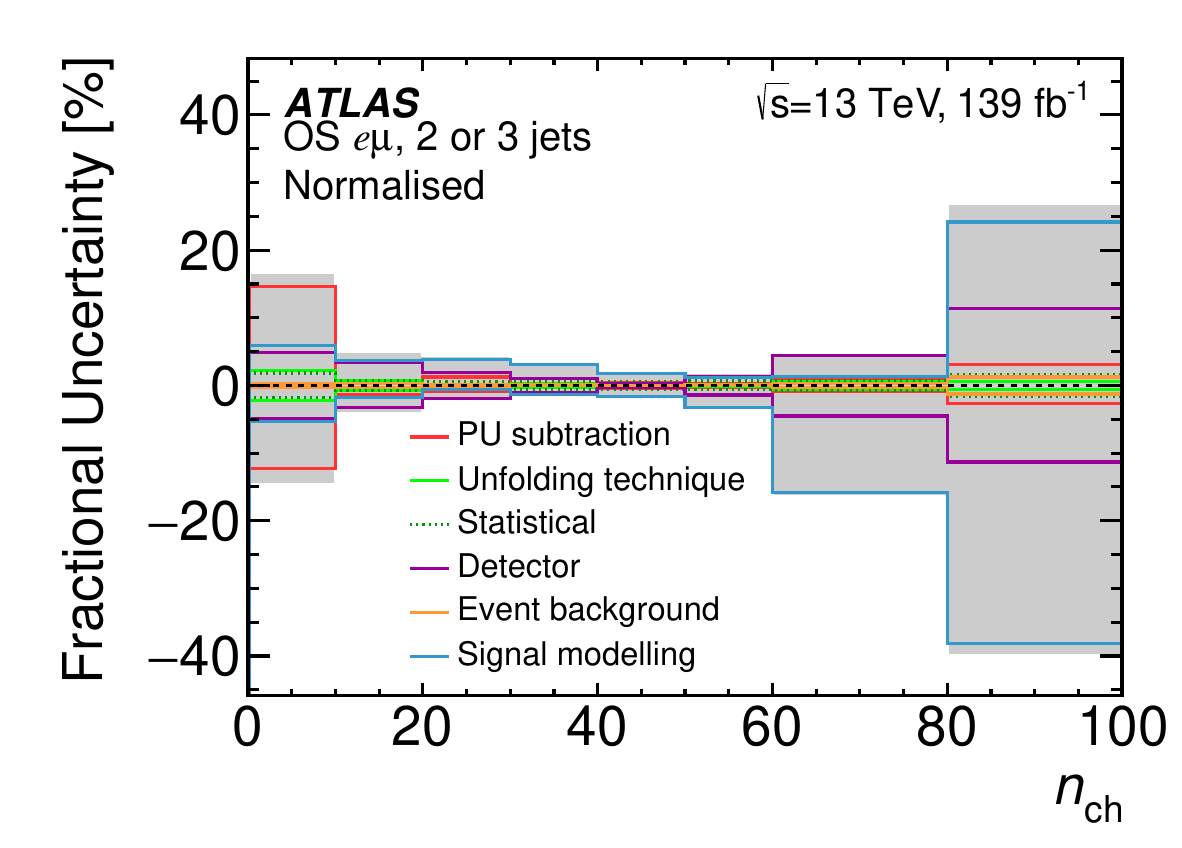}}
\subfigure[]{\includegraphics[width=0.48\textwidth]{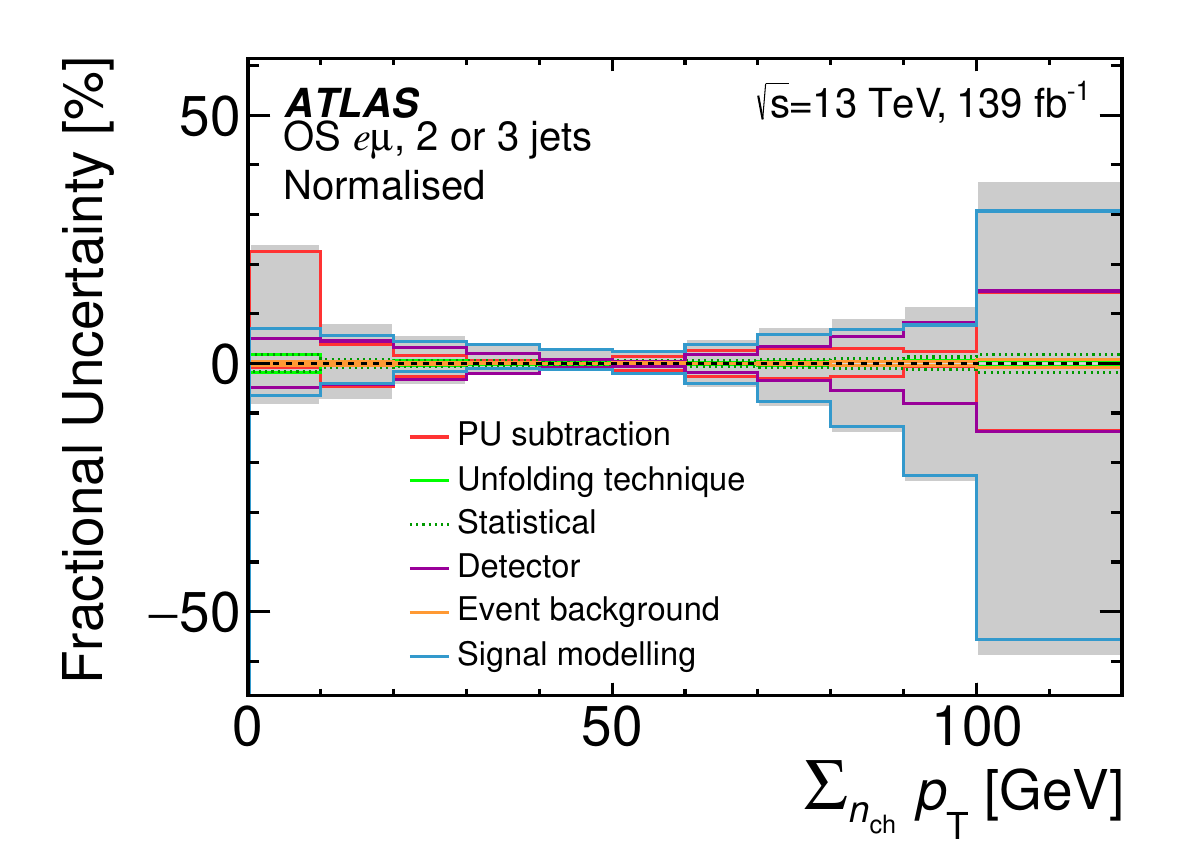}}
\subfigure[]{\hspace{-1cm}\includegraphics[width=\textwidth]{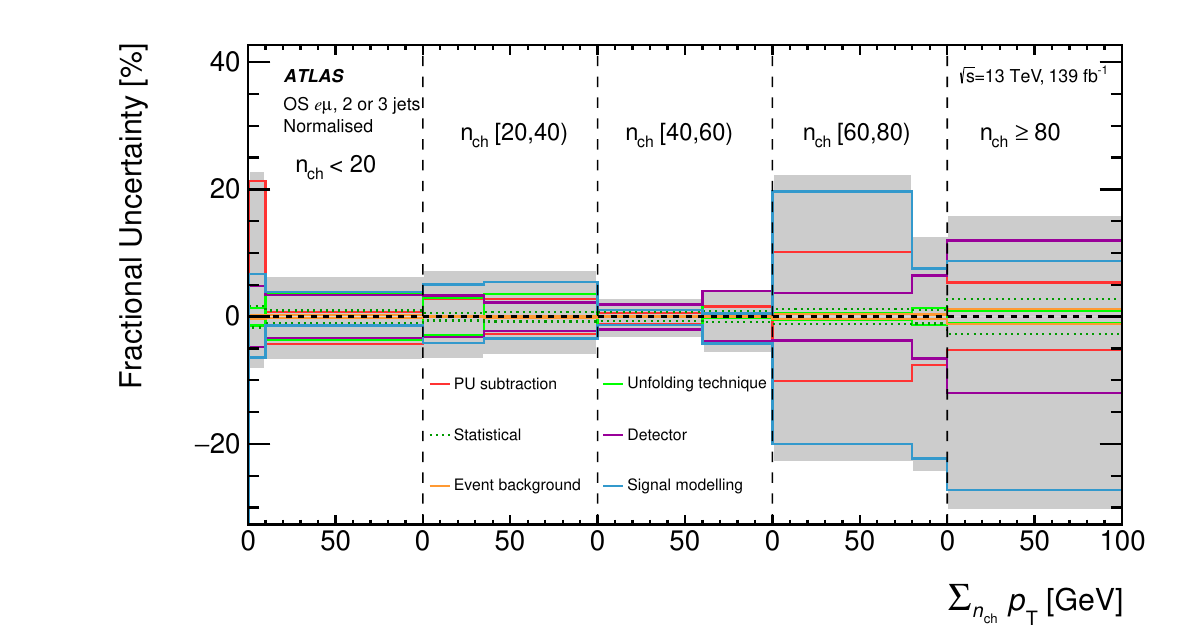}}
 
\caption{\label{fig:totalSyst_and_Statistical} Fractional uncertainties in the measured normalised differential cross-section as a function of (a) \nch , (b) \sumptch, (c)
\sumptch in bins of \nch for all systematic and statistical uncertainties.
The $x$-axis in (c) is split into five bins of \nch by the dashed vertical lines and \sumptch is presented in each bin.
The grey bands represent the sum in quadrature of the presented components. Events beyond the $x$-axis range are included in the last bin in (a) and (b), and in the corresponding last bin of each slice in (c).}
\end{figure}

\subsection{Impact of scale variations in \PYTHIA[8]}
\label{sec:result_scaleVariations}
In order to understand the sensitivity of the measured differential cross-section distributions to variations of  the $\POWHEGBOX\,{+}\,\PYTHIA[8]$ set-up which are not directly connected with CR or UE modelling, the differential cross-sections are compared with those with varied ISR/FSR and \hdamp parameter values.
 
Additional radiation can produce additional charged particles
either by leakage from high-\pT jets or by producing more soft jets. Thus,
the normalised differential cross-sections are compared in \Fig{\ref{fig:unfolded_radiation}} for variations of the \hdamp, ISR and FSR parameters in the nominal $\POWHEGBOX\,{+}\,\PYTHIA[8]$ set-up as described in \Sect{\ref{subsec:ttbarSyst_modelling}}.
The largest differences are seen for variations of  \hdamp. In the bulk of the \nch and \sumptch distributions the differences are only a few percent.
However, for low and high values of \nch and \sumptch, the differences are as large as \SI{15}{\%}.
The influence of FSR and ISR is much weaker and is only significant for low \sumptch values. It can also be seen that the effect of ISR is the exact opposite of that of FSR for these bins.
Based on these observations, it can be concluded that the measurement is only weakly sensitive to modelling of additional radiation in \ttbar events and that differences seen among the predictions from various generator set-ups can be attributed mainly to differences in the CR or UE modelling.
\begin{figure}[htpb]
\begin{center}
\subfigure[]{\includegraphics[width=0.46\textwidth]{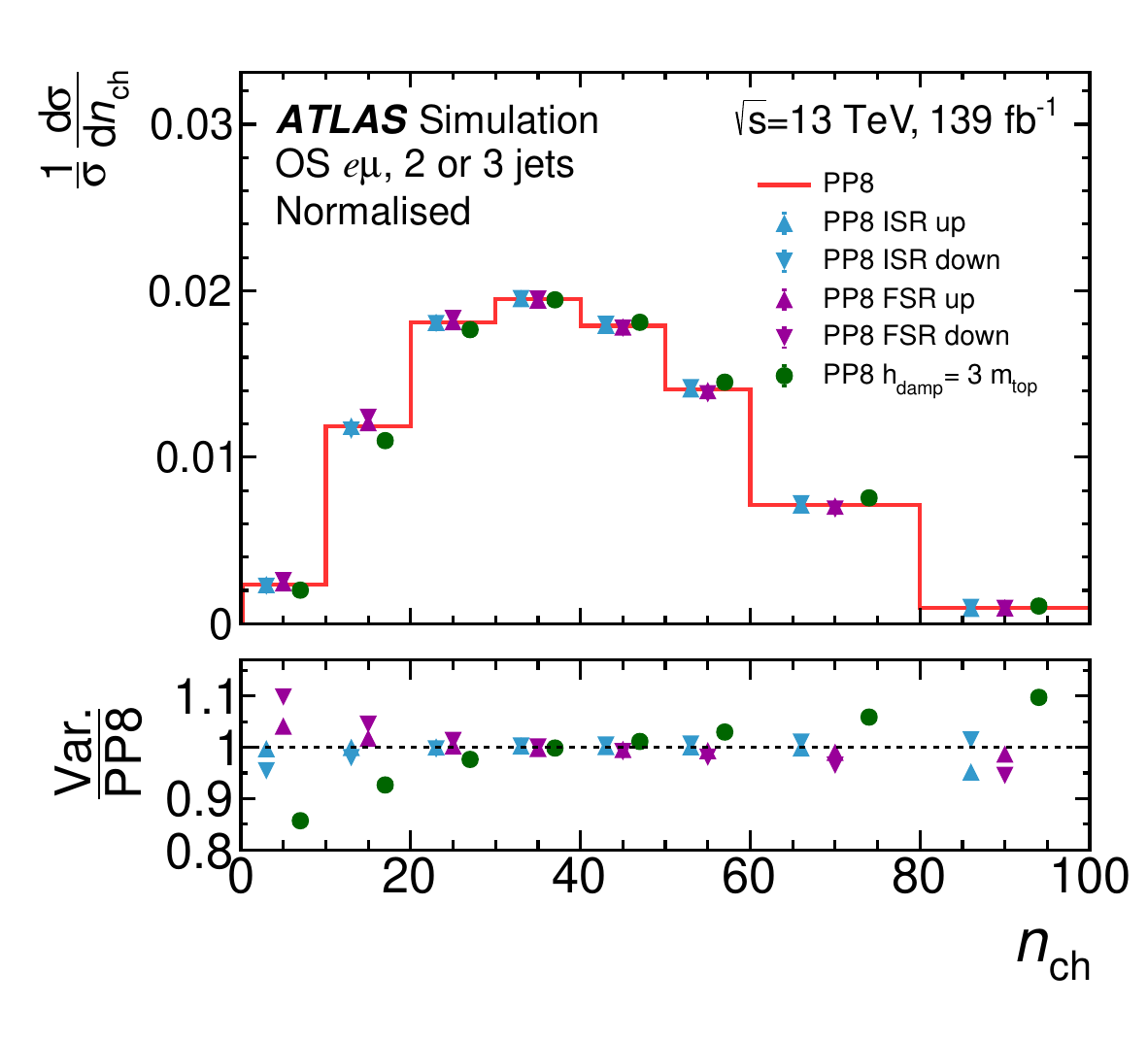}}
\subfigure[]{\includegraphics[width=0.46\textwidth]{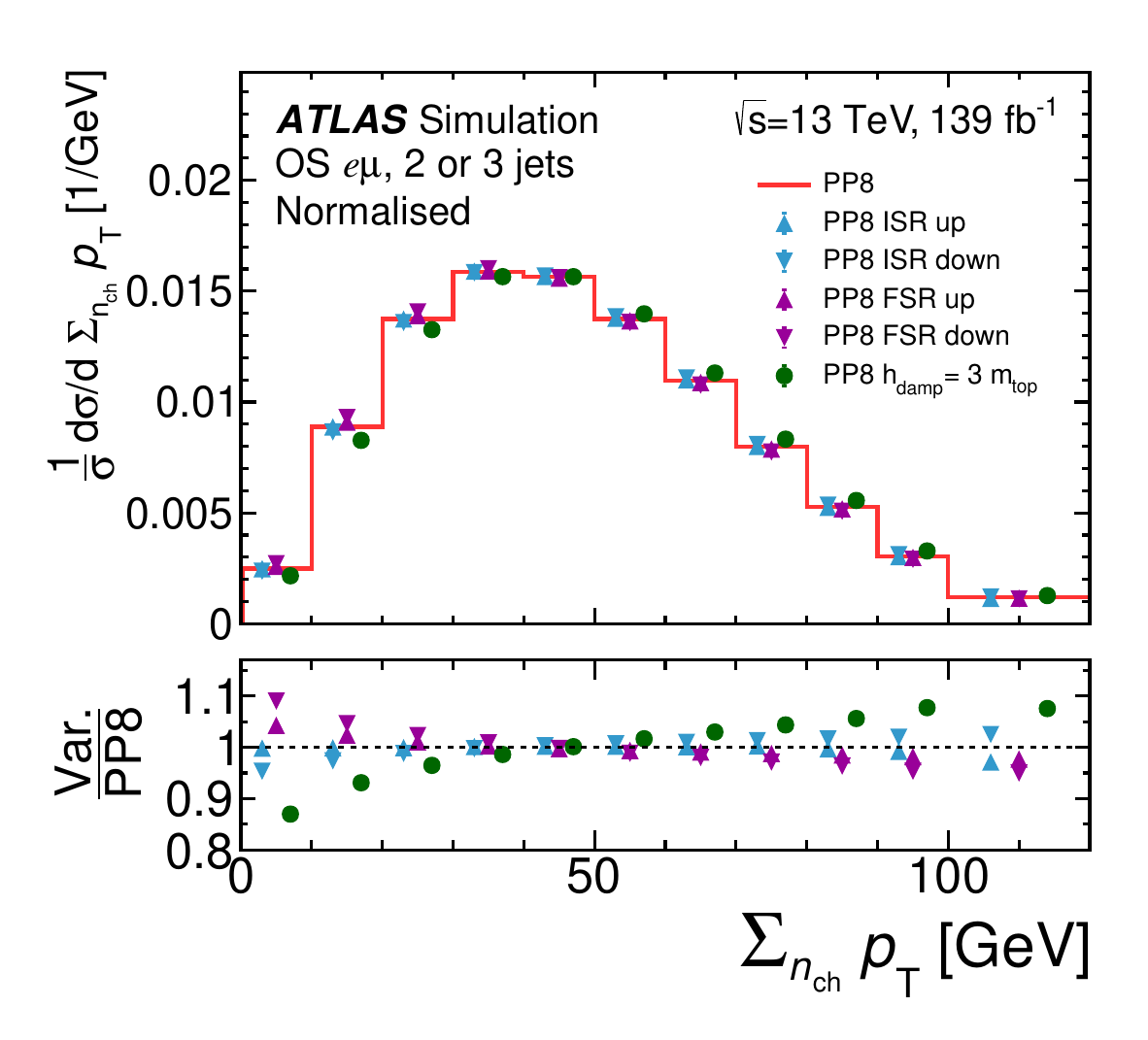}}
\subfigure[]{\includegraphics[width=\textwidth]{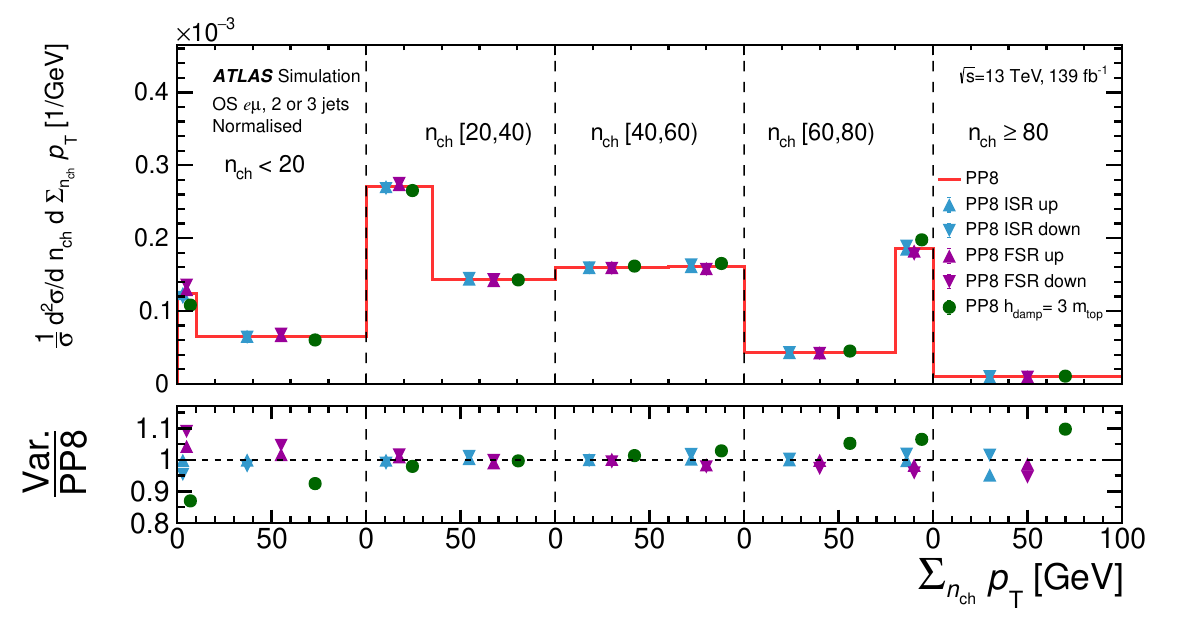}}
\caption{\label{fig:unfolded_radiation} Normalised differential cross-section as a function of (a) \nch, (b) \sumptch, and (c) \sumptch in bins of \nch.
The $x$-axis in (c) is split into five bins of \nch by the dashed vertical lines and \sumptch is presented in each bin.
The nominal \PYTHIA[8] prediction is compared with the predictions from different parameter variations in \PYTHIA[8].
Events beyond the $x$-axis range are included in the last bin.
The lower panels show the ratio of each parameter-varied prediction to the nominal prediction in each bin.}
\end{center}
\end{figure}
 
\FloatBarrier


\section{Results}
\label{sec:results}

In this section, the analysis results are presented by comparing the normalised differential cross-sections with predictions from various MC models with different colour reconnection models and settings. The absolute differential cross-sections can be used to calculate the inclusive cross-section in the fiducial phase-space, and then the normalised differential cross-sections.
At particle level the following requirements, as defined in \Sect{\ref{sec:selection}}, are imposed:
events are required to have exactly one electron and one opposite-sign muon with $\pt > \SI{25}{\GeV}$ and $|\eta| <2.5$ and contain two or three jets with $\pt>\SI{25}{\GeV}$ and $|\eta| < 2.5$, where exactly two of the jets must be \btagged. Requirements of $m_{\ell\ell}>\SI{15}{\GeV}$ and $\nch < 100$ are also applied. The resulting cross-section is:
\begin{equation*}
\sigma_{\ttbar}^{\mathrm{fid}} = 4.94^{+0.43}_{-0.41} \textrm{(syst)} \pm 0.03 \textrm{(stat)} \pm 0.08 \textrm{(lumi)} \, \textrm{pb}.
\end{equation*}
 
In order to quantify how well the measured differential cross-sections agree with the various model predictions, their respective $\chi^2$ values are calculated. The formula for calculating the $\chi^2$ value, given the unfolded data, the model, and the covariance matrix, is defined as follows:
\begin{equation*}
\chi^2 = V_{N_{\text{b}}}^{ \text{T}} \cdot \text{Cov}_{N_{\text{b}}}^{-1} \cdot V_{N_{\text{b}}},
\end{equation*}
where $N_{\text{b}}$ is the number of bins in the spectrum, $V_{N_{\text{b}}}$ is the vector of differences between the unfolded data and the prediction, and $\text{Cov}_{N_{\text{b}}}$ represents the covariance matrix.
 
The covariance matrix incorporates both the statistical and the systematic uncertainties.
The systematic uncertainties include experimental uncertainties, signal modelling and background-related
uncertainties as described in the previous chapter. In addition, uncertainties in the theoretical predictions,
which are independent of the CR modelling, are considered and referred to as `theory uncertainties'.
These theory uncertainties include scale variations in the matrix element and parton shower as well as the
\hdamp variation. The relative difference between the unfolded
variation and the nominal unfolded distribution is taken as uncertainty.
 
The covariance matrix is obtained by performing pseudo-experiments in the following way:
each bin of the measured distribution at the corrected level is varied according to a Poisson distribution
with the number of observed events in this bin as the mean value.
Gaussian-distributed shifts are coherently added for each detector-modelling systematic uncertainty and the MC statistical uncertainty by scaling each Poisson-fluctuated bin with the expected relative variation from the associated systematic uncertainty effect. In the case of a one-sided uncertainty,
the corresponding contribution is symmetrised.
The resulting distribution is then corrected with the nominal event-based background subtraction and passed
through the standard unfolding procedure using the nominal corrections.
Modelling uncertainties and an unfolding-procedure uncertainty are then coherently added with additional Gaussian-distributed shifts.
For the normalised differential cross-sections, $V_{N_{\text{b}}}$ is replaced by $V_{N_{\text{b}-1}}$,
which is the vector of differences between the unfolded data and the prediction, discarding one of the $N_{\text{b}}$ elements.
Consequently, $\text{Cov}_{N_{\text{b}-1}}$ is the sub-matrix derived from the full matrix of the normalised differential
cross-sections by discarding the corresponding row and column. Hence, the obtained sub-matrix is invertible and the $\chi^2$ does not depend on which element is discarded. In practice the last row and column is removed.
 
In addition, a covariance matrix including correlations of systematic uncertainties between the
two observables \nch and \sumptch is calculated and used to derive a so-called global $\chi^2$ value, denoted by `Global(\nch,\sumptch)'.
All of the covariance matrices have large off-diagonal elements, mainly due to the modelling uncertainties.
In the following comparisons of the differential cross-sections with predictions from different event generators using the nominal settings, predictions from CR models implemented in \PYTHIA[8] and CR models implemented in \HERWIG[7] are presented and discussed.
 
\subsection{Comparisons with predictions from different generators}
In \Fig{\ref{fig:unfolded_generator}} the normalised differential cross-sections are compared with the predictions obtained from \POWPY[8.230], \SHERPA[2.2.10], and
\POWHER[7] using three versions of \HERWIG[7], namely 7.0.4, 7.1.3 and 7.2.1. These three versions have different UE and CR tunes and partially modified MPI models.
\Tab{\ref{tab:chi2_gen_norm}} shows the $\chi^2/\text{NDF}$ values resulting from the comparisons of the measured differential cross-sections with the corresponding prediction from these generators for each observable.
\Fig{\ref{fig:unfolded_generator}} shows that the unfolded data disagree with the predictions from \SHERPA[2.2.10], which does not include CR effects, thus illustrating their importance.
The \nch measured cross-section is approximately equally well described by \PYTHIA[8.230] and \HERWIG[7], while the \sumptch measured normalised cross-section has a much better $\chi^2$ value for \HERWIG[7]. With increasing \HERWIG[7] version number, the predicted and observed \nch distributions show worse agreement, while for \sumptch the level of agreement remains about the same.
The distribution of \twod is best described by the three versions of \HERWIG[7] in comparison with \PYTHIA[8.230] and \SHERPA[2]. Relative to \PYTHIA[8.230], the lower $\chi^2$ value for \SHERPA[2] is due to the inclusion of the theory uncertainties.
Overall, the best agreement is achieved using \HERWIG[7.0.4].
 
\begin{table}[htbp]
\begin{center}
\caption{The $\chi^2$ and NDF for measured normalised differential cross-sections obtained by comparing the different predictions with the unfolded data. Global(\nch,\sumptch) denotes the scenario in which the covariance matrix is built including correlations of systematic uncertainties between the two observables \nch and \sumptch.}
\label{tab:chi2_gen_norm}
\begin{tabular}{lcccc}
\toprule
Observable         		&   \nch  	&   \sumptch 	&   Global(\nch,\sumptch)		&  \sumptch in bins of \nch                       \\
NDF					& 7		& 10			& ~~17					& ~~8       \\
\hline
Generator set-up         	        				&   \multicolumn{4}{c}{$\chi^2$}  \\
\hline
\POWPY[8.230] 	   	& 62	    	& 106~~		& 434					& 224~~	\\
CR0        				& 55		& 113~~	 	& 629					& 129~~	\\
CR1        			       	& 98		& 60	 		& 581					& 158~~	\\
CR2                        		& 58		& 179~~ 		& 402					& 238~~	\\
\POWHER[7.0.4]     		& 39	    	& 16			& 145					& 29		\\
\POWHER[7.1.3]     		& 53		& 42			& 188					& 41		\\
\POWHER[7.2.1] 	   	& 78		& 25			& 313					& 87		\\
\POWHER Baryonic CR    & 75		& 20	   		& 241					& 29		\\
\POWHER Stat CR        	& 23 		& 40	   		& 121					& 39		\\
\SHERPA[2.2.10] 		& 77		& 211~~		& 263					& 124~~	\\
\bottomrule
\end{tabular}
 
\end{center}
\end{table}
 
\begin{figure}[htpb]
\begin{center}
\subfigure[]{\includegraphics[width=0.46\textwidth]{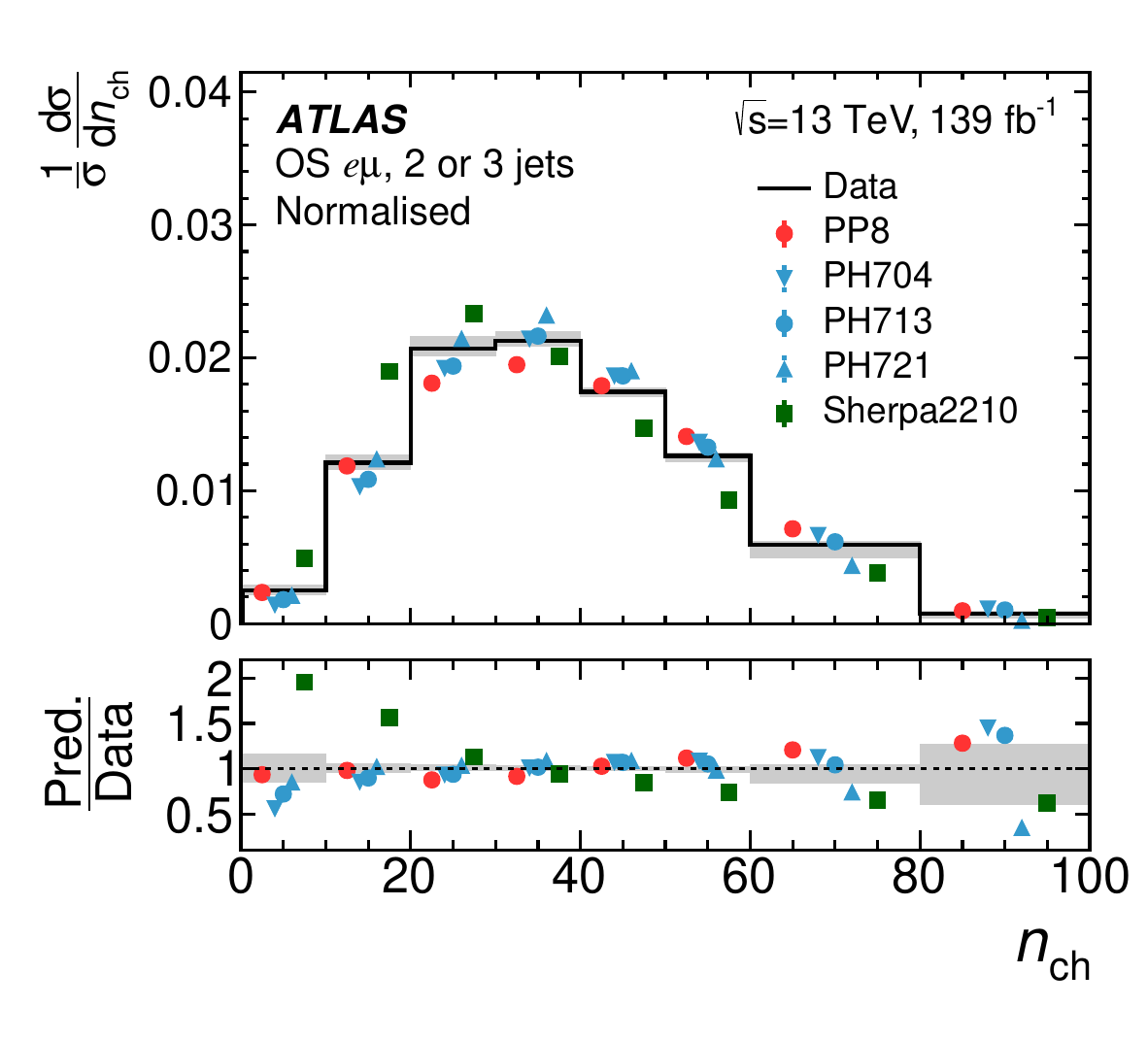}}
\subfigure[]{\includegraphics[width=0.46\textwidth]{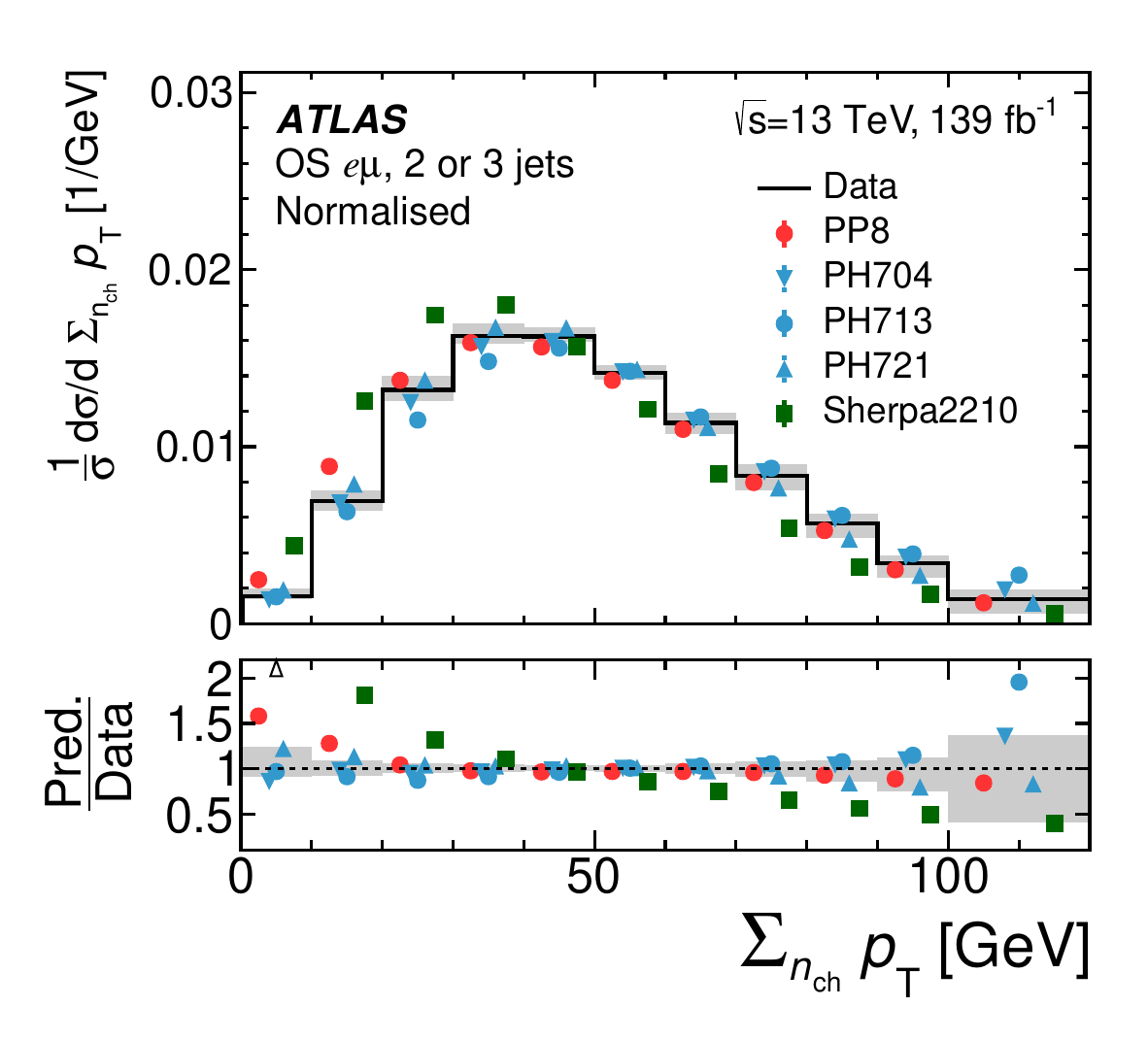}}
\subfigure[]{\hspace{-1cm}\includegraphics[width=\textwidth]{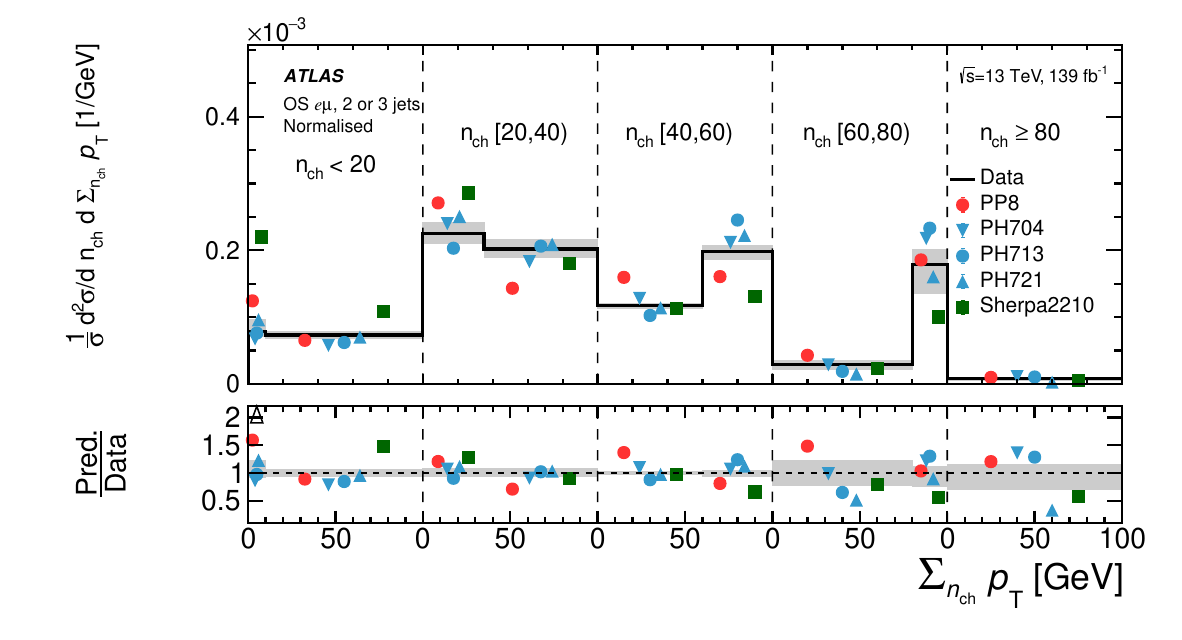}}
\caption{\label{fig:unfolded_generator} Normalised differential cross-section as a function of (a) \nch, (b) \sumptch, and (c) \sumptch in bins of \nch.
The $x$-axis in (c) is split into five bins of \nch by the dashed vertical lines and \sumptch is presented in each bin.
Unfolded data are shown as the black line with the grey band corresponding to the total uncertainty in each bin.
The results are compared with the predictions of different MC event generators.
Events beyond the $x$-axis range are included in the last bin.
The lower panels show the ratio of each prediction to data in each bin.
The black triangular markers in the lower panels point to prediction-to-data ratio values which lie beyond the shown $y$-axis range.
}
\end{center}
\end{figure}
 
\FloatBarrier
\subsection{Comparison with different CR models in \PYTHIA[8]}
 
In \Fig{\ref{fig:unfolded_atlas_CR}}, the normalised differential cross-sections are compared with predictions from different CR models implemented in \PYTHIA[8].
\Tab{\ref{tab:chi2_gen_norm}} shows $\chi^2/\text{NDF}$ values resulting from the comparisons between the measured differential cross-sections and the predictions.
The CR0 model is based on the same MPI-based CR model as the nominal \POWPY[8.230] sample. One can see that the unfolded data is in better agreement with the sample using the dedicated CR0 set of tuned parameters than with the nominal sample without a dedicated CR tune, but that the global $\chi^2$ is somewhat larger.
For the CR1 model, \sumptch is described best, while \nch is worst among the three CR models. In contrast, the \sumptch distribution, obtained with the CR2 model, is in worst agreement with data, while the other distributions have $\chi^2/\text{NDF}$ values comparable to those obtained from \POWPY[8.230] with the A14 set of tuned parameters.
It can also be seen that neither these CR models nor the A14 tune can describe the lower part of the \sumptch distribution.
 
\begin{figure}[htpb]
\begin{center}
\subfigure[]{\includegraphics[width=0.46\textwidth]{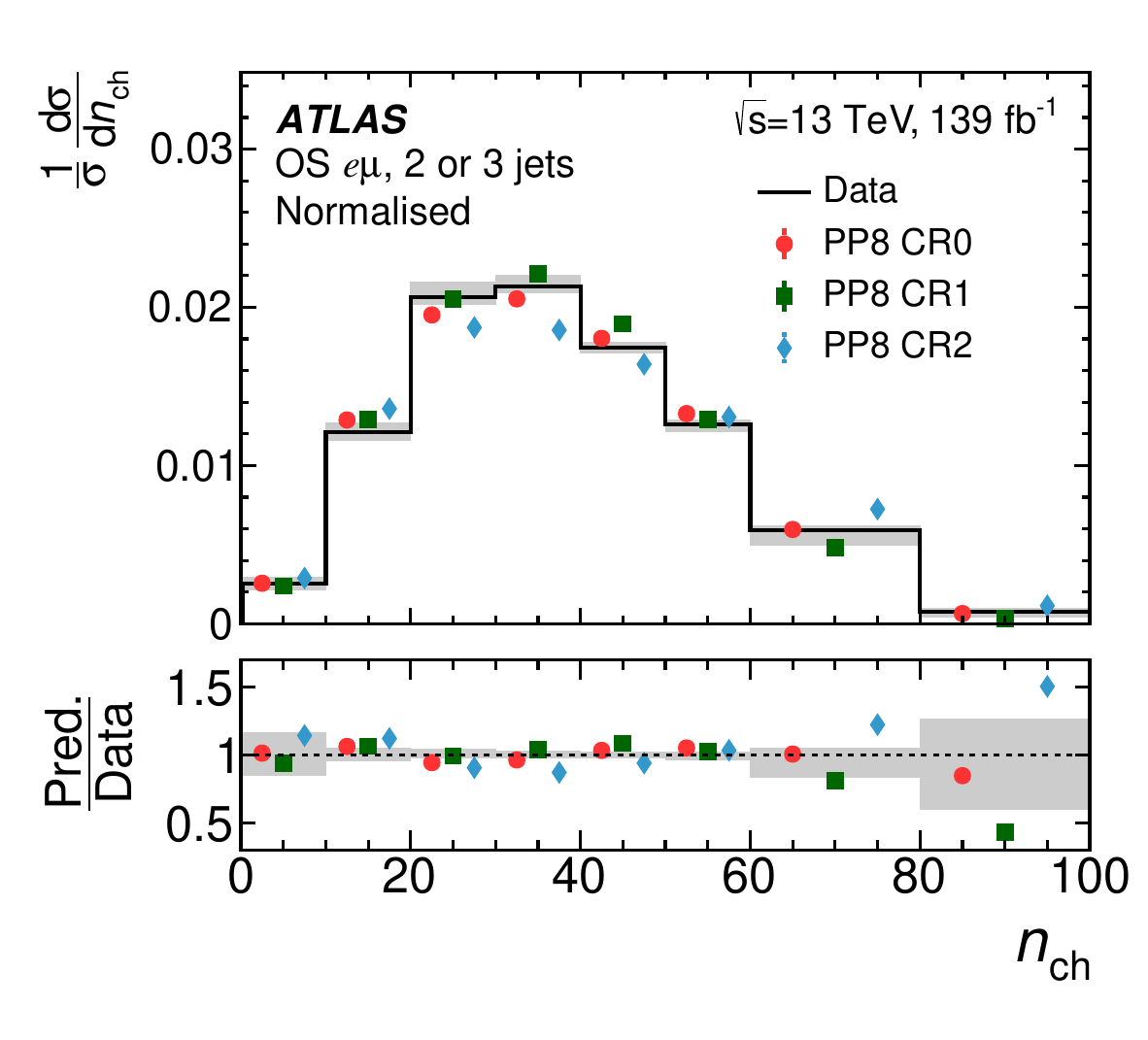}}
\subfigure[]{\includegraphics[width=0.46\textwidth]{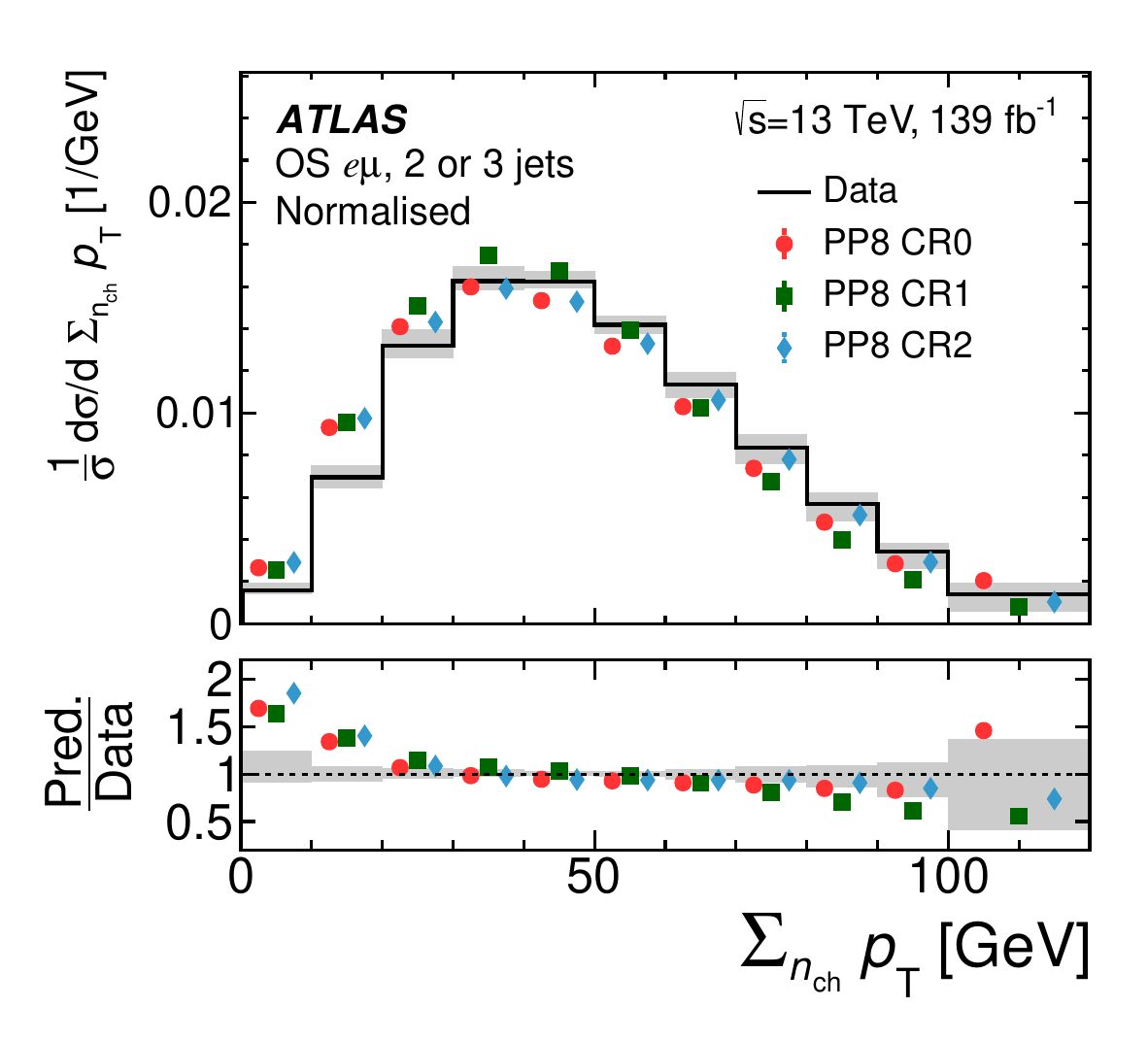}}
\subfigure[]{\hspace{-1cm}\includegraphics[width=\textwidth]{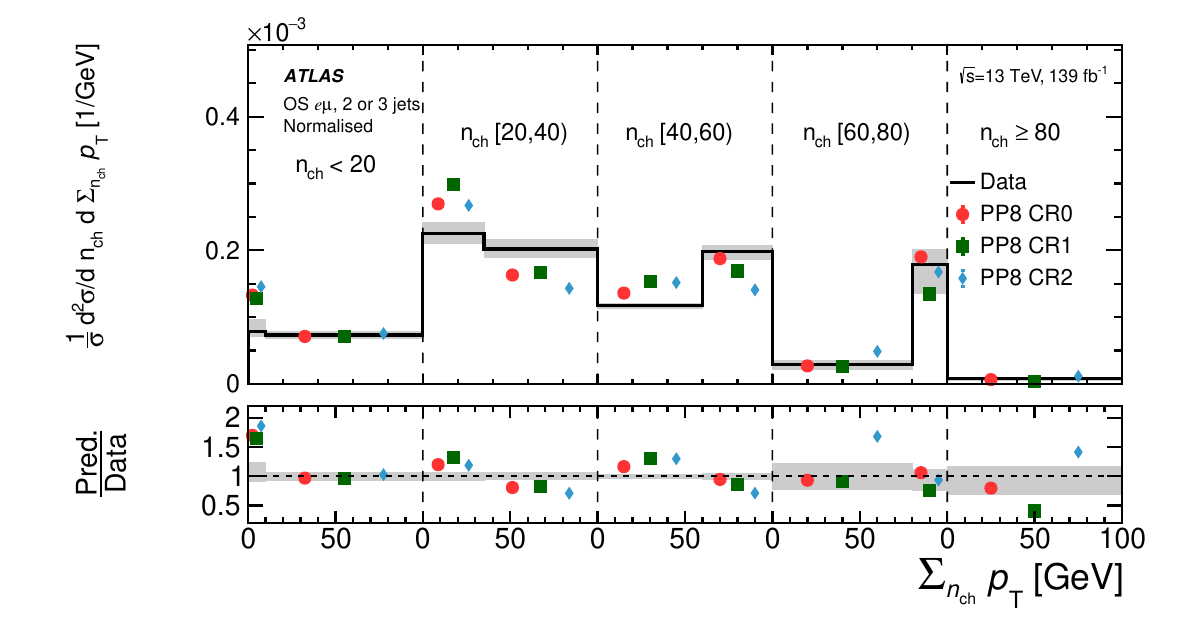}}
\caption{\label{fig:unfolded_atlas_CR} Normalised differential cross-section as a function of (a) \nch, (b) \sumptch, and (c) \sumptch in bins of \nch.
The $x$-axis in (c) is split into five bins of \nch by the dashed vertical lines and \sumptch is presented in each bin.
Unfolded data are shown as the black line with the grey band corresponding to the total uncertainty in each bin.
The results are compared with the predictions of different CR models in \PYTHIA[8]. Events beyond the $x$-axis range are included in the last bin.
The lower panels show the ratio of each prediction to data in each bin.}
\end{center}
\end{figure}
 
\subsection{\PYTHIA[8] CR and UE parameters}
 
In this section, the normalised differential cross-sections are compared with the predictions obtained from \PYTHIA[8] with variations of specific parameters connected with the CR and UE modelling. These variations are applied to the nominal \PYTHIA[8] set-up using CR0 and the A14 tune. The varied parameters are:
\begin{itemize}
\item CR range parameter $R_{\mathrm{rec}}$ is set to its maximal value of $R_{\mathrm{rec}}=10$, such that the reconnection probability reaches saturation (maxCR).
The default is $1.71$.
\item CR is switched off, i.e.\ $R_{\mathrm{rec}}=0$ (noCR).
\item MPI parameter $p_{\mathrm{T}0}^{\mathrm{ref}}$, by default set to 2.09, is lowered to 2.0 and raised to 2.2.
\item UE activity is varied by using the Var1 eigentune of the A14 tune~\cite{ATL-PHYS-PUB-2014-021}.
This eigentune includes variations of $\alphas$ and variations of $R_{\mathrm{rec}}$.
\end{itemize}
In \Fig{\ref{fig:unfolding_pythia_CR}} the normalised differential cross-sections are compared with both the predictions with these varied parameter values and the predictions of the nominal \PYTHIA[8] sample.
The predictions without CR are not compatible with the data, while predictions using a maximal probability for colour reconnection in the default \PYTHIA[8] CR0 model are still compatible with the measurement. Another observation is that the overall
\sumptch value is not sensitive to the CR strength, but it is sensitive within the \nch ranges.
Finally, a clear sensitivity to the variation of  $p_{\mathrm{T0}}^{\mathrm{ref}}$ is observed
and the presented measurement would suggest a value of $p_{\mathrm{T}0}^{\mathrm{ref}}$ lower than the one used in the A14 tune. Thus, it can be concluded that for further tuning of CR models both the parameters of the CR model itself and the parameters of the UE should be included.
\begin{figure}[htpb]
\begin{center}
\subfigure[]{\includegraphics[width=0.46\textwidth]{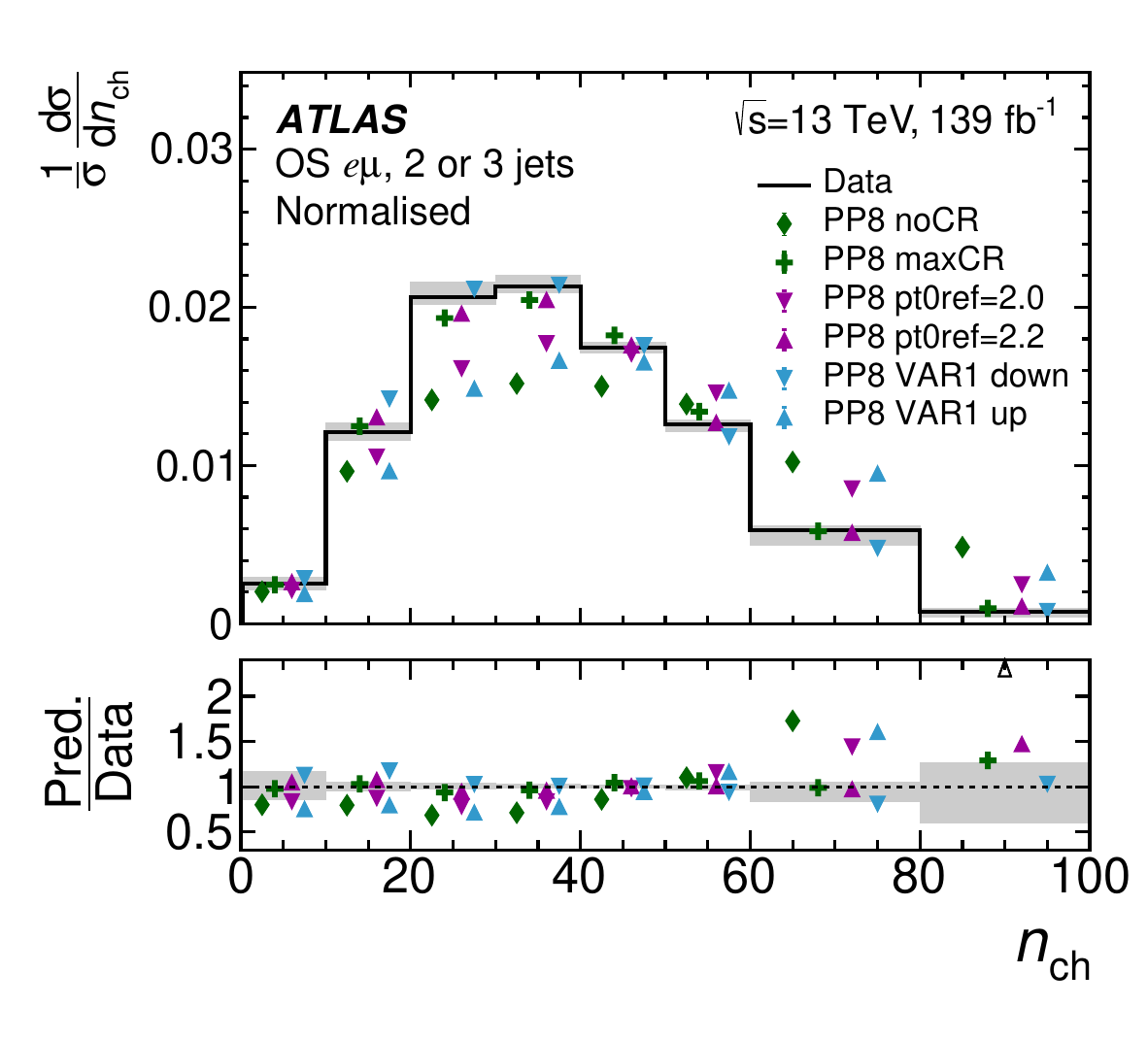}}
\subfigure[]{\includegraphics[width=0.46\textwidth]{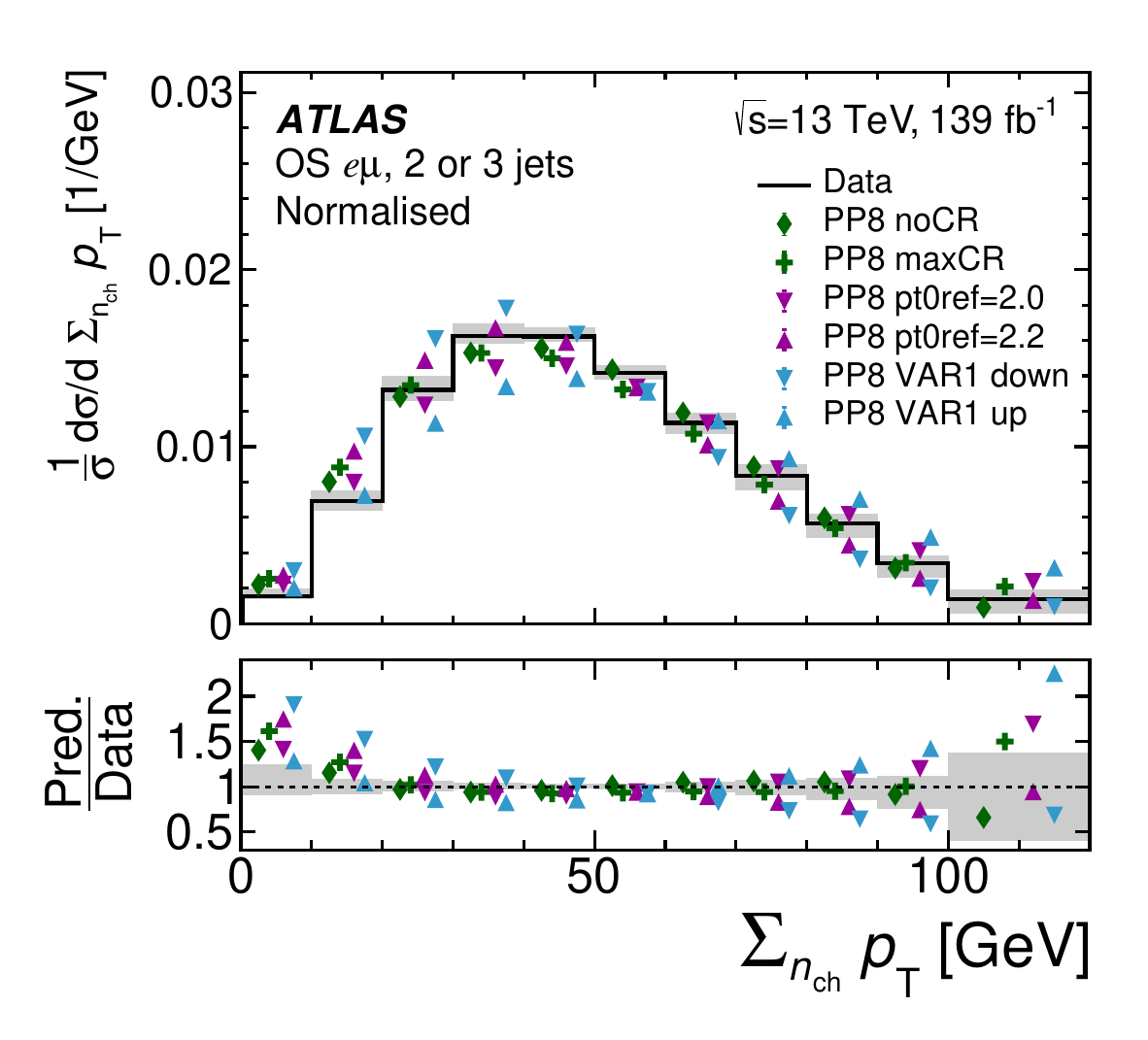}}
\subfigure[]{\hspace{-1cm}\includegraphics[width=\textwidth]{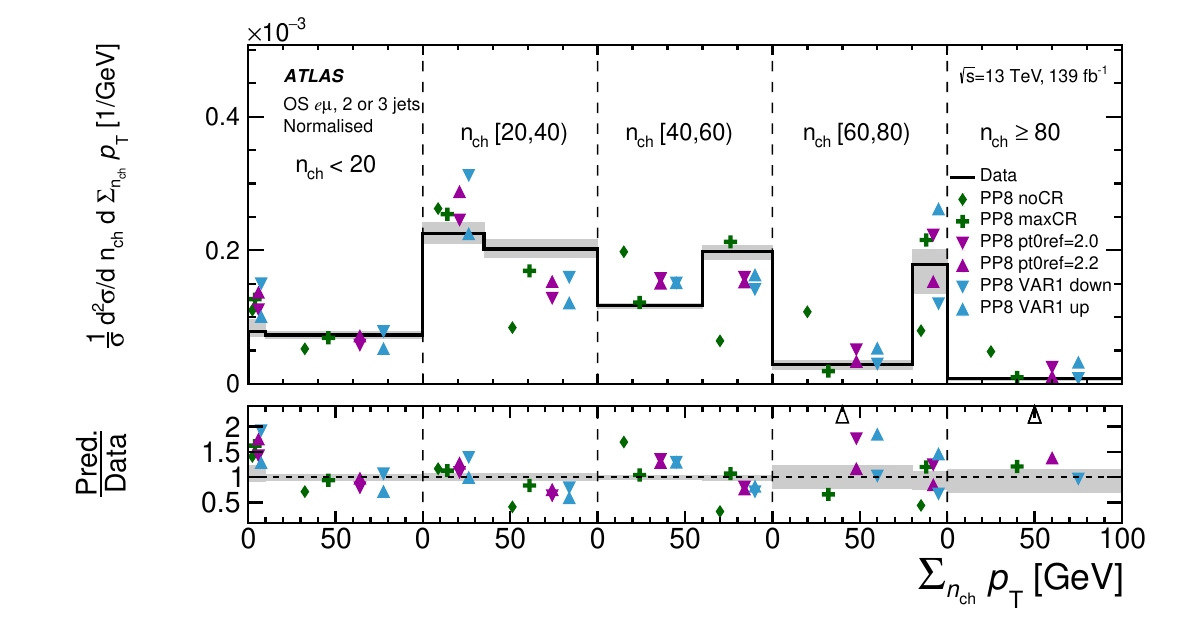}}
\caption{\label{fig:unfolding_pythia_CR} Normalised differential cross-section as a function of (a) \nch, (b) \sumptch, and (c) \sumptch in bins of \nch.
The $x$-axis in (c) is split into five bins of \nch by the dashed vertical lines and \sumptch is presented in each bin.
Unfolded data are shown as the black line with the grey band corresponding to the total uncertainty in each bin.
The results are compared with the predictions of different CR and UE parameter variations in \PYTHIA[8].
Events beyond the $x$-axis range are included in the last bin.
The lower panels show the ratio of each prediction to data in each bin.
The black triangular markers in the lower panels point to prediction-to-data ratio values which lie beyond the shown $y$-axis range.
}
\end{center}
\end{figure}
 
\FloatBarrier
\subsection{\HERWIG[7.2] CR models}
 
In this section, the normalised differential cross-sections are compared with the predictions obtained from different CR models implemented in \HERWIG[7.2], as described in~\Sect{\ref{subsec:CR_Herwig7}}. In \Fig{\ref{fig:unfolding_herwig_CR}} the
normalised differential cross-sections are compared with distributions from both the predictions of the different CR models and predictions without any CR.
Similarly to \PYTHIA[8], predictions without CR do not describe data well, but in the case of
\HERWIG[7.2] some deficiencies are compensated for by the UE modelling.
\Tab{\ref{tab:chi2_gen_norm}} shows $\chi^2$ values resulting from the comparisons between the
measured differential cross-sections and the predictions. For the \nch distribution the newly introduced statistical CR model performs the best, while for \sumptch the baryonic CR model agrees better with the observed distribution. Thus, even without dedicated retuning of the alternative CR models, they agree better with the measured data than the default plain CR model. In general, agreement with the \sumptch distribution is better than for any other generator considered in this analysis.
 
\begin{figure}[htpb]
\begin{center}
\subfigure[]{\includegraphics[width=0.46\textwidth]{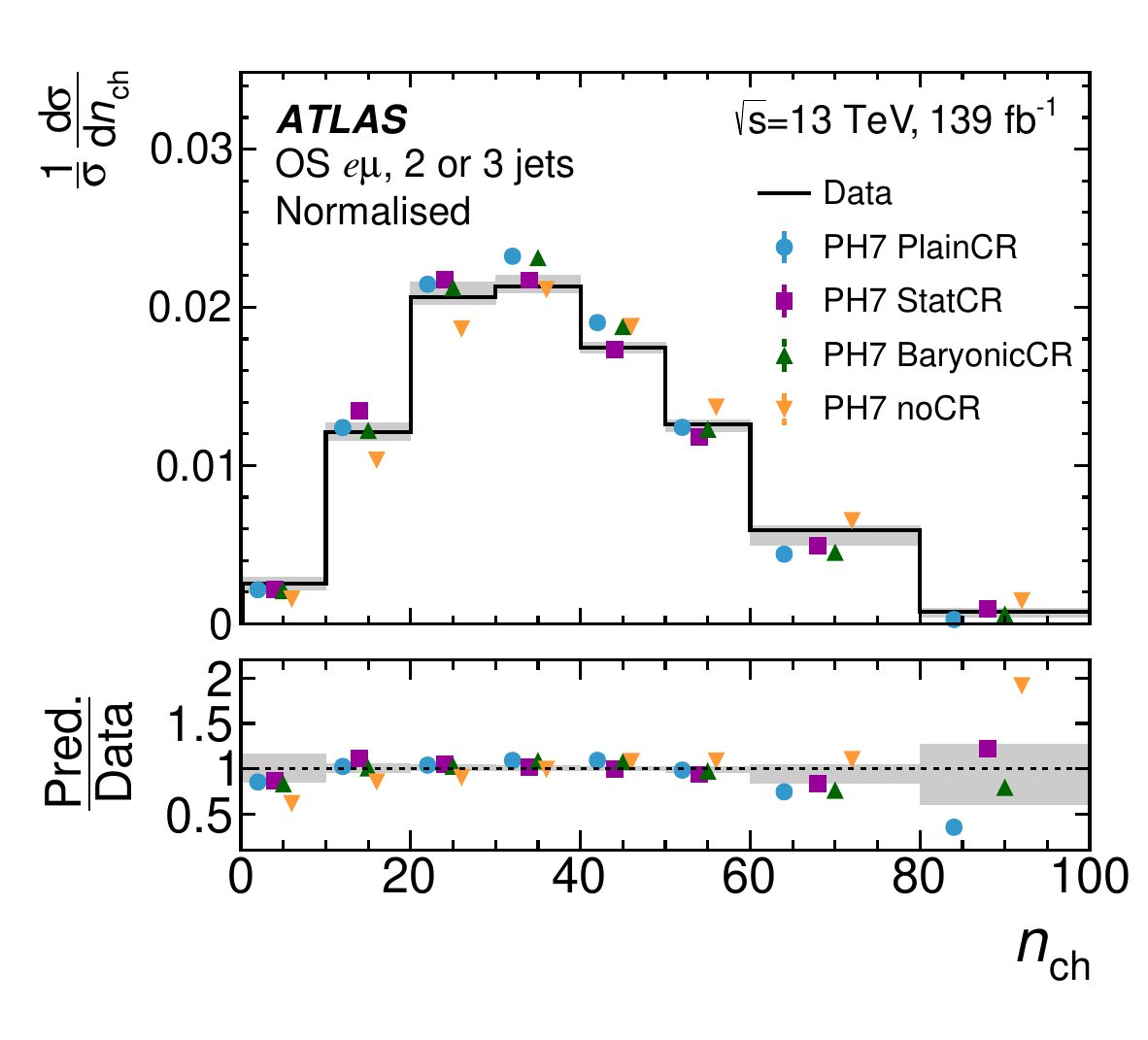}}
\subfigure[]{\includegraphics[width=0.46\textwidth]{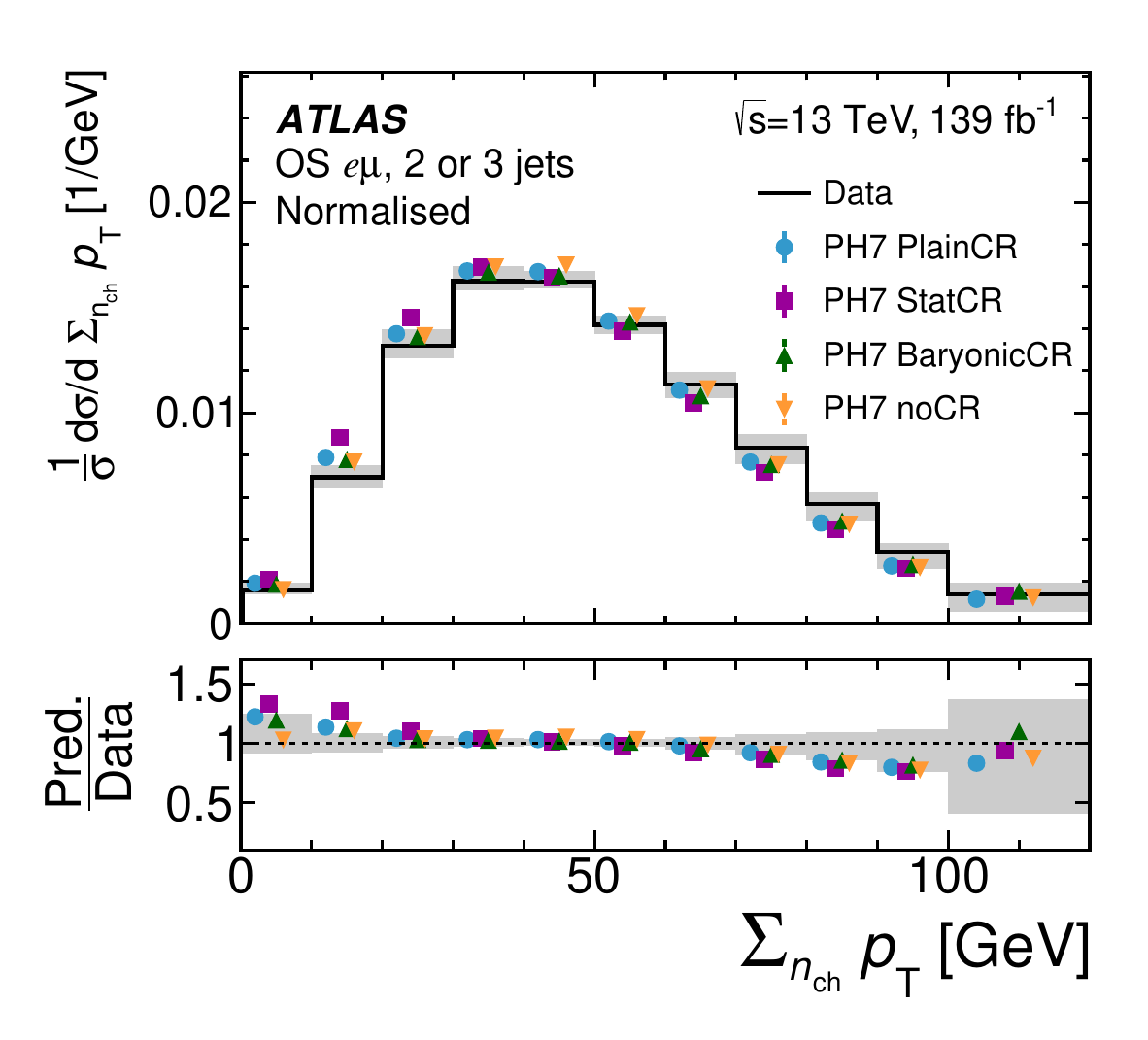}}
\subfigure[]{\hspace{-1cm}\includegraphics[width=\textwidth]{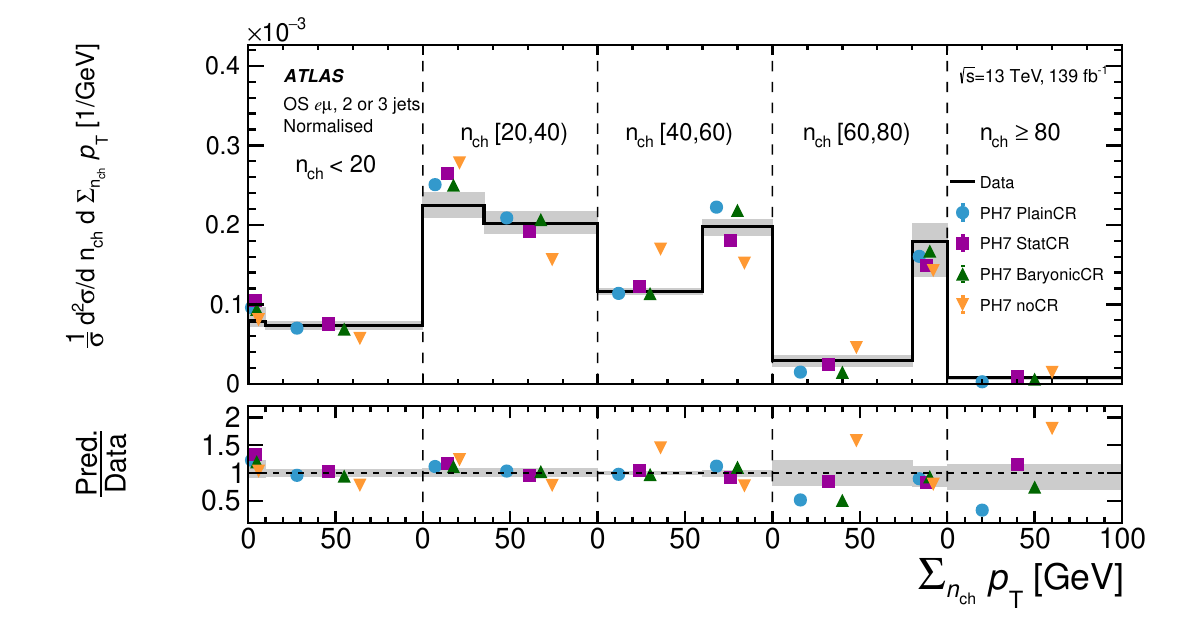}}
\caption{\label{fig:unfolding_herwig_CR}  Normalised differential cross-section as a function of (a) \nch, (b) \sumptch, and (c) \sumptch in bins of \nch.
The $x$-axis in (c) is split into five bins of \nch by the dashed vertical lines and \sumptch is presented in each bin.
Unfolded data are shown as the black line with the grey band corresponding to the total uncertainty in each bin.
The results are compared with the predictions of different CR models in \HERWIG[7.2].
Events beyond the $x$-axis range are included in the last bin.
The lower panels show the ratio of each prediction to data in each bin. }
\end{center}
\end{figure}
 
\subsection{Top-quark-specific CR models of \PYTHIA[8]}
 
In this section, the normalised differential cross-sections are compared with the
predictions obtained from different top-quark-specific CR models implemented in \PYTHIA[8],
as described in~\Sect{\ref{subsec:CR_Pythia8}}. In \Fig{\ref{fig:unfolding_TCR}} the
normalised differential cross-sections are compared with the distributions from the different predictions,
where the strength parameter ($s$) is set to either $s=0.1$ or $s=1.0$.
Distributions obtained using $s=1.0$ represent the maximal CR effect of the respective model.
The spread between the distributions obtained with $s=0.1$ and $s=1.0$ is therefore
a measure of the sensitivity of the observables to the corresponding model.
Of the five models studied, only TCR1 and TCR3 show sizeable differences between the two settings.

\begin{figure}[htpb]
\begin{center}
\subfigure[]{\includegraphics[width=0.46\textwidth]{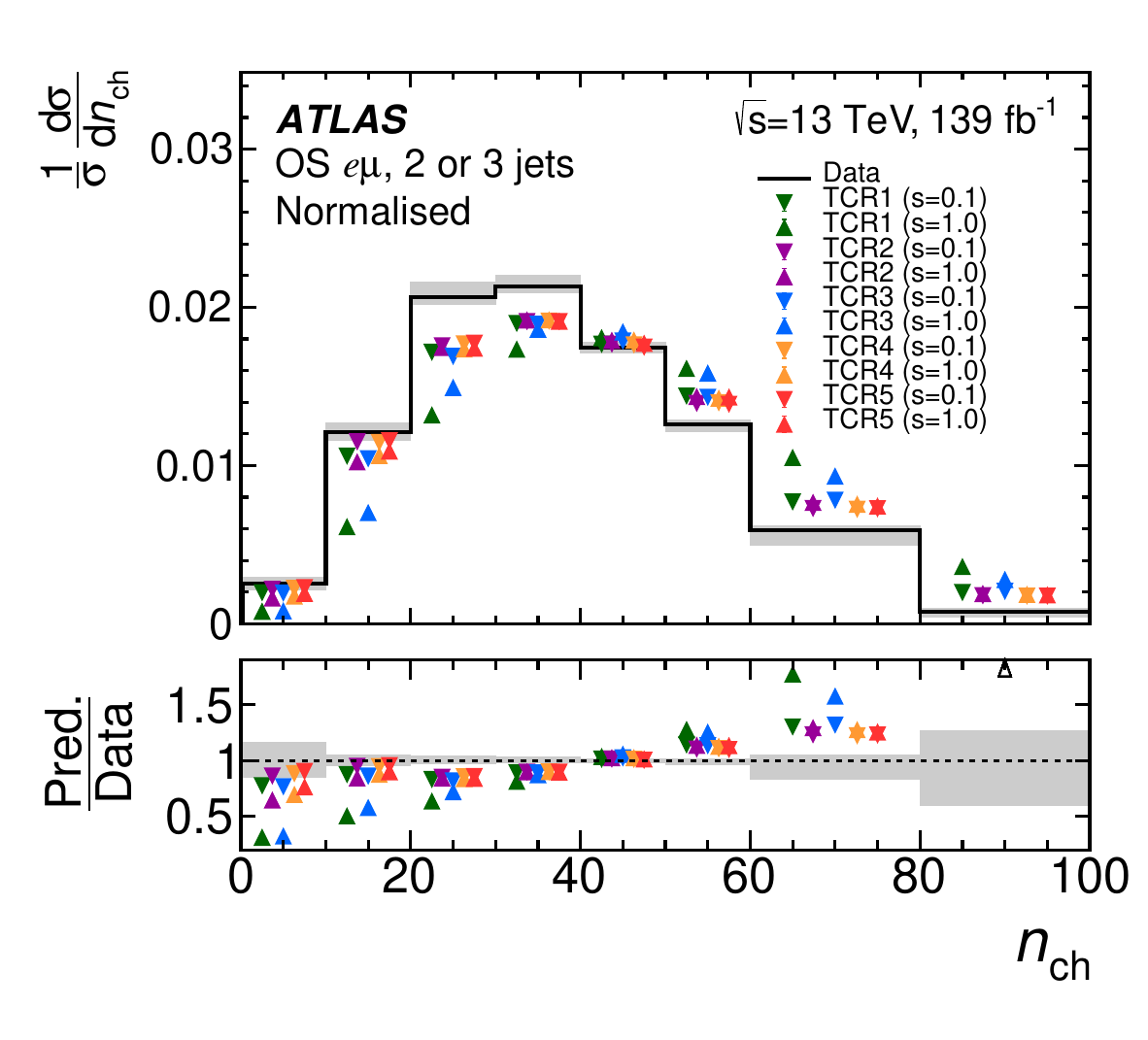}}
\subfigure[]{\includegraphics[width=0.46\textwidth]{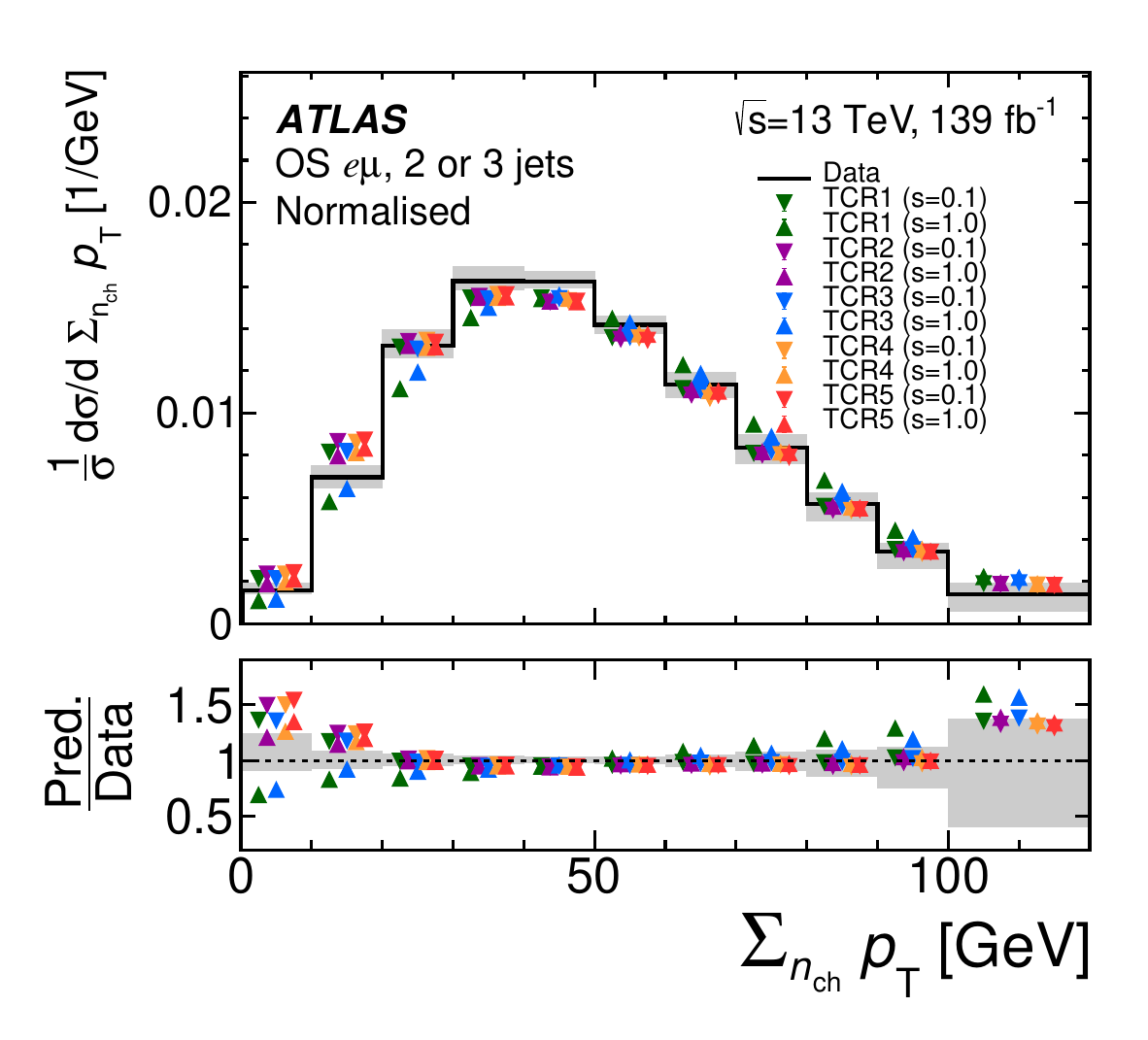}}
\subfigure[]{\hspace{-1cm}\includegraphics[width=\textwidth]{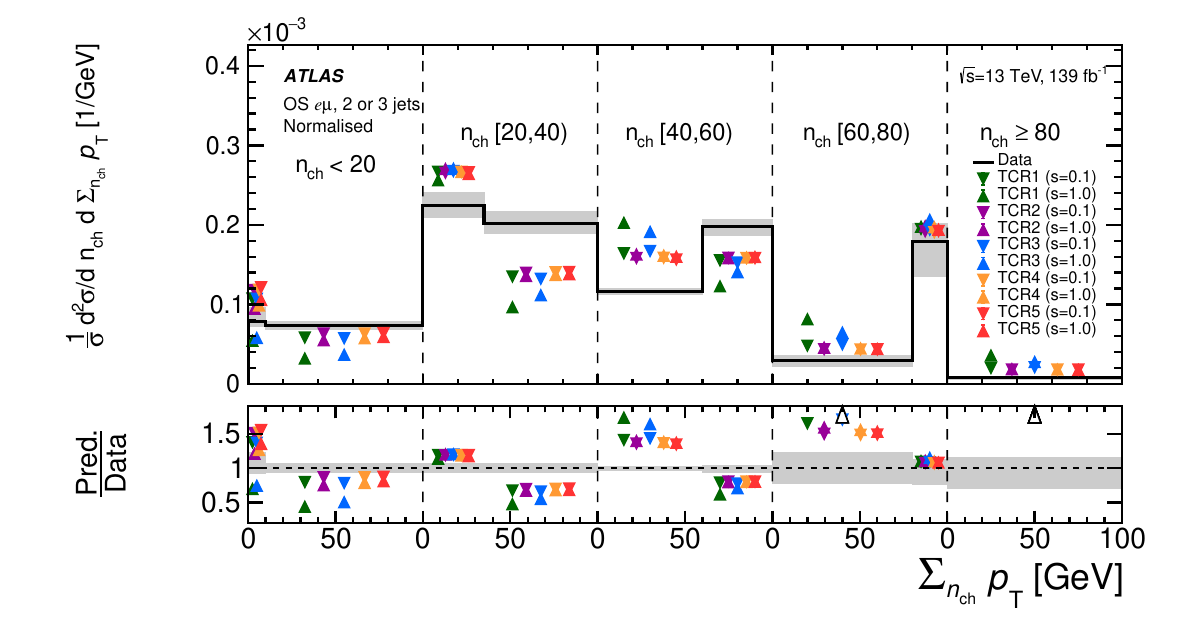}}
\caption{\label{fig:unfolding_TCR}  Normalised differential cross-section as a function of (a) \nch, (b) \sumptch, and (c) \sumptch in bins of \nch.
The $x$-axis in (c) is split into five bins of \nch by the dashed vertical lines and \sumptch is presented in each bin.
Unfolded data are shown as the black line with the grey band corresponding to the total uncertainty in each bin.
The results are compared with the predictions of different top-quark-specific CR models in \PYTHIA[8].
Events beyond the $x$-axis range are included in the last bin. The lower panels show the ratio of each prediction to data in each bin.
The black triangular markers in the lower panels point to prediction-to-data ratio values which lie beyond the shown $y$-axis range.
}
\end{center}
\end{figure}
 
\FloatBarrier


\section{Summary and conclusion}
\label{sec:conclusion}
 
A measurement of three observables sensitive to colour reconnection effects in
\ttbar events is presented using a data sample of \SI{139}{\ifb} recorded by the ATLAS experiment at the LHC in proton--proton collisions at a centre-of-mass energy of $\sqrt{s} = \SI{13}{\tev}$ from 2015 to 2018.
The charged-particle multiplicity, the scalar sum of their transverse momenta, and the scalar sum in bins of charged-particle multiplicity are measured in a fiducial phase space in events with exactly one isolated electron and one isolated muon with opposite charge, and two or three jets, where exactly two jets are required to be $b$-tagged.
The observables are unfolded to stable-particle level and they are compared with predictions from parton-shower programs.
The measurement has a precision of \SI{2}{\%}--\SI{10}{\%} in the central bins and up to \SI{50}{\%} in the outer bins.
The combination of the first two observables, and a double-differential measurement of the two quantities, provide stringent constraints on the colour reconnection models in mainstream Monte Carlo generators.
Using these results as a benchmark, more precise models can be developed in a dedicated tuning effort.
Distributions of \sumptch obtained from \HERWIG[7.2] are able to describe data well, while predictions from \PYTHIA[8.2] for \nch are better than those for \sumptch.
The event generator \SHERPA[2.2.10] does not include a colour reconnection model, which should lead to a higher particle multiplicity because the suppression mechanism from CR is missing. Instead, it predicts too few particles, and also its  \sumptch distribution is softer than measured in data.
Among the studied \PYTHIA[8] colour reconnection models the best agreement with data is achieved with the default MPI-based set of tuned parameters and the largest disagreement is found for the set of parameters based on the gluon-move model.
In the case of the MPI-based model, variations of the colour-reconnection range parameter to its minimal and maximal values are investigated and, similarly to \SHERPA, a model without colour reconnection is excluded, while a maximal strength is still compatible with the measurement. Finally, top-quark-specific colour reconnection models are compared with the measured distributions,
and sensitivity to strength-parameter variations is mainly seen in the distribution of \twod for two of the five models.
Variations of the \PYTHIA[8] MPI model parameter $p_{\mathrm{T0}}^{\mathrm{ref}}$ showed that a value lower than the one used in the A14 set of tuned parameters is suggested by the measurement.
In the case of \HERWIG[7.2] the newly introduced statistical
colour-reconnection model agrees best for the \nch distribution, while for the \sumptch distribution the baryonic CR model agrees best.
 
The effects of varying the colour reconnection models can be used to develop estimates of a colour reconnection uncertainty. The chosen observables are not only sensitive to parameters related to colour reconnection models, but also to the MPI parameters. Thus, these results could be used as input to future tuning of Monte Carlo generators for both colour reconnection and MPI parameters, which should be handled simultaneously.
 
\section*{Acknowledgements}


We thank CERN for the very successful operation of the LHC, as well as the
support staff from our institutions without whom ATLAS could not be
operated efficiently.
 
We acknowledge the support of
ANPCyT, Argentina;
YerPhI, Armenia;
ARC, Australia;
BMWFW and FWF, Austria;
ANAS, Azerbaijan;
CNPq and FAPESP, Brazil;
NSERC, NRC and CFI, Canada;
CERN;
ANID, Chile;
CAS, MOST and NSFC, China;
Minciencias, Colombia;
MEYS CR, Czech Republic;
DNRF and DNSRC, Denmark;
IN2P3-CNRS and CEA-DRF/IRFU, France;
SRNSFG, Georgia;
BMBF, HGF and MPG, Germany;
GSRI, Greece;
RGC and Hong Kong SAR, China;
ISF and Benoziyo Center, Israel;
INFN, Italy;
MEXT and JSPS, Japan;
CNRST, Morocco;
NWO, Netherlands;
RCN, Norway;
MEiN, Poland;
FCT, Portugal;
MNE/IFA, Romania;
MESTD, Serbia;
MSSR, Slovakia;
ARRS and MIZ\v{S}, Slovenia;
DSI/NRF, South Africa;
MICINN, Spain;
SRC and Wallenberg Foundation, Sweden;
SERI, SNSF and Cantons of Bern and Geneva, Switzerland;
MOST, Taiwan;
TENMAK, T\"urkiye;
STFC, United Kingdom;
DOE and NSF, United States of America.
In addition, individual groups and members have received support from
BCKDF, CANARIE, Compute Canada and CRC, Canada;
PRIMUS 21/SCI/017 and UNCE SCI/013, Czech Republic;
COST, ERC, ERDF, Horizon 2020 and Marie Sk{\l}odowska-Curie Actions, European Union;
Investissements d'Avenir Labex, Investissements d'Avenir Idex and ANR, France;
DFG and AvH Foundation, Germany;
Herakleitos, Thales and Aristeia programmes co-financed by EU-ESF and the Greek NSRF, Greece;
BSF-NSF and MINERVA, Israel;
Norwegian Financial Mechanism 2014-2021, Norway;
NCN and NAWA, Poland;
La Caixa Banking Foundation, CERCA Programme Generalitat de Catalunya and PROMETEO and GenT Programmes Generalitat Valenciana, Spain;
G\"{o}ran Gustafssons Stiftelse, Sweden;
The Royal Society and Leverhulme Trust, United Kingdom.
 
The crucial computing support from all WLCG partners is acknowledged gratefully, in particular from CERN, the ATLAS Tier-1 facilities at TRIUMF (Canada), NDGF (Denmark, Norway, Sweden), CC-IN2P3 (France), KIT/GridKA (Germany), INFN-CNAF (Italy), NL-T1 (Netherlands), PIC (Spain), ASGC (Taiwan), RAL (UK) and BNL (USA), the Tier-2 facilities worldwide and large non-WLCG resource providers. Major contributors of computing resources are listed in Ref.~\cite{ATL-SOFT-PUB-2021-003}.


\appendix
\part*{Appendices}
\addcontentsline{toc}{part}{Appendices}
 
\section{Migration matrices}

 
\Fig{\ref{fig:migration}} shows the migration matrices used as input to the unfolding method for the observables \nch, \sumptch, and \sumptch in bins of \nch.
The corrected-level observable is shown on the $x$-axis, while the particle-level observable is shown on the $y$-axis. The values are
normalised so that each row sums to unity.
 
\begin{figure}[htbp]
\begin{center}
\subfigure[]{\includegraphics[width=0.46\textwidth]{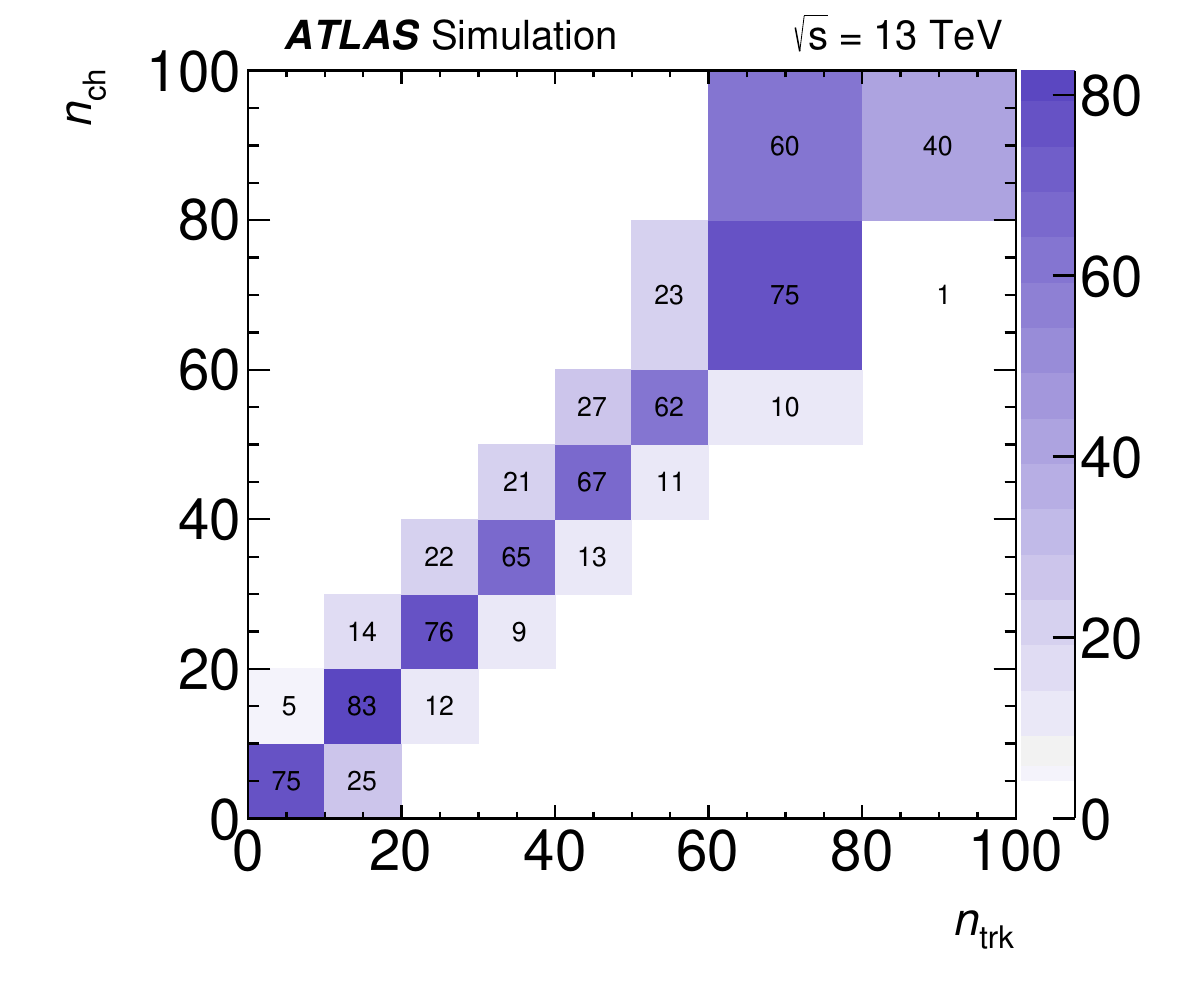}}
\subfigure[]{\includegraphics[width=0.46\textwidth]{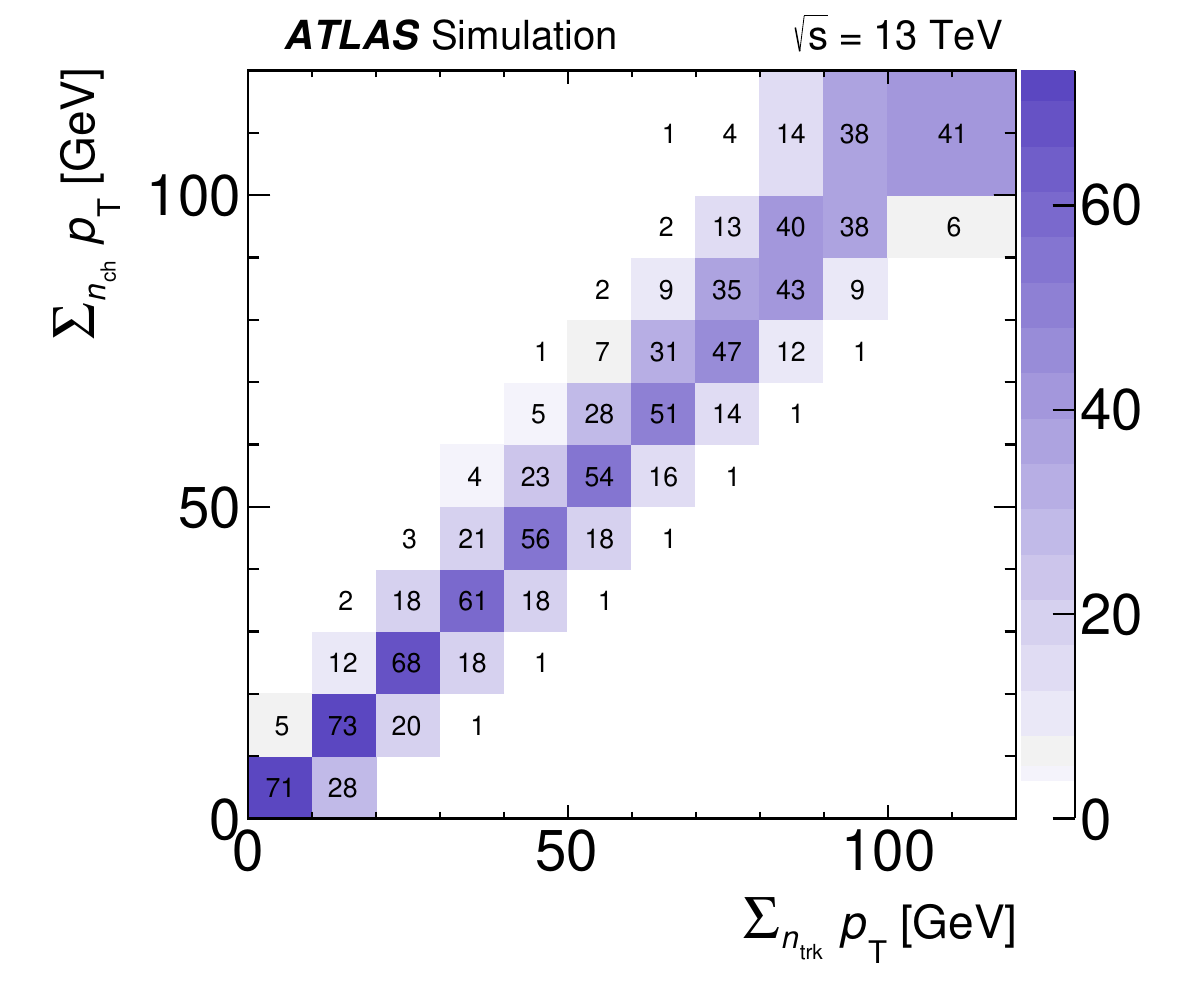}}
\subfigure[]{\includegraphics[width=0.7\textwidth]{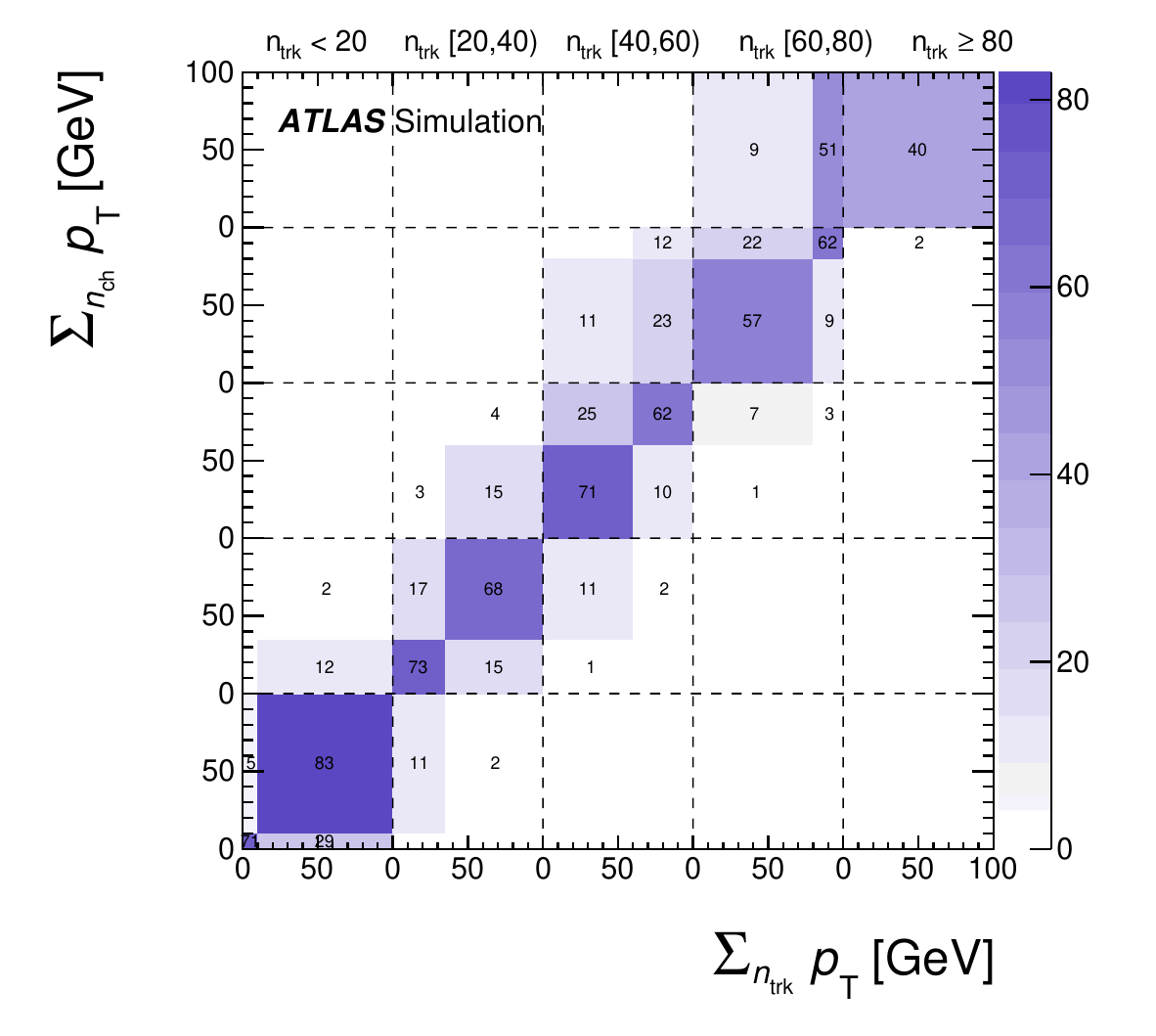}}
\caption{\label{fig:migration} Migration matrix of the observables (a) \ntrk, (b) \sumpttrk and (c) \sumpttrk in bins of \ntrk.
The values are normalised per row and shown as percentages. The numerical value of the bin is displayed if it is > 1\%.
Events beyond the $x$- or $y$-axis range are included in the last bin.}
\end{center}
\end{figure}


\section{Absolute differential cross-sections}

Absolute differential cross-sections are compared with predictions from various MC event generators and presented in \Figrange{\ref{fig:unfolded_generator_abs}}{\ref{fig:unfolding_herwig_CR_abs}}.

\begin{figure}[htpb]
\begin{center}
\subfigure[]{\includegraphics[width=0.46\textwidth]{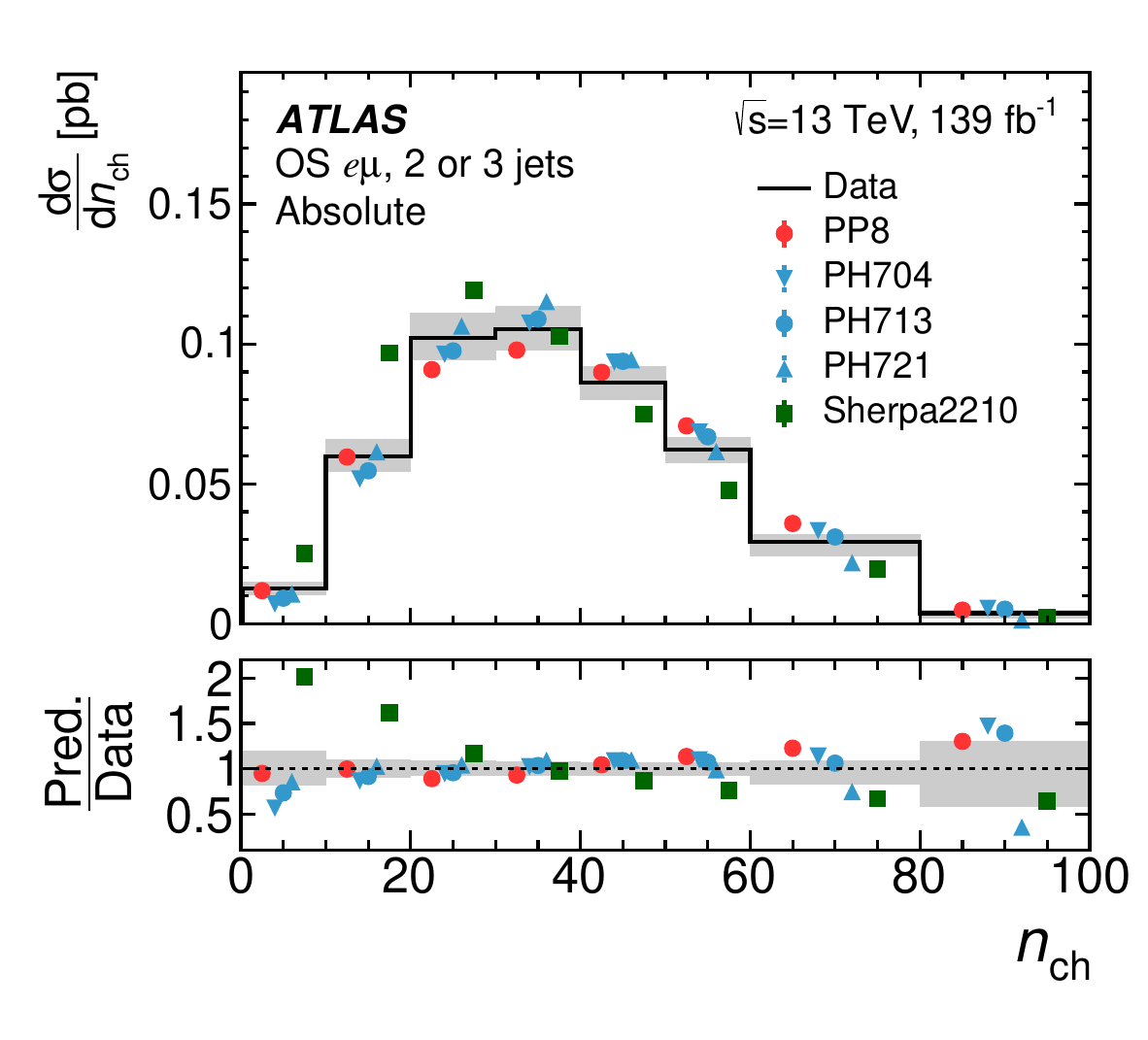}}
\subfigure[]{\includegraphics[width=0.46\textwidth]{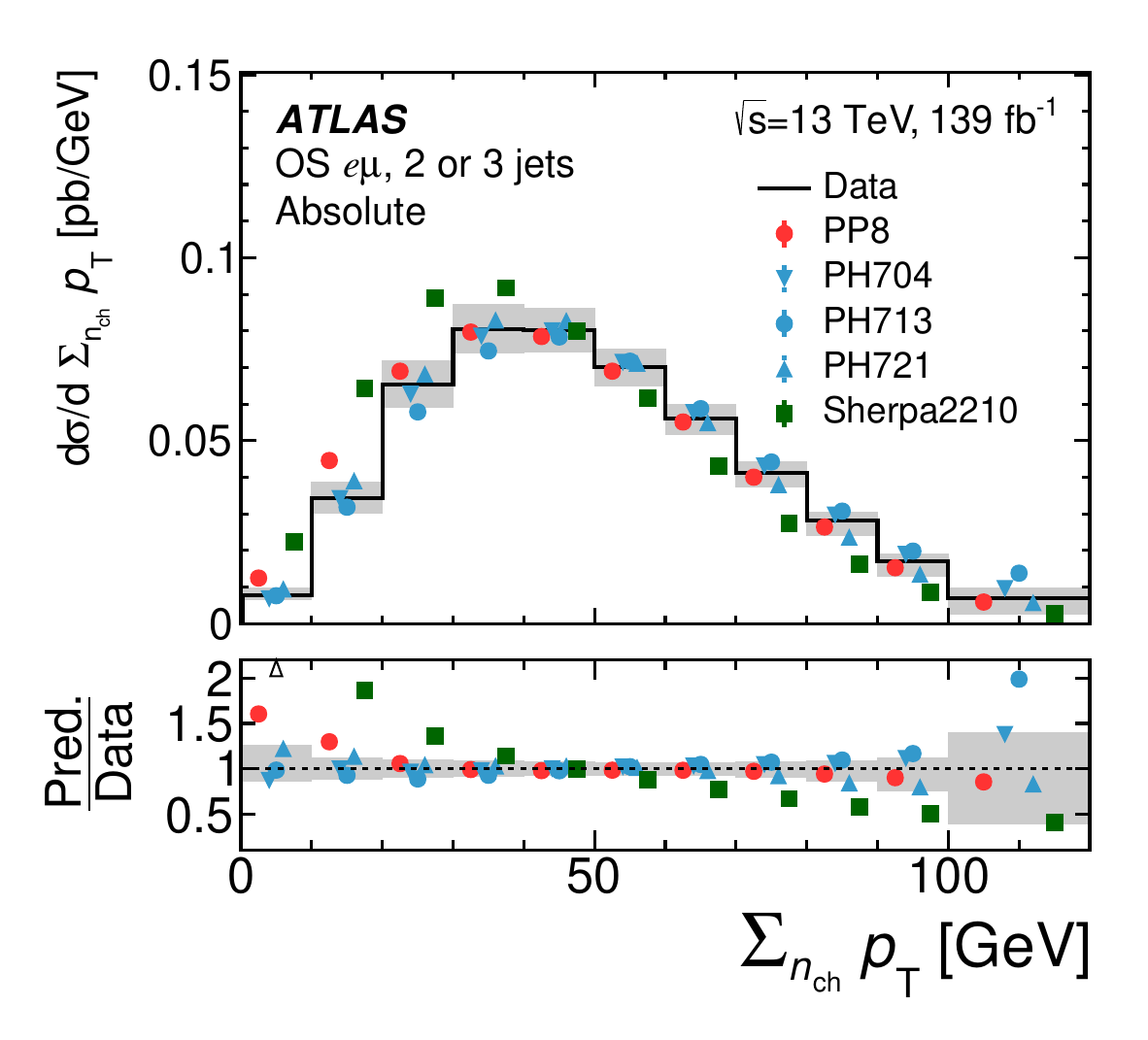}}
\subfigure[]{\hspace{-1cm}\includegraphics[width=\textwidth]{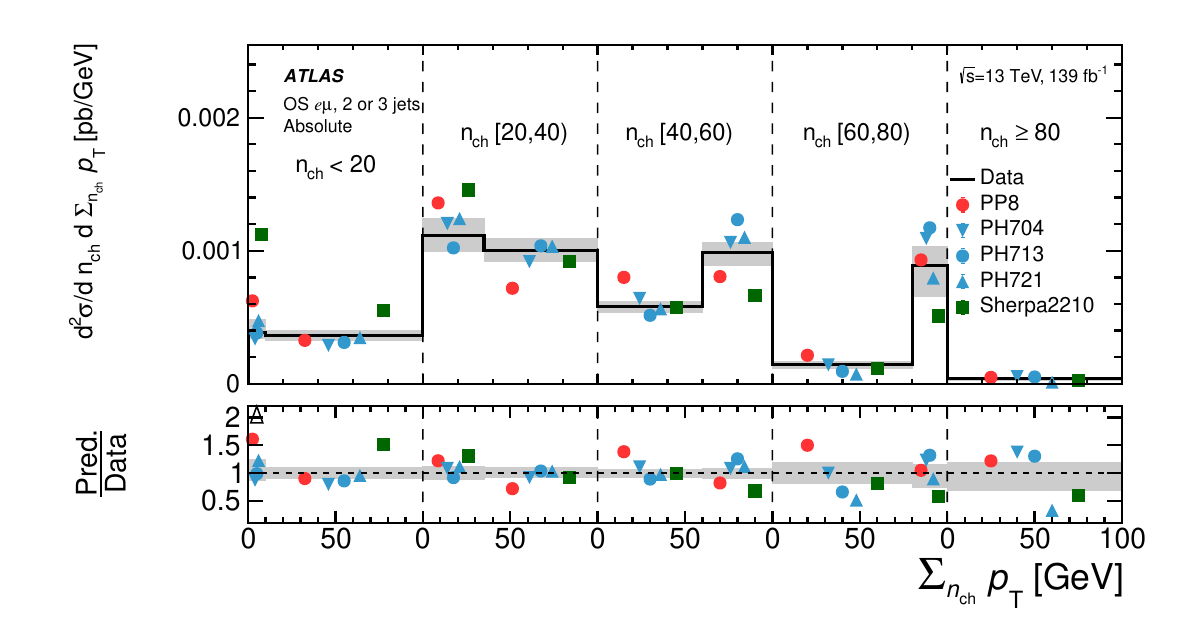}}
\caption{\label{fig:unfolded_generator_abs} Absolute differential cross-section as a function of (a) \nch, (b) \sumptch, and (c) \sumptch in bins of \nch.
Unfolded data are shown as the black line with the grey band corresponding to the total uncertainty in each bin. The results are compared with the predictions of different MC event generators. Events beyond the $x$-axis range are included in the last bin. The lower panels show the ratio of each prediction to data in each bin. The black triangular markers in the lower panels point to prediction-to-data ratio values which lie beyond the shown $y$-axis range.}
\end{center}
\end{figure}
 
\begin{figure}[htpb]
\begin{center}
\subfigure[]{\includegraphics[width=0.46\textwidth]{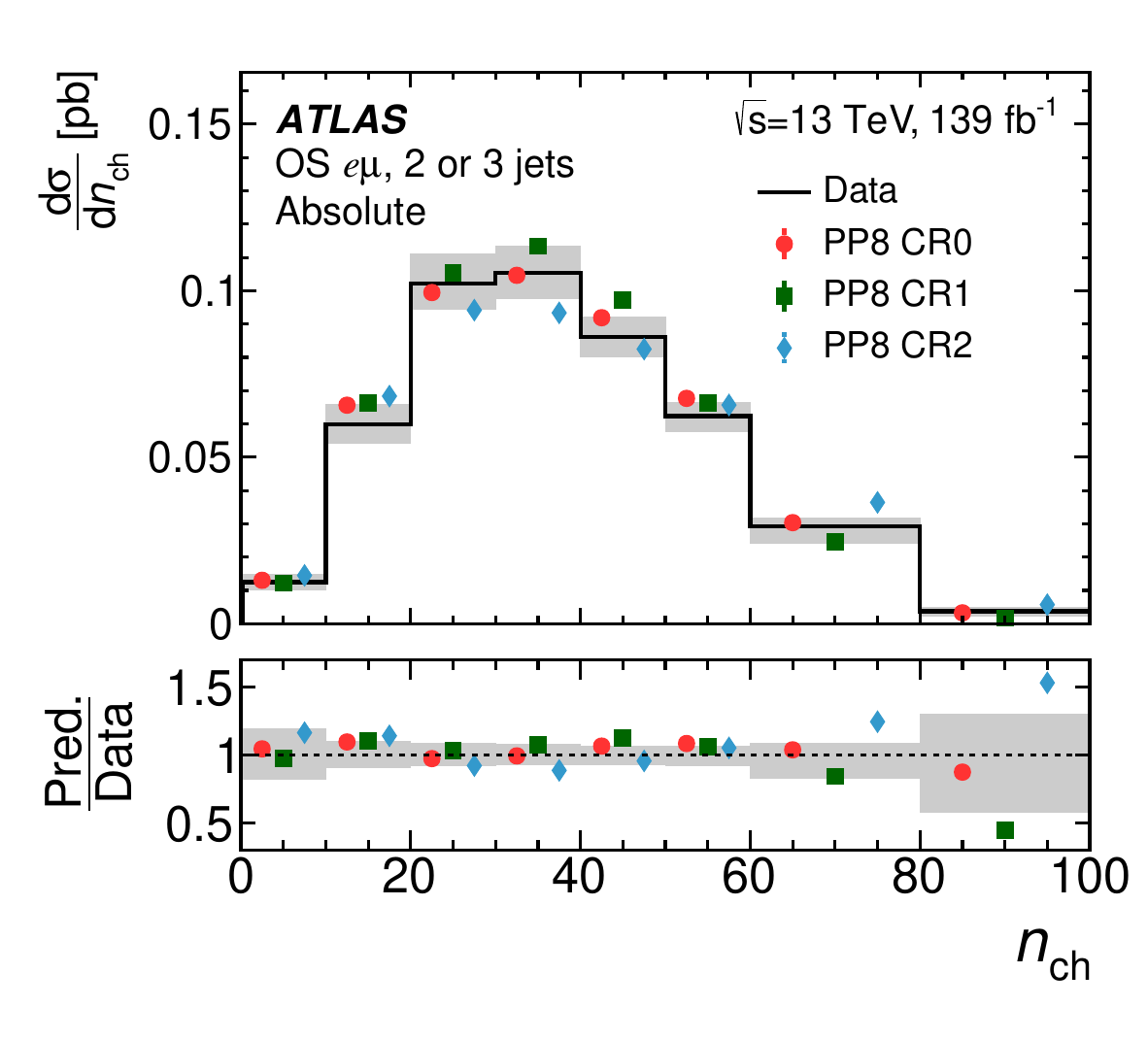}}
\subfigure[]{\includegraphics[width=0.46\textwidth]{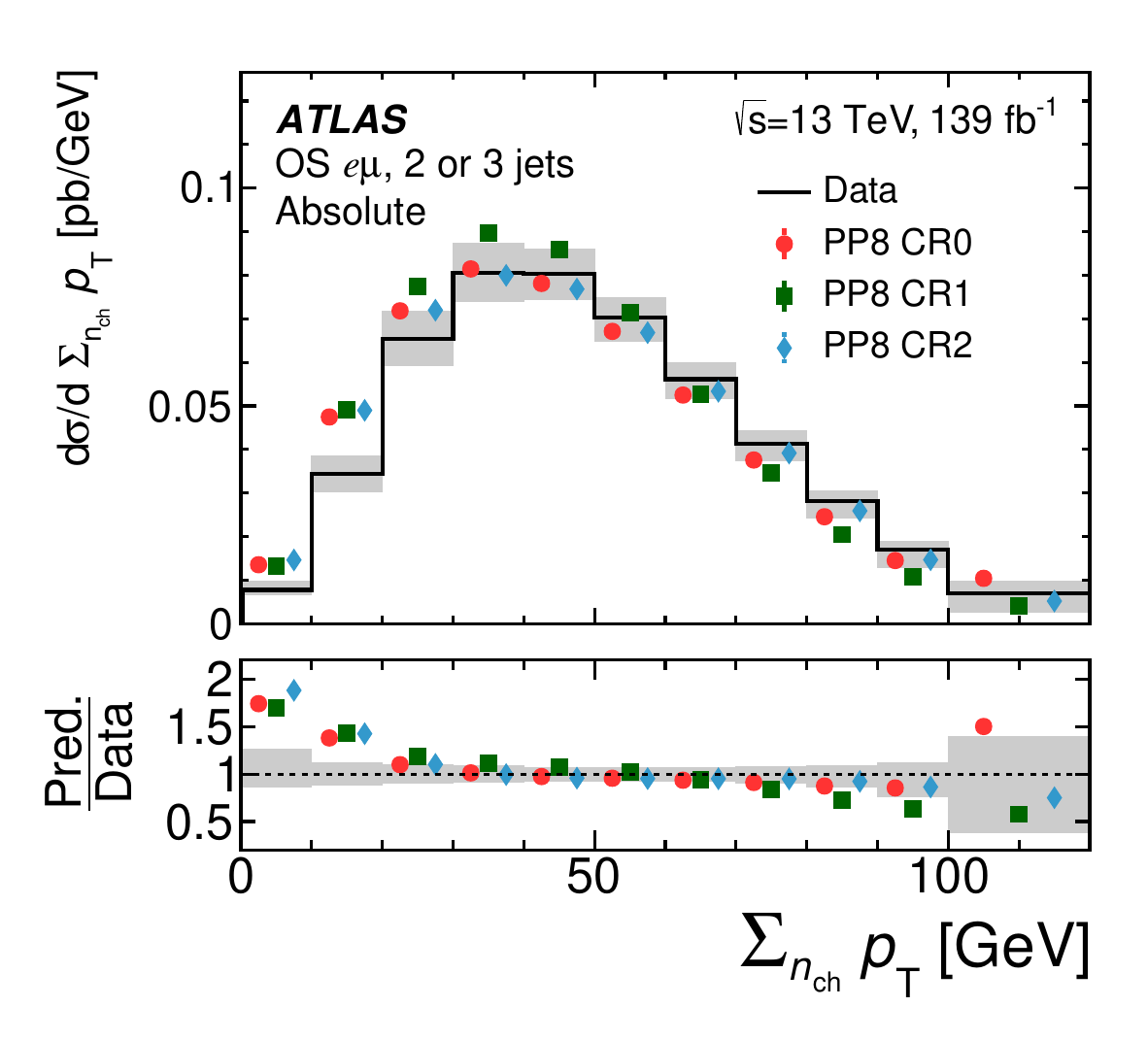}}
\subfigure[]{\hspace{-1cm}\includegraphics[width=\textwidth]{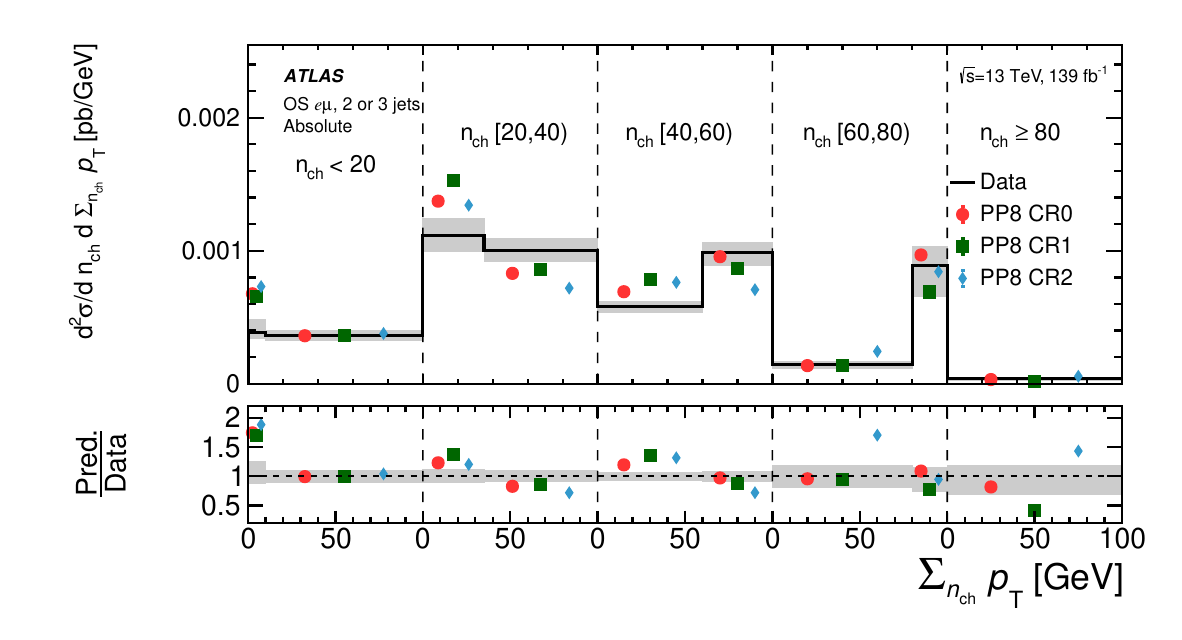}}
\caption{\label{fig:unfolded_atlas_CR_abs} Absolute differential cross-section as a function of (a) \nch, (b) \sumptch, and (c) \sumptch in bins of \nch.
Unfolded data are shown as the black line with the grey band corresponding to the total uncertainty in each bin.  The results are compared with the predictions of different CR models in \PYTHIA[8]. Events beyond the $x$-axis range are included in the last bin. The lower panels show the ratio of each prediction to data in each bin.}
\end{center}
\end{figure}

\begin{figure}[htpb]
\begin{center}
\subfigure[]{\includegraphics[width=0.46\textwidth]{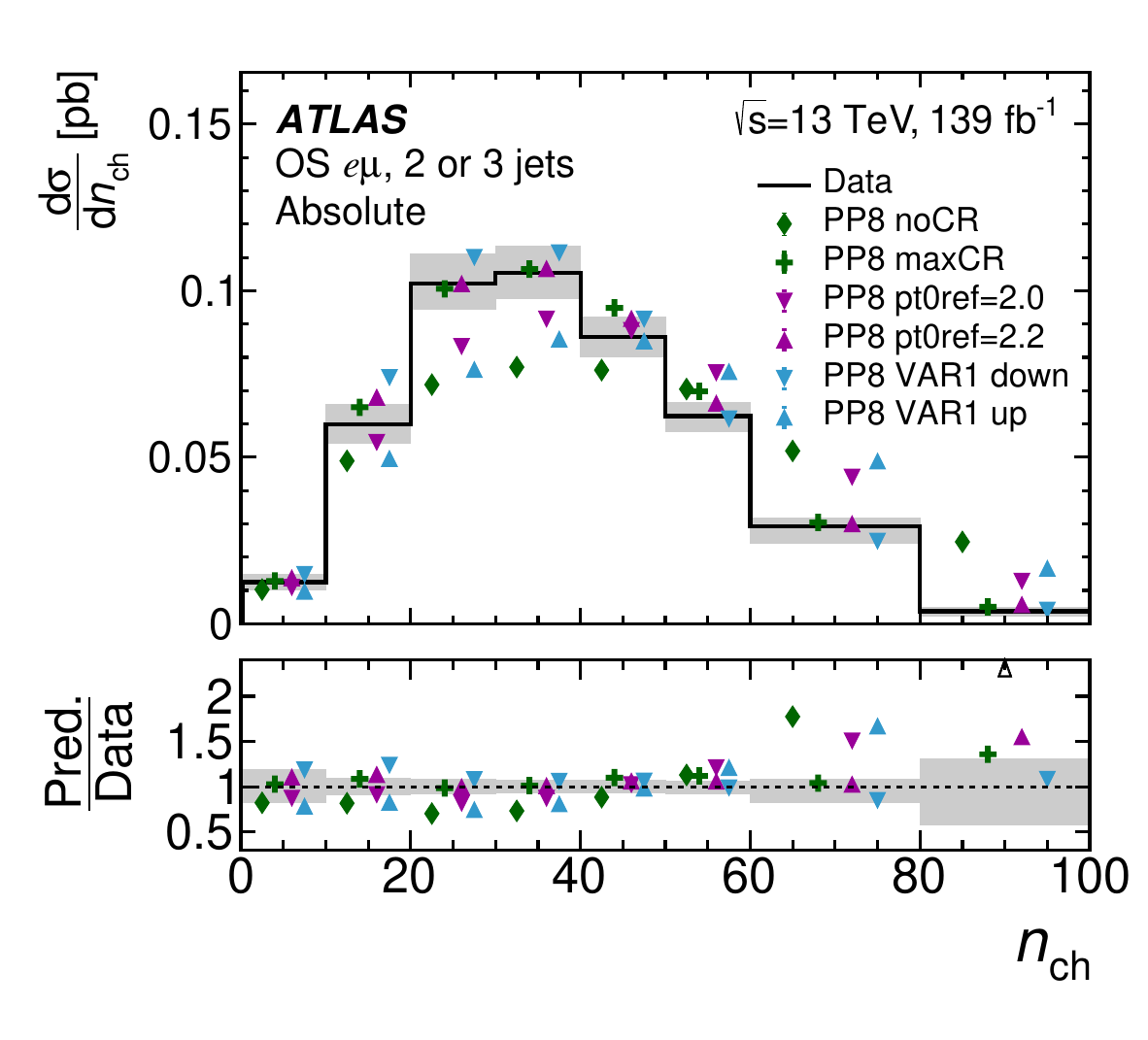}}
\subfigure[]{\includegraphics[width=0.46\textwidth]{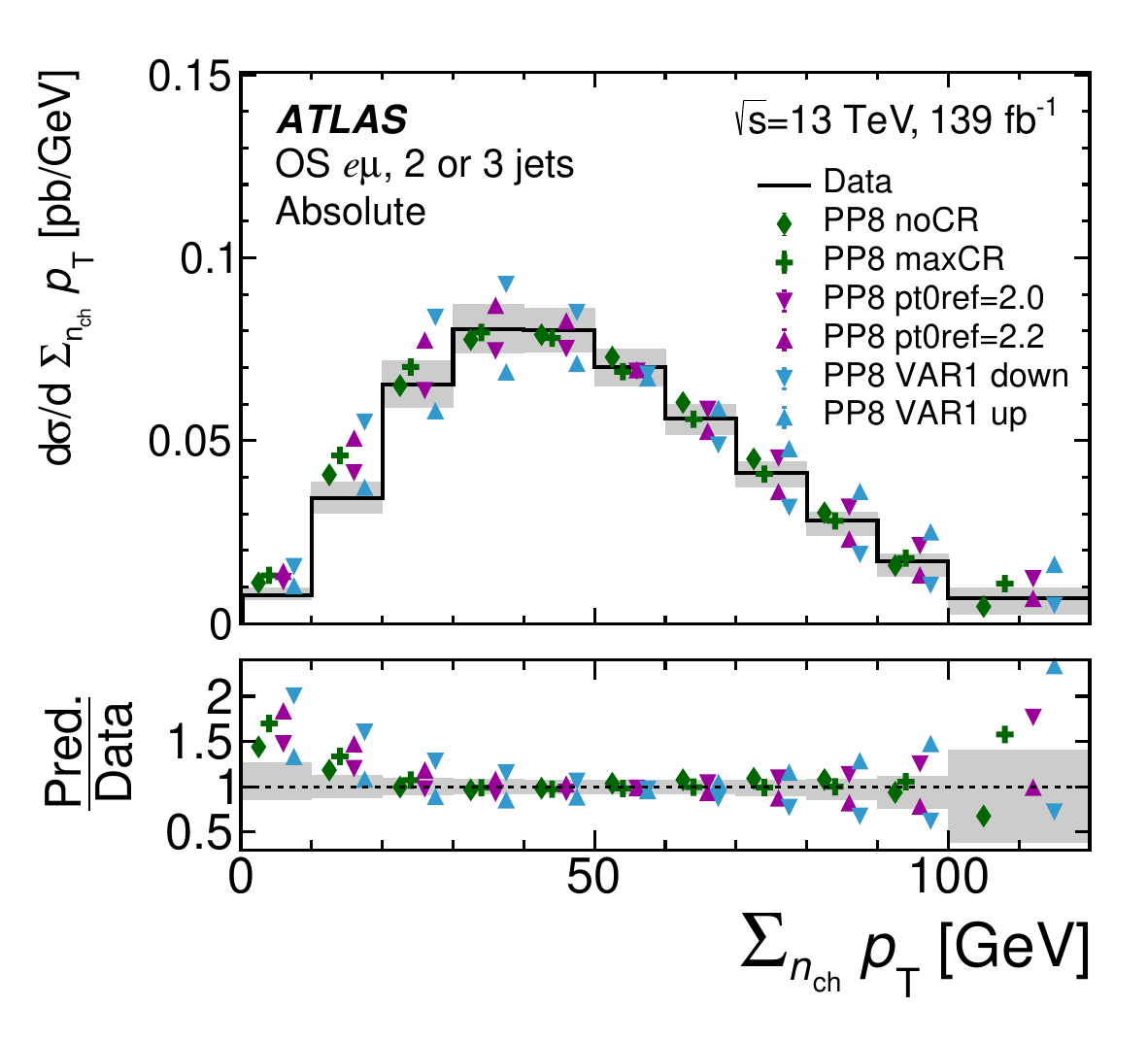}}
\subfigure[]{\hspace{-1cm}\includegraphics[width=\textwidth]{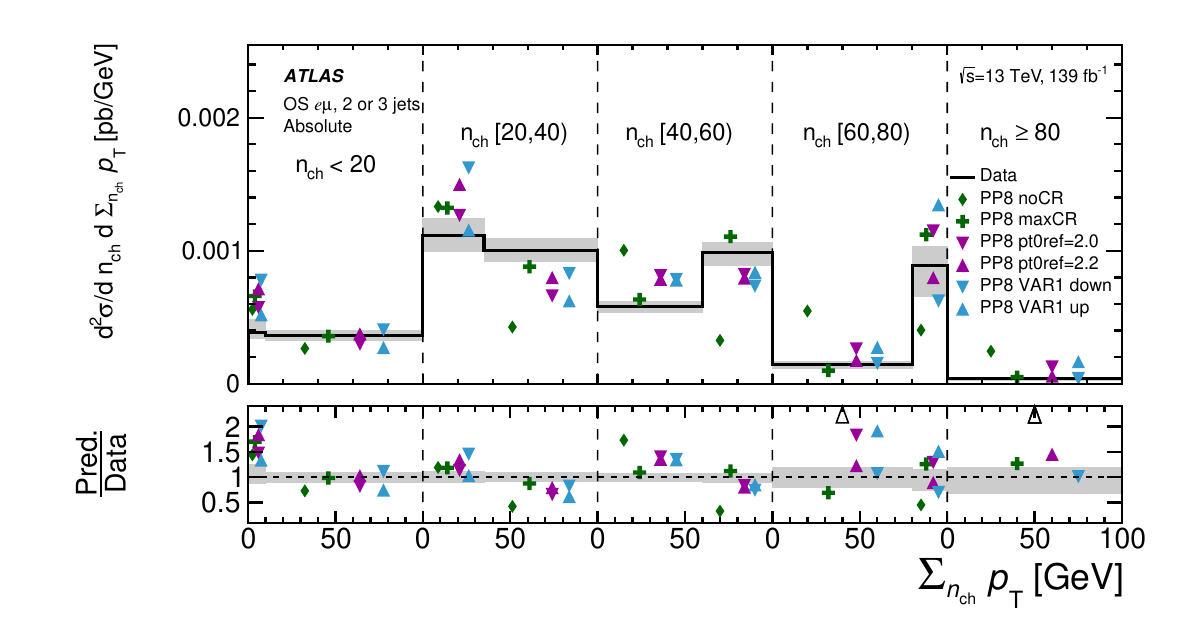}}
\caption{\label{fig:unfolding_pythia_CR_abs} Absolute differential cross-section as a function of (a) \nch, (b) \sumptch, and (c) \sumptch in bins of \nch.
Unfolded data are shown as the black line with the grey band corresponding to the total uncertainty in each bin. The results are compared with the predictions of different CR and UE parameter variations in \PYTHIA[8]. Events beyond the $x$-axis range are included in the last bin. The lower panels show the ratio of each prediction to data in each bin. The black triangular markers in the lower panels point to prediction-to-data ratio values which lie beyond the shown $y$-axis range.
}
\end{center}
\end{figure}
 
\begin{figure}[htpb]
\begin{center}
\subfigure[]{\includegraphics[width=0.46\textwidth]{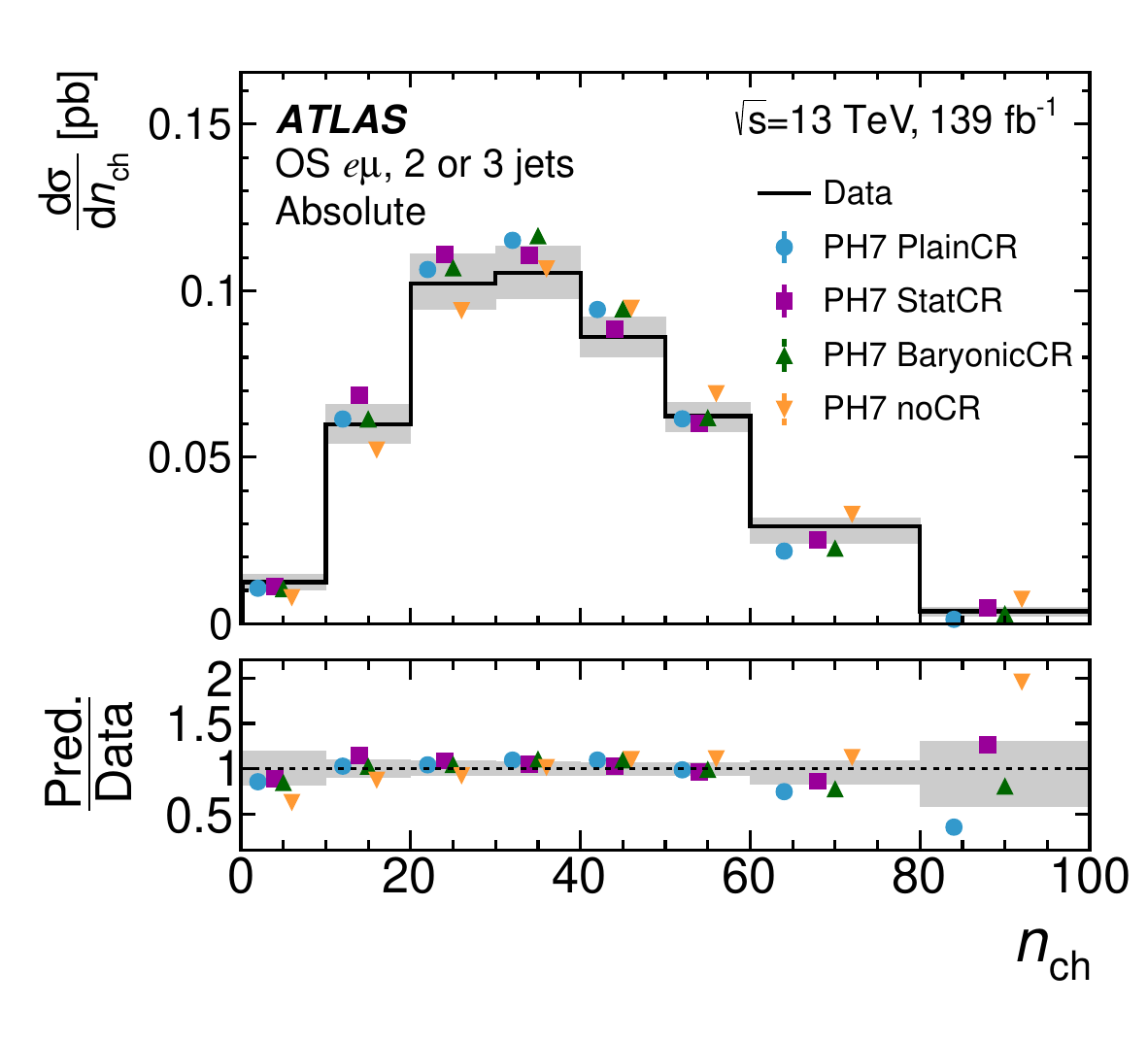}}
\subfigure[]{\includegraphics[width=0.46\textwidth]{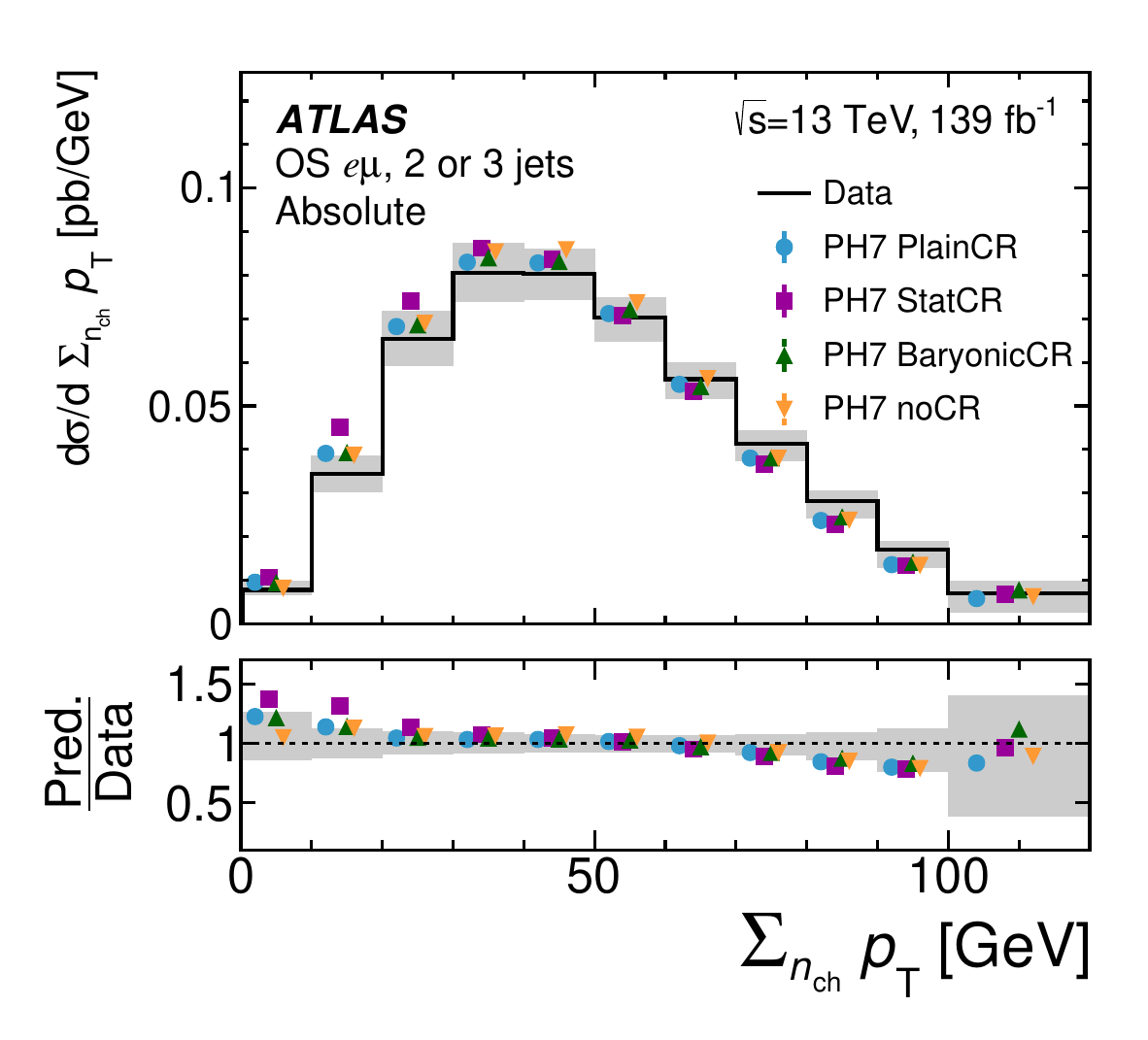}}
\subfigure[]{\hspace{-1cm}\includegraphics[width=\textwidth]{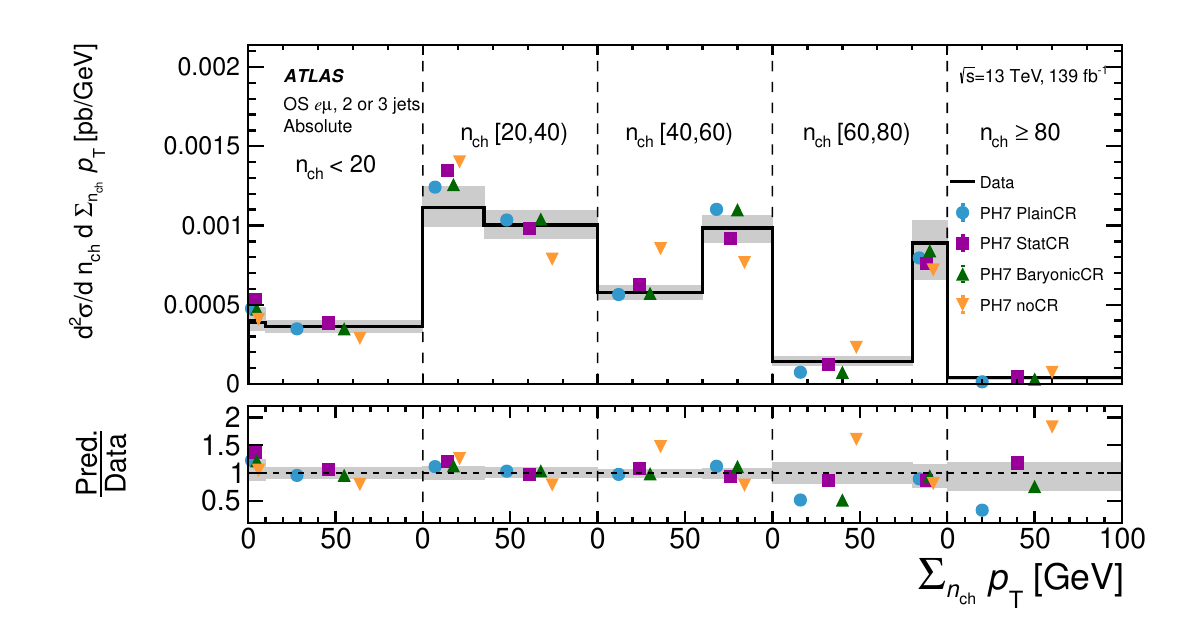}}
\caption{\label{fig:unfolding_herwig_CR_abs} Absolute differential cross-section as a function of (a) \nch, (b) \sumptch, and (c) \sumptch in bins of \nch.
Unfolded data are shown as the black line with the grey band corresponding to the total uncertainty in each bin. The results are compared with the predictions of different CR models in \HERWIG[7.2]. Events beyond the $x$-axis range are included in the last bin. The lower panels show the ratio of each prediction to data in each bin. }
\end{center}
\end{figure}


\section{Additional material for the measured distributions}

In addition to the quantitative comparisons of normalised differential cross-sections with predictions from different MC models, this appendix presents two alternative $\chi^2$ scenarios:
 
\begin{itemize}
 
\item Using only uncertainties described in~\Sect{\ref{sec:sys}}, i.e.\  modelling uncertainties are only included as non-closure uncertainties, while the extra theory uncertainties defined in~\Sect{\ref{sec:results}} are omitted.\\
This scenario is referred to as `Total' in the following tables.
 
\item Using the detector covariance matrix, and adding modelling uncertainties, scale variations in the matrix element and parton shower as well as the \hdamp variation only to the diagonal elements of the covariance matrix. This scenario is referred to as `De-correlate modelling' in the following tables.
\end{itemize}
 
The $\chi^2$ values corresponding to these two scenarios for the normalised differential cross-sections are shown in \Tab{\ref{tab:chi2_variations_norm}}, while
the $\chi^2$ values for the absolute differential cross-sections are shown in \Tab{\ref{tab:chi2_variations_absolute}}. In both tables, a clear reduction of the $\chi^2$ values can be seen in the `De-correlate modelling' scenario compared to the `Total' scenario. This clearly shows that the modelling uncertainty correlations between the bins are the reason for the large $\chi^2$ values in the case of the `Total' scenario. The results for the
two scenarios follow similar trends. The \nch measured cross-section is approximately equally well described by \PYTHIA[8.230] and \HERWIG[7], while the \sumptch measured normalised and absolute differential cross-sections are in much better agreement with \HERWIG[7].
With increasing \HERWIG[7] version number, the \nch distribution agreement gets worse, while the \sumptch agreement with remains at a similar level.
The distribution of \twod is described best by the three versions of \HERWIG[7], not so well by \PYTHIA[8.230], and worst by \SHERPA[2].
Overall, the best agreement is achieved using \HERWIG[7.0.4].

\begin{table}[htbp]
\begin{center}
\caption{The $\chi^2$ and NDF for the measured normalised differential cross-sections obtained by comparing the different predictions with the unfolded data.
The values corresponding to the `Total' scenario are given first, followed by the `De-correlate modelling' scenario.}
\label{tab:chi2_variations_norm}
\begin{tabular}{lcccc}
\toprule
Observabls         		   		&   \nch  		&   \sumptch 	&   Global(\nch,\sumptch)		&  \sumptch in bins of \nch                       \\
NDF							& ~~7		& 10			& ~~17					& ~~~~8       \\
\hline
Generator set-up         	        				&   \multicolumn{4}{c}{$\chi^2$}  \\
\hline
\hline
Total &&&& \\
\POWPY[8.230]     				& 73      		&  293~~    	& 573					& 307 	    \\
CR0          					& 69	  		&  296~~    	& 791					& 254   	\\
CR1          					& 96     		&  131~~    	& 635					& 610   	\\
CR2          					& 151~~    	&  570~~    	& 880					& 424   	\\
\POWHER[7.0.4]     				& 80    		&  17     		& 206					& 184  	    \\
\POWHER[7.1.3]     				& 84    		&  44     		& 241					& ~~61  	    \\
\POWHER[7.2.1]     				& 91    		&  27     		& 380					& 405  	    \\
\POWHER Baryonic CR     		& 107~~		&  22  		& 288					& 118       \\
\POWHER Stat CR         			& 23			&  76  		& 145					& 117       \\
\SHERPA[2.2.10]		   	   	& 459~~		& 950~~     	& 1409~~					& 1453~~ 	    \\
\hline
De-correlate modelling &&&& \\
\POWPY[8.230] 	   			& 24			& 18			& ~~63					& 155	    \\
CR0       						& ~~8		& 37			& ~~53					& ~~53			\\
CR1       						& 18			& 23			& ~~49					& 107		\\
CR2       						& 47    		& 40			& 106					& 125		\\
\POWHER[7.0.4]    				& 19			& ~~2		& ~~21					& ~~28		\\
\POWHER[7.1.3]     				& 14			& 11			& ~~20					& ~~27		\\
\POWHER[7.2.1] 	   			& 32			& ~~5		& ~~37					& ~~19		\\
\POWHER Baryonic CR     		& 24			& ~~5		& ~~30					& ~~16		\\
\POWHER Stat CR         			& ~~7  		& 13			& ~~18					& ~~10		\\
\SHERPA[2.2.10] 		  	 	& 148~~		& 155~~		& 250					& 137	\\
\bottomrule
\end{tabular}
 
\end{center}
\end{table}
 
\begin{table}[htbp]
\begin{center}
\caption{The $\chi^2$ and NDF for the measured absolute differential cross-sections obtained by comparing the different predictions with the unfolded data.
The nominal values are first shown followed by the `Total' scenario and the `De-correlate modelling' scenario.}
\label{tab:chi2_variations_absolute}
\begin{tabular}{lcccc}
\toprule
Observabls         		   		&   \nch  		&   \sumptch 	&   Global(\nch,\sumptch)		&  \sumptch in bins of \nch                       \\
NDF							& ~~8		& 11			& ~~19					& ~~9       \\
\hline
Generator set-up         	        				&   \multicolumn{4}{c}{$\chi^2$}  \\
\hline
\hline
\POWPY[8.230]     				& 88			& 121~~		& 486					& 270~~	\\
CR0    						& 88			& 143~~		& 738					& 153~~	\\
CR1    						& 176~~		& 77			& 702					& 216~~	\\
CR2    						& 96			& 209~~		& 468					& 279~~	\\
\POWHER[7.0.4]     				& 56			& 22			& 162					& 45		\\
\POWHER[7.1.3]     				& 80			& 44			& 198					& 54		\\
\POWHER[7.2.1]     				& 128~~		& 33			& 342					& 99		\\
\POWHER Baryonic CR     		& 115~~		& 24			& 262   					& 36     \\
\POWHEG+\HERWIG Stat CR      	& 30			& 55			& 142   					& 54     \\
\SHERPA[2.2.10]	           		& 144~~		& 275~~		& 414					& 144~~	\\
\hline
Total &&&& \\
\POWPY[8.230]     				& 97      		& 327~~     	& 606					& 325~~      \\
CR0   						& 101~~		& 337~~    	& 888					& 278~~    \\
CR1   						& 169~~		& 153~~    	& 762					& 642~~    \\
CR2   						& 176~~		& 643~~    	& 899					& 525~~   \\
\POWHER[7.0.4]     				& 98      		& 21      		& 227					& 191~~     \\
\POWHER[7.1.3]     				& 106~~     	& 52      		& 262					& 72       \\
\POWHER[7.2.1]     				& 128~~     	& 32     		& 402					& 420~~     \\
\POWHER Baryonic CR     		& 139~~      	& 25			& 307					& 125~~       \\
\POWHER Stat CR         			& 30        	   	& 90      		& 163					& 156~~       \\
\SHERPA[2.2.10]		   	   	& 536~~     	& 1093~~~~   	& 1615~~ 					& 1624~~~~     \\
\hline
De-correlate modelling &&&& \\
\POWPY[8.230] 	   			& ~~8  		& ~~9  		& ~~22					& 72	\\
CR0   						& ~~2  		& 17			& ~~23					& 28	\\
CR1   						& ~~8		& 22			& ~~31					& 53	\\
CR2   						& 11			& 19			& ~~35					& 83	\\
\POWHER[7.0.4]     				& ~~8		& ~~1		& ~~~~9					& 11	\\
\POWHER[7.1.3]     				& ~~5		& ~~6		& ~~10					& 19	\\
\POWHER[7.2.1] 	   			& 10			& ~~4		& ~~13					& 20	\\
\POWHER Baryonic CR     		& ~~7  		& 10			& ~~11					& 15	\\
\POWHER Stat CR         			& ~~4		& 10			& ~~15					& ~~9	\\
\SHERPA[2.2.10] 		   		& 54			& 100~~		& 127					& 87	\\
\bottomrule
\end{tabular}
 
\end{center}
\end{table}


\clearpage
\printbibliography
 
\clearpage
 
\begin{flushleft}
\hypersetup{urlcolor=black}
{\Large The ATLAS Collaboration}

\bigskip

\AtlasOrcid[0000-0002-6665-4934]{G.~Aad}$^\textrm{\scriptsize 101}$,
\AtlasOrcid[0000-0002-5888-2734]{B.~Abbott}$^\textrm{\scriptsize 119}$,
\AtlasOrcid[0000-0002-7248-3203]{D.C.~Abbott}$^\textrm{\scriptsize 102}$,
\AtlasOrcid[0000-0002-1002-1652]{K.~Abeling}$^\textrm{\scriptsize 55}$,
\AtlasOrcid[0000-0002-8496-9294]{S.H.~Abidi}$^\textrm{\scriptsize 29}$,
\AtlasOrcid[0000-0002-9987-2292]{A.~Aboulhorma}$^\textrm{\scriptsize 35e}$,
\AtlasOrcid[0000-0001-5329-6640]{H.~Abramowicz}$^\textrm{\scriptsize 150}$,
\AtlasOrcid[0000-0002-1599-2896]{H.~Abreu}$^\textrm{\scriptsize 149}$,
\AtlasOrcid[0000-0003-0403-3697]{Y.~Abulaiti}$^\textrm{\scriptsize 116}$,
\AtlasOrcid[0000-0003-0762-7204]{A.C.~Abusleme~Hoffman}$^\textrm{\scriptsize 136a}$
% ...
\bigskip
\\

% ...
$^{29}$Physics Department, Brookhaven National Laboratory, Upton NY; United States of America.\\
% ...
$^{35}$$^{(a)}$Facult\'e des Sciences Ain Chock, R\'eseau Universitaire de Physique des Hautes Energies - Universit\'e Hassan II, Casablanca;$^{(b)}$Facult\'{e} des Sciences, Universit\'{e} Ibn-Tofail, K\'{e}nitra;$^{(c)}$Facult\'e des Sciences Semlalia, Universit\'e Cadi Ayyad, LPHEA-Marrakech;$^{(d)}$LPMR, Facult\'e des Sciences, Universit\'e Mohamed Premier, Oujda;$^{(e)}$Facult\'e des sciences, Universit\'e Mohammed V, Rabat;$^{(f)}$Institute of Applied Physics, Mohammed VI Polytechnic University, Ben Guerir; Morocco.\\
% ...
$^{55}$II. Physikalisches Institut, Georg-August-Universit\"{a}t G\"ottingen, G\"ottingen; Germany.\\
% ...
$^{101}$CPPM, Aix-Marseille Universit\'e, CNRS/IN2P3, Marseille; France.\\
$^{102}$Department of Physics, University of Massachusetts, Amherst MA; United States of America.\\
% ...
$^{116}$Department of Physics, New York University, New York NY; United States of America.\\
% ...
$^{119}$Homer L. Dodge Department of Physics and Astronomy, University of Oklahoma, Norman OK; United States of America.\\
% ...
$^{136}$$^{(a)}$Departamento de F\'isica, Pontificia Universidad Cat\'olica de Chile, Santiago;$^{(b)}$Millennium Institute for Subatomic physics at high energy frontier (SAPHIR), Santiago;$^{(c)}$Instituto de Investigaci\'on Multidisciplinario en Ciencia y Tecnolog\'ia, y Departamento de F\'isica, Universidad de La Serena;$^{(d)}$Universidad Andres Bello, Department of Physics, Santiago;$^{(e)}$Instituto de Alta Investigaci\'on, Universidad de Tarapac\'a, Arica;$^{(f)}$Departamento de F\'isica, Universidad T\'ecnica Federico Santa Mar\'ia, Valpara\'iso; Chile.\\
% ...
$^{149}$Department of Physics, Technion, Israel Institute of Technology, Haifa; Israel.\\
$^{150}$Raymond and Beverly Sackler School of Physics and Astronomy, Tel Aviv University, Tel Aviv; Israel.\\
% ...
$^{a}$ Also Affiliated with an institute covered by a cooperation agreement with CERN.\\
$^{b}$ Also at Borough of Manhattan Community College, City University of New York, New York NY; United States of America.\\
$^{c}$ Also at Bruno Kessler Foundation, Trento; Italy.\\
$^{d}$ Also at Center for High Energy Physics, Peking University; China.\\
$^{e}$ Also at Centro Studi e Ricerche Enrico Fermi; Italy.\\
$^{f}$ Also at CERN, Geneva; Switzerland.\\
% ...

\end{flushleft}


\end{document}